# ICORE: Image Co-addition with Optional Resolution Enhancement

## Algorithms and Software Usage

**Version 3.8.2, 05/13/2013**

*Frank Masci, IPAC - Caltech*

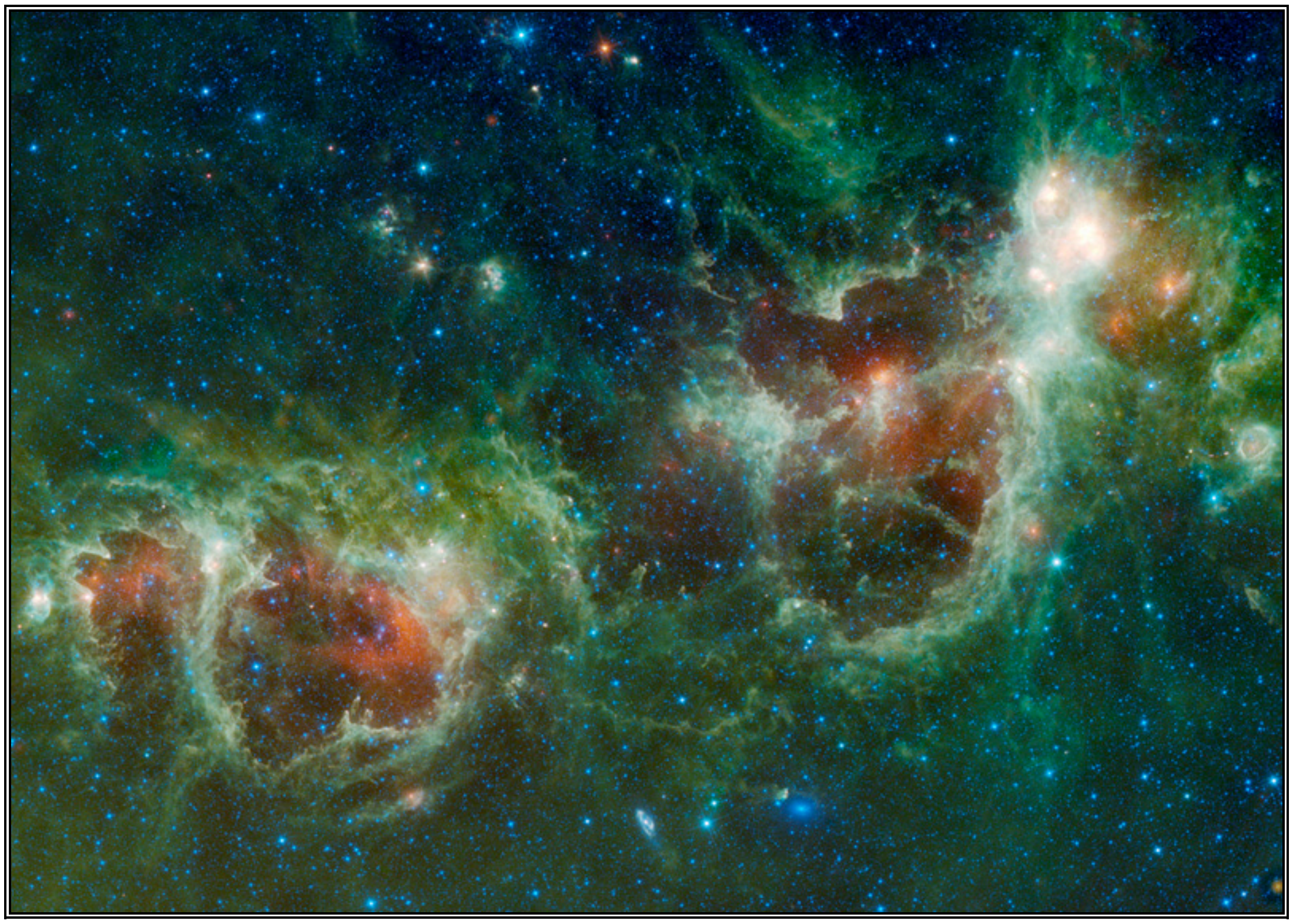

*The Heart & Soul Nebulae in Cassiopeia observed by WISE (Feb 2010) and constructed using ICORE*

*Disclaimer*

*There is no warranty that the software described herein is free of bugs. This includes proper graceful program termination on bad input data. Please report all bugs and suggestions for future improvement to the author.*

*California Institute of Technology*

*fmasci@caltech.edu*
*http://web.ipac.caltech.edu/staff/fmasci/home/icore.html*

## Revision History

| Date | Version | Author | Description |
|---|---|---|---|
| August 20, 2007 | 1.0 | Frank Masci | Initial version. |
| November 1, 2007 | 1.1 | Frank Masci | Optimized when creating a simple (unHIRES'd) co-add. |
| November 2, 2007 | 1.2 | Frank Masci | New command-line inputs: -rf, -ct, -if. |
| November 29, 2007 | 1.3 | Frank Masci | Made specification of output co-add pixel scale flexible. |
| February 12, 2008 | 1.4 | Frank Masci | * Allow specification of output *square* co-add pixel scales in absolute arcsec units (-pa);<br>* Read in bad-pixel FITS image masks with up to 32-bit integers per pixel;<br>* Allow specification of a bit-string template for conditional flagging of pixels (-m);<br>* Replaced bilinear interpolation option with a more robust area-overlap weighting method (-if). |
| April 16, 2008 | 1.5 | Frank Masci | * Added option (-sc) to perform "overlap-area" weighted co-add;<br>* Option to generate standard deviation co-add of input pixel stacks (-o4) for -sc case. |
| July 10, 2008 | 1.6 | Frank Masci | *Replaced cfitsio header reading function with ffhdr2str function (this properly stores all WCS and distortion keys).<br>* Implemented auto-tiling option to support memory management for outlier detection.<br>* Implemented partitioning of frame lists to support memory management for outlier detection.<br>* Implemented more robust |
| | | | frame background estimation when extended structure present. |
| August 19, 2008 | 1.7 | Frank Masci | Included ringing suppression algorithm to support HiRes. |
| September 1, 2008 | 1.8 | Frank Masci | Implemented more robust frame background estimation when extended structure present. To support b'gnd matching and HiRes. |
| September 5, 2008 | 1.9 | Frank Masci | * Included -mcmprod switch to generate CFV and intensity cell images at each MCM iteration.<br>* Also made box sizes for bmatch step more robust and avoid NaNs when computing extended source metrics. |
| September 8, 2008 | 2.0 | Frank Masci | Added AWOD pad parameter: –pb_odet. |
| October 25, 2008 | 2.1 | Frank Masci | Generated QA meta table and compute metrics. |
| November 17, 2008 | 2.2 | Frank Masci | * Implemented QA plot files.<br>* Included –wf option to control inverse variance weighting. When on, leads to flux under-estimation for for bright srcs. |
| February 13, 2009 | 2.3 | Frank Masci | * Changed some for-loop notations in *partitionme* subroutine since syntax appears to conflict with PDL.<br>* Added subroutine to randomize input frame/mask lists prior to partitioning for outlier detection if "-partition" is set. |
| February 23, 2009 | 2.4 | Frank Masci | * Combined mask sub-mosaics into final mask if partitioning was triggered; update MAGZP and MAGZPUNC keys in input frame headers with global co-add values if tmatch performed.<br>* Update MAGZP and MAGZPUNC keys in input frame headers with global co-add values if tmatch performed |
| May 20, 2009 | 2.6 | Frank Masci | Included new CL options -ip_odet and -is_odet as per AWOD vsn 1.6. |
| June 26, 2009 | 2.7 | Frank Masci | Compute median of input MAGZPUNCs instead of mean to protect against outliers; don't update MAGZPUNC in modified intensity frames since formally incorrect. |
| July 7, 2009 | 2.8 | Frank Masci | Included top-hat mosaic creation flag for use as MCM prior (-fp_coad); added -o9 output (cellcors from all iters) from first pass run of awaic, generated if -mcmprod option specified. |
| July 11, 2009 | 3.0 | Frank Masci | Include -ns_odet CL param. |
| July 12, 2009 | 3.1 | Frank Masci | Included subroutine to expand or blanket outlier regions. New CL params: -nei_odet, -nsz_odet, -exp_odet, -expodet |
| September 30, 2009 | 3.2 | Frank Masci | Added -ms_coad and -om_coad options to create mosaic mask and tag saturated and bad/unsaturated pixels. |
| October 1, 2009 | 3.3 | Frank Masci | Included NaN filtering in plane fitting for background matching. |
| October 20, 2009 | 3.4 | Frank Masci | Implemented drizzle option: -d_coad to match awaic vsn 4.5 |
| November 2, 2009 | 3.5 | Frank Masci | Extensive HiRes functionality, new CL inputs: -h_coad; -oi_coad; -of_coad; -snu_coad; -snc_coad; -siggrid; also changed coadd exec to awaico. |
| November 2, 2009 | 3.6 | Frank Masci | Fix error-checking bug whereby creation of mask mosaic for n=1 is bypassed if any hires-related product or functionality is desired. Also added verbosity to SVB sub. |
| December 24, 2009 | 3.7 | Frank Masci | Fixed OutlierHist.svg QA plot when no outliers detected. Const. values caused crash. |
| January 18, 2010 | 3.7.1 | Frank Masci | * Renamed framecoadderpub to awaic, must never build into global bin directory. |
| | | | * Filter NaNs before compute block median for SVB. |
| February 13, 2010 | 3.7.2 | Frank Masci | Fixed cases when all int pixels of a QA partition are NaN'd and metrics croak. |
| March 6, 2010 | 3.7.3 | Frank Masci | Changed bmatch from switch to binary argument for easier calling in csh script. |
| March 11, 2010 | 3.7.4 | Frank Masci | * added gausize, gausigm and flxbias as 1\|0 binary CL options.<br>* added –bgrid param to compute robust frame background when extended structure detected in bmatch. |
| March 20, 2010 | 3.7.5 | Frank Masci | Use more robust header-keyword PDL reading function for NAXIS1,2 (~line 2879 in awaic script) |
| April 15, 2010 | 3.7.6 | Frank Masci | * removed linking against sqllite macros.<br>* fixed truncation bug with CD matrix in FITS headers<br>* new awod v1.8 with new params -h, -k, -q.<br>* included PowFit2D C++ module for bckgnd matching.<br>* new script params: -h_odet, -k_odet, -q_odet, -mg_odet, -tn_odet, -tg_odet, -tsat_odet, -crep_coad, -cmin_coad, -bmeth, -order, -bfgrid, -clsig.<br>* optimized WISE processing params in test scripts.<br>* added new bigger W4 PSF for HiRes. Original (small) PSF still available.<br>* new test frames for IC342 in test dir.<br>* changed nmaxodet from 150 to 200 in all test scripts.<br>* included param -ms_coadd in HiRes test scripts to support saturated frame flagging. |
| August 16, 2010 | 3.7.7 | Frank Masci | * moved the frame-background estimation and subtraction step |
| | | | for HiRes to after outlier rejection<br>* fixed sporadic bug when combining sublist mask mosaics from odet<br>* fixed unc rescaling bug for cases when median unc is computed from image with NaNs and comes out zero. Fix is to force scaling factor to 1<br>* fixed anomaly when cell-imge from HiRes -flxbias processing has NaNs after 1st pass. NaNs are filled in with median so second pass of awaic that uses cell image as prior won't choke<br>* if -tmatch not specified, explicitly copy uncert frames to outdir and make new image list to avoid original list being updated by flagframes routine downstream |
| May 5, 2012 | 3.8.0 | Frank Masci | * renamed AWAIC to ICORE!<br>* new command-line params: -reimg,-refac,-edgw, -modfilt<br>* new background matching method (-bmeth=2)<br>* faster frame-background estimation; option to use mode filter instead of median filter<br>* updated modules so can handle gzipped input frames<br>* redefined units of -gausize and –gausigm<br>* streamlined all test scripts<br>* use 'small' PSFs in WISE HiRes test scripts for speed |
| Jan 4, 2013 | 3.8.1 | Frank Masci | * multithreaded CPU-intensive tasks for co-addition and HiRes'ing. Yea! Now runs faster on multi-core processors |
| | | | * added functionality to co-add and/or HiRes in the reference frame of a moving object, with example scripts<br>* added ability to use a prior image (i.e., with higher resolution from another wavelength) to assist with HiRes processing. Also included options to regularize input prior before use<br>* made background matching method (option -bmeth=2) more robust against bad pixels<br>* better checking and regularization of offset solutions obtained from bckgnd matching using –bmeth=2 option<br>* pick reference image for bckgnd matching using –bmeth=2 option as that whose median level is equal to closest to median levels of all input images<br>* made keyword names of photometric zero-points to use from input FITS headers user-specifiable<br>* more robust rescaling of output uncertainty images<br>* added new example/test script to HiRes Herschel SPIRE 250 and 500μm maps |
| May 13, 2013 | 3.8.2 | Frank Masci | * included icore parameter "–arbconst" to add constant to image inputs to ensure internal positivity for HiRes'ing, then removed at the end<br>* don't write photometric ZP info to output products if an input image was missing its ZP. Otherwise, outputs will be |
| | | | wrong due to bad throughput matching. Warning given if so<br><br>* don't use zero-valued CFV-derived (HiRes) pixel uncertainties when rescaling these using the intensity RMS. Otherwise, output CFV uncerts could be biased<br><br>* only read and use FPALOCX,Y keywords from PRF headers if multiple (FPA-dependent) PRFs are actually provided in input list. For a single input PRF, these keywords are now not needed<br><br>* added: "double ** T_DOUBLEP" to src/Astro-WCS-LibWCS-0.93/typemap so get clean build on some Ubuntu systems<br><br>* suppress building of unnecessary PDL fortran library: "minuitlib"<br><br>* made bckgnd overlap matching method 2 (bmeth=2) more robust by (i) redefining internal reference image; (ii) better checking and regularization of internal offset solutions; (iii) ensure positivity in frame levels after solutions applied to not wreck havoc with outlier rejection (under ip_odet=1) and HiRes downstream. Also retuned "–offtol" parameter |

# Table of Contents

# 1. INTRODUCTION

ICORE is a generic co-addition, mosaicking, and resolution enhancement (HiRes) tool for creating science quality products from image data in FITS format and with World Coordinate System (WCS) information following the FITS-WCS standard in astronomy. Distortion, if present, is expected to follow the SIP (Simple Imaging Polynomial) convention. ICORE was written to support the creation of a digital Image Atlas from exposures acquired with the Wide-field Infrared Survey Explorer (WISE), but is generic enough for use on any astronomical image data. The software modules are written in ANSI-compliant C and wrapped into a Perl script.

Here's a summary of ICORE's features, most of which can turned on/off using command-line parameters and switches.

1. *Preprocessing/preparation steps*:
    a. throughput or gain matching (using input photometric calibration zero-point information) and rescaling to a common target zero-point;
    b. background level offset-matching between input frames using a number of robust methods;
    c. pixel outlier detection and masking using robust statistics on stacks of input frames if a sufficient number of overlaps are available;
    d. entire frame-flagging (and rejection from co-addition/HiRes'ing) by thresholding on number of pixel-outliers and/or saturated pixels therein;
    e. optional spatial expansion of pixel outliers for complete blanketing (e.g., for masking extended latents, moving objects etc.).

2. *Coaddition/mosaicking methods*:
    a. Simple overlap-area interpolation with optional drizzling;
    b. Point Response Function (PRF) weighted averaging to create "matched filtered" products optimized for point source detection;
    c. For either methods a. or b., have option to co-add in the reference frame of a moving object (e.g., an asteroid), or on static inertial coordinates (with fixed RA, Dec);
    d. bad pixel avoidance using input image masks;
    e. signal-to-noise (S/N) map generation to support source detection;
    f. optional inverse variance weighting using input priors;
    g. output images of the depth-of-coverage, outlier locations, and tagged saturated pixels;
    h. uncertainty image products: both from propagating priors or derived *a posteriori* from image-stack statistics;
    i. pixel processing is multithreaded for speed;

3. *Resolution enhancement (HiRes'ing)*:
    a. based on an extension of the Richardson-Lucy deconvolution algorithm, also referred to as the Maximum Correlation Method (MCM);

b. performed in conjunction with co-addition to take advantage of the full pixel sampling provided by all overlapping input frames;
c. allows for non-isoplanatic (spatially varying) PRFs;
d. includes a ringing suppression algorithm based on flux positivity constraints and local background removal. Subsequent restoration of image noise and background;
e. statistically motivated convergence criteria based on the chi-square metric;
f. noise suppression and convergence acceleration in noisy regions;
g. specifiable number of iterations;
h. internal generation of an overlap-area weighted co-add to use as a prior, otherwise, a flat internal image prior is the default;
i. generic specification of any input image prior (e.g., same scene from another wavelength or instrument with higher resolution) to assist with reconstruction;
j. a powerful diagnostic (the Correction Factor Variance - CFV) to locate inconsistencies in frame overlap measurements, assess quality of instrumental calibration, PRFs, HiRes solutions, and degree of convergence;
k. uncertainty estimation per output HiRes pixel using either input priors or derived a posteriori through the CFV;
l. optional rescaling of uncertainty image(s) using robust noise metrics;
m. signal-to-noise ratio maps using either priors or uncertainties computed *a posteriori*;
n. a wealth of (optional) diagnostic output image products at every iteration: MCM correction factors, intensities, and CFV estimates;
o. optional inverse-variance weighting of input data using priors;
p. generic preprocessing steps outlined above are optional (throughput matching, background/overlap matching, and outlier detection and masking);
q. option to either co-add and HiRes in the reference frame of a moving object (e.g., an asteroid), or on static inertial coordinates (with fixed RA, Dec);
r. pixel processing is multithreaded for speed;

4. *Quality Assurance (QA) metrics*:
   a. standard and robust image metrics quantifying backgrounds, spatial noise, and depth-of-coverage within regions of an N × N grid over footprint of intensity and uncertainty image products;
   b. uncertainty image consistency checks using pseudo *chi-square* computed from spatial noise metrics as a function of depth-of-coverage;
   c. statistics of number of pixels masked or detected as outliers per frame;
   d. metrics on quality of background-level matching;

## 1.1 How to use this Document

It is assumed you have read the README file included in the ICORE distribution[1], installed the software, and executed one of the examples according to the instructions therein. We encourage you start with one of these examples and tailor it to your needs. If you're in a hurry to get started, you can follow the quick recipe outlined in §10 (also contained in the README). Table 1 (§3.1) may also be interest if you're looking for the meaning of a parameter. This table contains references to relevant sections of this document where the parameter is further described. You may need to backtrack to get the full algorithmic details. Also, be aware of the Advisories and Caveats in §12 before diving in. If having trouble with an acronym, see §16.

Overall, this document describes the algorithms, software I/O, and recipes for optimizing parameters to create co-adds and HiRes'd products. The statistical robustness and performance of algorithms are also addressed. It is broken up into the following themes:

- *I/O and software assumptions* (§3);
- *preparatory steps* (§4);
- *outlier detection* (§5);
- *co-addition* (§6, §7);
- *HiRes'ing* (§8);
- *Quality Assurance* (§9);
- *examples and recipe* (§10);
- *disk, memory usage, and runtime* (§11);
- *advisories and caveats* (§12);
- *aperture photometry and uncertainties* (§13);

## 1.2 Papers, Presentations and Websites that used ICORE (formerly AWAIC)

At the time of writing, here are some publications and web-media sites that used ICORE:

- Koenig, X. P., et al., 2012, *Wide-field Infrared Survey Explorer Observations of the Evolution of Massive Star-forming Regions*, ApJ, 744, 130

- Noriega-Crespo, A. & Raga, A. C., 2012, *Spitzer Observations of the HH 1/2 System: The Discovery of the Counterjet*, ApJ, 750, 101

- Ressler, M. E., et al., 2010, *The Discovery of Infrared Rings in the Planetary Nebula NGC 1514 During the WISE All-sky Survey*, AJ, 140, 1882

- Jarrett, T. H., Cohen, M., Masci, F., et al., 2011, *The Spitzer-WISE Survey of the Ecliptic Poles*, ApJ, 735, 112

[1] *http://web.ipac.caltech.edu/staff/fmasci/home/icore.html*

- Meech, K. J., et al, 2013, *The demise of Comet 85P/Boethin, the first EPOXI mission target*, Icarus, 222, issue 2, 662

- Bauer, J. M., et al., 2011, *WISE/NEOWISE Observations of Comet 103P/Hartley 2*, ApJ, 738, 171

- Bauer, J. M., et al., 2012, *WISE/NEOWISE Observations of Active Bodies in the Main Belt*, ApJ, 747, 49

- Jarrett, T. H., Masci, F., et al., 2012, *Constructing a WISE High Resolution Galaxy Atlas*, AJ, 144, 68

- Jarrett, T. H., Masci, F., et al., 2013, *Extending the Nearby Galaxy Heritage with WISE: First Results from the WISE Enhanced Resolution Galaxy Atlas,* AJ, 145, 6

- Masci, F. J. & Fowler, J. W., 2009, *AWAIC: A WISE Astronomical Image Co-adder*, in ASP Conf. Ser. 411, ADASS XVIII, ed. D.A. Bohlender, D. Durand, & P. Dowler, (San Francisco: ASP), 67

- Cutri, R., ..., Masci, F., et al., 2012, Explanatory Supplement to the WISE All-Sky Data Release Products, On-line publication: *http://wise2.ipac.caltech.edu/docs/release/allsky/expsup/*

- Herschel Map-Making Workshop presentation, Shupe, D., 01/29/2013, ESAC: *http://herschel.esac.esa.int/2013MapmakingWorkshop/presentations/SPIRE_HiRes_DShupe_MapMaking2013_DPWS.pdf*

- WISE "image of the week" public website (a majority of the images were created using ICORE): *http://wise.ssl.berkeley.edu/gallery_images.html*

- Invited Talk on ICORE at ADASS XVIII, Masci, F., 11/04/2008, Quebec City: *http://web.ipac.caltech.edu/staff/fmasci/home/wise/adass08_talk.pdf*
  and accompanying paper:
  *http://wise2.ipac.caltech.edu/staff/fmasci/awaic_adass08.pdf*

## 2 OVERVIEW

The goal of image co-addition is to optimally combine a set of (usually dithered) exposures to create an accurate representation of the sky, given that all instrumental signatures, glitches, and cosmic-rays have been properly removed. By "optimally", we mean a method which maximizes the signal-to-noise ratio (SNR) given prior knowledge of the statistical distribution of the input measurements. Figure 1 gives an overview of the main steps in ICORE. The co-addition and optional HiRes'ing step is shown by the red box. All steps are expanded further below.

It is assumed that the input science frames (specified by –imglist) have been preprocessed to remove instrumental signatures and their pointing refined in some WCS using an astrometric catalog. Accompanying bad-pixel masks (–msklist) and prior-uncertainty frames (–unclist) are optional. The frames are assumed to overlap with some predefined footprint (or tile) on the sky. This also defines the dimensions of the co-add products. The uncertainty frames store 1-σ values for the flux in each pixel. These are expected to be initiated upstream, e.g., from a noise model specific to the detector and then propagated and updated as all instrumental calibrations are applied. The uncertainties are used for optional inverse-variance weighting of the input measurements and for computing co-add flux uncertainties. If bad-pixel masks are specified, a bit-string template (–m_coad) is used to select which conditions to mask against. The corresponding pixels in the science frames are then omitted from co-addition.

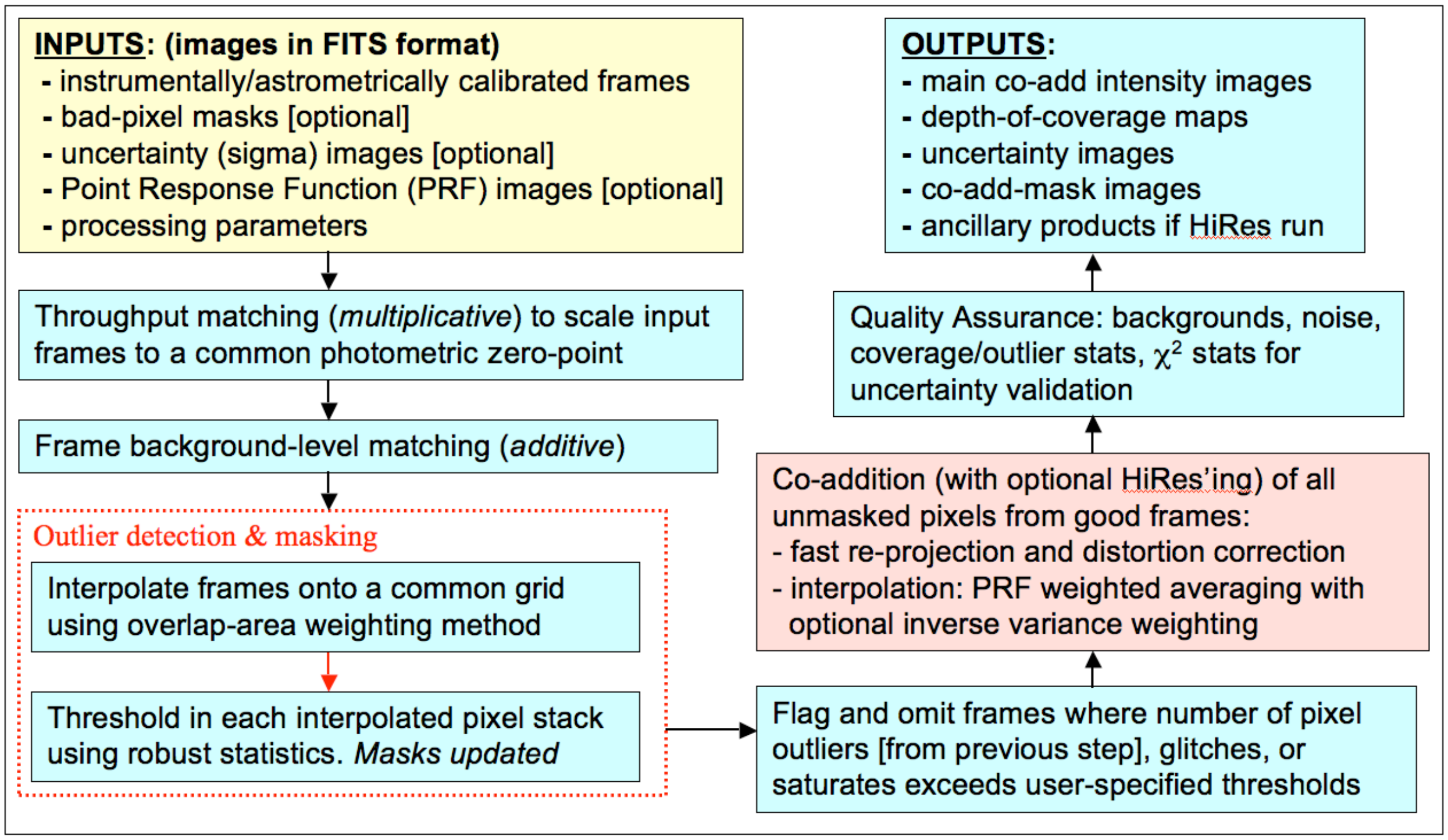

**Figure 1: processing flow in ICORE**

The first (optional) step is to scale the frame pixel values to a common user-specified photometric zero-point using calibration zero-point information in the input FITS image headers (if present). The common (or target) zero-point is then written to the FITS headers of co-add products to enable the calibration of photometric measurements. Frame overlap matching (or background-level regularization) is then performed. These steps are described in §4. Following these preparatory steps, pixel-outlier rejection is performed and outlying pixels are tagged in masks. This is described in §5. Since these initial steps modify the input frame and mask pixel values, local copies of the frames and masks are made to avoid overwriting the originals. After single pixel-outliers have been tagged, *entire* frames containing an excessive number of outliers are tagged for rejection from co-addition (or HiRes'ing). During co-addition or HiRes'ing, all "good" (unmasked) pixels are reprojected and interpolated onto an upsampled output grid. The reprojection uses a fast input-to-output plane coordinate transformation that implicitly corrects for focal plane distortion if represented in the input FITS headers. The Simple Imaging Polynomial (SIP) convention for distortion is assumed (Shupe et al. 2005). Details of co-addition are described in §§6,7 and HiRes'ing in §8.

The primary outputs from ICORE are the main intensity image (–o1_coad), a depth-of-coverage map (–o2_coad), a 1-$\sigma$ uncertainty image based on input *priors* (–o3_coad), an image of the outlier locations (–om_odet), and optionally if the overlap-area interpolation method was used, an image of the data-derived uncertainty estimated from the standard deviation in each interpolated pixel stack and appropriately scaled by the depth-of-coverage (–o4_coad). Additional ancillary products are generated under HiRes'ing. ICORE also produces a wealth of Quality Assurance (QA) metrics over pre-specified regions of the co-add footprint. These include background noise estimates, coverage and outlier statistics, and metrics to validate co-add flux uncertainties using $\chi^2$ tests. A summary of the QA outputs is given in §9.

# 3 INPUT/OUTPUT SUMMARY AND ASSUMPTIONS

## 3.1 I/O Specification

*icore* is a Perl script that takes all of its input from the command-line. This command-line can be set-up and executed via a shell script (see the examples provided in the *icore* distribution, and also summarized in §10). Prior to parsing the command-line, default values for the optional input parameters are assigned. All parameters are checked for validity and that they're specified in the correct combination. Assumptions on input data formats are summarized in §3.2.

Table 1 summarizes all command-line inputs, their purpose, data-type and units where applicable, default assignments, and section(s) in this document where you can find more information. Command-line inputs suffixed by "_odet" in Table 1 are specific to outlier detection. Inputs suffixed by "_coad" are specific to co-addition and/or HiRes'ing. All other inputs are generic to overall processing. If *icore* is executed with no command-line arguments, an I/O synopsis summarizing the contents of Table 1 is printed to the screen.

| **Option** | **Description** | **Data-type, [units]** | **Default** | **More info.** |
|---|---|---|---|---|
| -imglist | Input text file name containing list of pre-calibrated 32-bit / pixel FITS intensity images; may be gzipped | Char*256 | Required input | §2, §3.2 |
| -unclist | Input text file name containing list of 32-bit / pixel FITS (1-sigma) uncertainty images; may be gzipped | Char*256 | None used | §2, §3.2 |
| -msklist | Input text file name containing list of 32-bit / pixel (long int) FITS mask images; may be gzipped | Char*256 | None used | §2, §3.2 |
| -psflist | Input text file name containing list of focal-plane dependent PSF images. Only searched for if <-psfdir> below not specified. | Char*256 | <psfdir> and <-basepsf> first used if specified. If <-psflist> also absent, area-overlap method <-sc_coad> is used | §3.2, §6 |
| -psfdir | Input pathname containing PSF files | Char*256 | <-psflist> used | §3.2 |
| -basepsf | Input generic base filename of PSF for *N* x *N* frame grid, for constructing focal-plane dependent PSFs | Char*256 | <-psflist> used | §3.2 |
| -outdir | Pathname for intermediate files and working directory | Char*256 | Required input | §4.1, §5.2.2, §5.2.3, §8.1, §8.2, §8.3 |
| -qameta | Output filename for QA meta-data. | Char*256 | meta-coadd.tbl | §9 |
| -qagrid | Number of partitions *N* along an | I*2 int | 3 | §9 |
| | axis of co-add footprint for computing QA metrics within *N* x *N* square regions | | | |
| -sizeX | E-W mosaic dimension for crota2 = 0 | R*4 float [degrees] | Required input | §3.2, §5.2 |
| -sizeY | N-S mosaic dimension for crota2 = 0 | R*4 float [degrees] | Required input | §3.2, §5.2 |
| -ra | Right Ascension of mosaic center | R*4 float [degrees] | Required input | §3.2, §5.2 |
| -dec | Declination of mosaic center | R*4 float [degrees] | Required input | §3.2, §5.2 |
| -rot | Mosaic position angle in terms of CROTA2: +Y axis W of N: 0 ≤ rot < 360 | R*4 float [degrees] | 0.0 | §3.2, §5.2 |
| -mov | Switch to co-add and/or HiRes in the reference frame of a moving object | Null | 0 | §8.9 |
| -stpX | Size of the input image cutout (stamp) along the X-axis; used to speed-up processing for moving co-addition/hires'ing | R*4 float [#input img pixels] | Required if <-mov> set. Must be odd positive integer ≥ 5 pixels | §8.9 |
| -stpY | Size of the input image cutout (stamp) along the Y-axis; used to speed-up processing for moving co-addition/hires'ing | R*4 float [#input img pixels] | Required if <-mov> set. Must be odd positive integer ≥ 5 pixels | §8.9 |
| -pa_odet | Output interpolation grid pixel scale for outlier detection; recommend: ≥ 0.5 * input native pixel scale | R*4 float [arcsec] | Required if <-odet> set | §5.2 |
| -pb_odet | Additional padding to add around mosaic grid for outlier detection; recommend: of order PRF size for co-addition | R*4 float [arcsec] | 0 => possible outliers outside nominal mosaic boundary may contribute to co-add | §5.2 |
| -nx_odet | Number of tiles along X dimension of mosaic for outlier detection | I*2 int | 1 => whole mosaic | §5.2, §5.2.2, §5.2.3 |
| -ny_odet | Number of tiles along Y dimension of mosaic for outlier detection | I*2 int | 1 => whole mosaic | §5.2, §5.2.2, §5.2.3 |
| -tl_odet | Lower-tail threshold in number of sigma for outlier detection | R*4 float | 8 | §5.1, §5.2.2, §5.2.3, §8.6 |
| -tu_odet | Upper-tail threshold in number of sigma for outlier detection | R*4 float | 8 | §5.1, §5.2.2, §5.2.3, §8.6 |
| -ts_odet | Minimum "(median bckgnd) / sigma" value for stack above which to rescale <-tl> and <-tu> thresholds by <-r_odet> factor | R*4 float | 1.0E+30 | §5.1, §5.2.2 |
| -ta_odet | Maximum out/in pixel area ratio below which to use nearest-neighbor and max-overlap area weighted interpolation; used for outlier detection | R*4 float | 0.26 | §5.2.2 |
| -r_odet | Scaling factor for upper/lower- | R*4 float | 1 | §5.1, §5.2.2 |
| | tail thresholds in stacks satisfying <-ts_odet>; for outlier detection | | | |
| -s_odet | Scaling factor for $\sigma_{MAD}$ estimates for outlier detection | R*4 float | 1.5 | §5.2.1 |
| -b_odet | Smooth $\sigma_{MAD}$ images using median filter? 0=>no; 1=>yes; used for outlier detection | I*2 int | 1 | §5.2.1 |
| -w_odet | Square window side length in pixels for median filter; must be odd integer; used for outlier detection | I*2 int | 3 | §5.2.1 |
| -d_odet | Omit input frame from outlier-detection step if it doesn't overlap with tile footprint? 0=>no; 1=>yes | I*2 int | 1; advise setting to 0 for co-addition in the reference frame of a moving object | §5.2.1 |
| -ip_odet | Use signed 2-byte integer storage with scaled log transform for stack estimators in outlier detection? 0=>no; 1=>yes | I*2 int | 0 | §5.2.3 |
| -is_odet | Scaling factor to support 2-byte integer storage option -ip_odet 1; require < 32767/Log_e[max] for expected max interp pix value | R*4 float | 2000.0 | §5.2.3 |
| -ns_odet | Minimum number of stack samples for reliable outlier detection | I*2 int | 5 => hard lower limit for $\sigma_{MAD}$ method | §5.1, §5.2.1 |
| -nei_odet | Minimum number of neighbors around a pixel with <-m_odet> set such that if > than this, expansion over <-exp_odet> x <-exp_odet> region will be triggered if <-expodet> switch was set | I*2 int | 10 | §5.2.3 |
| -nsz_odet | Assume pixel region of size <-nsz_odet> x <-nsz_odet> centered on outlier pixel when counting outlier neighbors; must be odd number. | I*2 int | 5 | §5.2.3 |
| -exp_odet | Desired <-exp_odet> x <-exp_odet> pixel region to expand (or force) into outliers centered on original outlier pixel; must be odd number. | I*2 int | 7 | §5.2.3 |
| -expodet | Switch to expand detected outliers (or blanket) over <-exp_odet> x <-exp_odet> region after thresholding on min. number of outlier neighbors <-nei_odet> within region <-nsz_odet> x | Null | 0 | §5.2.3 |
| | <-nsz_odet> | | | |
| -m_odet | Mask bit to set for temporal outlier detection in input masks. Specified as decimal equivalent | I*4 int | 134217728 => bit 27 in WISE frame masks; value 0 => no mask updating | §5.1, §5.2.2 |
| -om_odet | Output FITS filename of 8-bit mosaic showing temporal outlier locations | Char*256 | None generated | §2, §5.2.2, §5.2.3 |
| -h_odet | Homogenize sigma_MAD images by setting to median value? 0=>no; 1=>yes. Used for outlier detection. | I*2 int | 1 | §5.2.1 |
| -k_odet | Minimum S/N in coadd tile below which to homogenize sigma_MAD values under -h_odet processing. | R*4 float | 3.0 | §5.2.1 |
| -q_odet | Number of spatial sigma in sigmad to homogenize sigma_MAD values under -h_odet processing. | R*4 float | 5.0 | §5.2.1 |
| -mg_odet | Mask bit corresponding to spike-glitch detection in input masks from ical for omitting entire frames using tg_odet threshold. | I*4 int | 268435456 => tuned to bit 28 in WISE frame masks; 0 => thresholding not triggered. | §5.2.4 |
| -tn_odet | Threshold for max tolerable number of temporal outliers per frame above which entire frame will be omitted from co-addition | I*2 int [pixels] | 1E+09 | §5.2.4 |
| -tg_odet | Threshold for max tolerable number of spike-glitch pixels per frame above which entire frame will be omitted from co-addition | I*2 int [pixels] | 1E+09 | §5.2.4 |
| -tsat_odet | Threshold for max tolerable number of saturated pixels per frame above which entire frame will be omitted from co-addition | I*2 int [pixels] | 1E+09 | §5.2.4 |
| -nmaxodet | Maximum number of input frames to trigger partitioning of input lists if -partition switch is set. Must be greater than 100 | I*2 int | 300 | §5.2.3 |
| -partition | Switch to partition input image list for outlier detection if number of input frames exceeds maximum specified by -nmaxodet. In this mode, -nx_odet and -ny_odet are reset to 1 | Null | 0 | §5.2.3 |
| -pri_coad | Input image filename in FITS format specifying image prior to support HiRes processing, e.g., a higher resolution image from | Char*256 | Internal flat-image prior used unless -fp_coad=1 | §8.2 |
| | another instrument or wavelength. Must contain valid WCS | | | |
| -rat | Right Ascension of target position in cell-grid prior image about which to retain pixel signals within radius <-isorad> and set pixels outside this to 1 | R*4 float [degrees] | Required if –isolate switch is set | §8.2 |
| -dect | Declination of target position in cell-grid prior image about which to retain pixel signals within radius <-isorad> and set pixels outside this to 1 | R*4 float [degrees] | Required if –isolate switch is set | §8.2 |
| -isorad | Radius about target position <-rat>, <-dect> in cell-grid prior image within which to retain pixel signals and set all pixels outside this to 1 | R*4 float [arcsec] | Required if –isolate switch is set | §8.2 |
| -isolate | Switch to isolate pixel signals within radius <-isorad> of target position <-rat>, <-dect> in cell-grid prior image by setting all pixels outside this radius to 1 | Null | 0 | §8.2 |
| -clipt | Pixel signal-to-noise threshold below which to set pixel signals in cell-grid prior image to zero and then add 1 to all pixels to regularize for MCM/HiRes | R*4 float [#sigma] | 0 | §8.2 |
| -pa_coad | Output co-add or HiRes'd image pixel scale | R*4 float [arcsec] | 1.375 arcsec | §3.2, §6 |
| -pc_coad | Ratio of internal linear cell pixel size to output mosaic pixel size; must equal input PSF pixel size | R*4 float [arcsec] | 0.5 | §3.2, §6 |
| -m_coad | Mask template bitstring specifying conditions for omitting pixels from co-add | I*4 int | 0 => no pixel flagging and omission | §2, §3.2 |
| -ms_coad | Bitstring specifying saturated pixels in an input frame mask to tag in output mask (-om_coad); latter only possible for -n_coad = 1 co-adds | I*4 int | 0 => no saturated pixel tagging in output co-add mask | §3.2 |
| -ct_coad | Maximum tolerance for difference between cell-grid pixel size <-pc_coad> and input PSF pixel size | R*4 float [arcsec] | 0.0001 | §3.2, §6 |
| -wf_coad | Combine input pixels in single interp frame and stack using inverse variance weighting? 0=>no, 1=>yes. If yes, requires input uncertainties <-unclist> | I*2 int | 0 | §6 |
| -sf_coad | Scale output pixel flux with pixel size? 0=>no, 1=>yes | I*2 int | 1 => e.g., if input frame pixels are in DN. | §3.2 |
| -sc_coad | Create simple co-add/mosaic | I*2 int | 0 | §7 |
| | using exact overlap-area weighting? 0=>no, 1=>yes. If yes, PSFs not needed. | | | |
| -d_coad | Input pixel linear drizzle factor, = ratio: new pix scale/native pix scale (<= 1); only used for simple coadds <-sc_coad 1> | R*4 float | 1.0 => no drizzling of input pixels | §7, §13 |
| -n_coad | Number of MCM (HIRES) iterations | I*2 int | 1 => PRF-interpolated coadd with no resolution enhancement | §3.2, §6, §8.1, §8.4.1, §8.5, §8.6 |
| -n_coadn | Number of additional MCM (HIRES) iterations to support flux-bias (ringing suppression) processing | I*2 int | 0 => additional iterations under "-flxbias" mode not performed | §8.5 |
| -rf_coad | Rotate PSF when projecting input frame pixels? 0=>no, 1=>yes. Recommended for -n_coad > 1 | I*2 int | 0 | §3.2, §6 |
| -if_coad | Method for interpolating PSF onto cell-grid: 0=>nearest neighbor, 1=>area-overlap weighting. 1 only possible for "–n_coad 1" | I*2 int | 0 | §3.2, §6 |
| -fp_coad | Create and use overlap area-weighted co-add for MCM starting model; 0=>no, 1=>yes; Default=0 => use flat prior image of 1's; only used in first MCM pass if <-flxbias> set | I*2 int | 0 | §8.1 |
| -h_coad | Minimum tolerance for % change in real CFV from iteration n -> n+1 (initial <-n_coad> only) below which MCM pixel-cell arrays get frozen at n; assists in noise-suppression; may need to also use aggressive outlier rejection to assist in suppressing noise spikes and keeping CFV low; activated if -n_coad > 1 and -o5_coad,-o6_coad specified; -oi_coad output also useful | R*4 float [percent] | 0.0 => no noise suppression | §8.6 |
| -oi_coad | Output mosaic of position-dependent ending iteration numbers; only generated if -o5_coad, -o6_coad specified | Char*256 | None generated | §8.6 |
| -of_coad | Output mosaic of first iteration MCM intensities in down-sampled frame; only generated for -n_coad > 1 | Char*256 | None generated | §8.1 |
| -crep_coad | Replace int and unc coadd pixels with covs < cmin_coad with NaN? 0=>no, 1=>yes | I*2 int | 0 | §8.8 |
| -cmin_coad | Minimum depth-of-coverage below which to replace int and unc coadd pixels with NaN if crep_coad=1 | R*4 float | 4 | §8.8 |
| -nt_coad | Number of concurrent threads to run in co-addition or HiRes'ing steps; set to number of CPU cores available on machine. **Note:** some CPUs may support nt_coad = 2 × the number of physical cores available | I*2 int | 1 (1 => non-optimal if you indeed have a multi-core processor) | §10.1, §11.3 |
| -o1_coad | Output mosaic intensity image FITS filename | Char*256 | Required if <-coadd> set | §2, §6 |
| -o2_coad | Output mosaic coverage map FITS filename | Char*256 | Required if <-coadd> set | §2, §6 |
| -o3_coad | Output uncertainty mosaic FITS filename: valid for all -n >= 1 and "–sc_coad 1" co-adds; based on input prior uncerts. Rescaled (using robust intensity RMS) for -n > 1 | Char*256 | None generated | §2, §3.2, §6, §8.4.2 |
| -o4_coad | Output standard deviation mosaic FITS filename; can only be generated for simple area-overlap co-add: "-sc_coad 1"; will not account for correlated noise like -o3_coad output. Latter accounts for movement of noise-power to lower spatial frequencies | Char*256 | None generated | §2, §3.2, §7 |
| -o5_coad | Output mosaic of MCM correction factors; really only valid for -n_coad > 1 products; | Char*256 | None generated | §8.1 |
| -o6_coad | Output mosaic of data-derived MCM-uncertainties estimated from Correction Factor Variance (CFV); really only valid for -n_coad > 1 products; rescaled (using robust intensity RMS) for all -n >= 1 | Char*256 | None generated | §3.2, §8.4.1 |
| -snu_coad | S/N ratio coadd product where 1-sigma uncertainties are from priors (-o3_coad output) and rescaled for -n_coad > 1; only generated if -o3_coad output specified; for -n_coad = 1, this product is a matched-filter optimized for detecting point (PRF-like) sources; cannot be generated when -sc_coad = 1 | Char *256 | None generated | §8.4.3 |
| -snc_coad | S/N ratio coadd product where 1-sigma uncertainties are purely data-derived (from CVF: - | Char *256 | None generated | §8.4.3 |
| | o6_coad output) and rescaled; only generated if -o6_coad output specified; really only applicable for HiRes with -n_coad (and optionally -n_coadn) > 1 and obviously cannot be made under -sc_coad = 1 mode | | | |
| -om_coad | Output 8-bit mosaic mask name showing locations of bad and saturated pixels that occur at least once in the stack; only applicable to -n_coad=1 co-adds | Char* 256 | None generated | §3.2 |
| -ratmax | Value of ratio: [84ptile – med ] / [med – 16ptile ] above which a frame is suspected to contain bright extended structure. Used to make the -bmatch step more robust | R*4 float | 2.0 | §4.2.1, §8.4.3, §8.5 |
| -bgrid | Number of partitions along an axis of native input frame for background (median) estimation within a partition when extended structure is detected under -bmatch processing using -ratmax parameter | I*2 int | 8 | §4.2.1 |
| -siggrid | Number of partitions along an axis of final coadd product for finding smallest robust RMS to support uncertainty coadd rescaling both from priors (-o3_coad) and/or CFV (-o6_coad) | I*2 int | 8 | §8.4.1 |
| -modfilt | Switch to use mode filter instead of default median filter when computing slowly varying background in partitioned -svbgrid grid over input frame. Triggered if number of non-NaN'd pixels in a partition is >= 1000, otherwise median is used | Null | 0 | §8.4.3, §8.5 |
| -svbgrid | Number of partitions along an axis of native input frame for median SVB computation to support HiRes with -flxbias processing and generation of S/N coadds: -snu_coad and/or -snc_coad for -n_coad >= 1 | I*2 int | 3 | §8.4.3, §8.5 |
| -gausize | Linear size of Gaussian smoothing kernel; must be odd number | R*4 float [frame pixels] | 21 | §8.4.3 |
| -gausigm | Linear sigma of Gaussian smoothing kernel | R*4 float [frame pixels] | 5 | §8.4.3 |
| -arbconst | Arbitrary global constant to add to all input image pixels to ensure positivity for HiRes processing (–n_coad > 1) regardless if the –flxbias switch was set. Only operative if the –tmatch switch was set | R*4 float | 10 | §8.5 |
| -magzp | Global photometric zero-point value to scale to | R*4 float [mag] | Required if <-tmatch> set | §4.1 |
| -magzpk | Name of keyword in input FITS images containing the photometric zero-point value | Char *8 | MAGZP | §4.1 |
| -magzpuk | Name of keyword in input FITS images containing the photometric zero-point uncertainty value | Char *8 | MAGZPUNC | §4.1 |
| -tmatch | Switch to perform throughput (gain) matching of input frames | Null | 0 | §4.1 |
| -bmatch | Optional background (offset) matching of input frames; 0=>no, 1=>yes; | I*2 int | 0 | §4.2, §8.4.3, §8.5 |
| -bmeth | Background-matching / regularization method if -bmatch was set: 0 => robust planar fit; 1 => higher order surface fit to clipped data; 2 => global minimization of offsets between all frame overlaps (recommended) | I*2 int | 2 | §4.2.2 |
| -order | Order of polynomial to fit for -bmeth=1 | I*2 int | 2 | §4.2.2 |
| -bfgrid | Size of bfgrid x bfgrid for partitioning frame and computing median pixel values for input into surface fitting routine for bmeth=1 | I*2 int | 9 | §4.2.2 |
| -clsig | Number (N) of robust sigma to clip high tail from mode and replace with mode+N*sigma before partitioning and fitting surface under bmeth=1 | R*4 float | 0.5 | §4.2.2 |
| -reimg | Internally rebin each input image first to speed up background matching for bmeth=2 | Null | 0 | §4.2.3 |
| -refac | If -reimg switch is on; this is the down-sampling factor to use for each axis of an input image for bmeth=2 (must be ≥ 2) | I*2 int | 4 | §4.2.3 |
| -edgw | Image edge-width to use [in native or rebinned frame pixels if -reimg was set] to support background matching for | I*2 int [pixels] | 5 | §4.2.3 |
| | bmeth=2 (must be > 3) | | | |
| -offtol | Tolerance on ratio of a frame bckgnd offset solution from bmeth=2 to robust sigma of all offset solutions above which to reset (outlying) offset solution to median of all offsets | R*4 float | 1000 | §4.2.3 |
| -cpmsk | Switch to copy frame masks to <-outdir> and update with outlier detection results therein. Avoids corrupting original input masks | Null | 0 | §5.2.2 |
| -odet | Switch to perform outlier detection | Null | 0 | §5 |
| -coadd | Switch to execute co-adder and create co-add products | Null | 0 | §3.2 |
| -flxbias | Invoke flux bias (ringing suppression) processing for HiRes mode; only possible for -n_coad > 1; 0=>no, 1=>yes; | I*2 int | 0 | §3.2, §8.5 |
| -addbck | Switch to add background to HIRES'd products after first MCM processing pass under -flxbias mode | Null | 0 | §8.5 |
| -mcmprod | Switch to generate ancillary products for each MCM iteration: intensity, Correction Factor Variance (CFV), and correction factor images in internal cell grid frame | Null | 0 | §3.2, §8.1, §8.3, §8.6 |
| -qa | Switch to generate QA metrics on co-add products | Null | 0 | §9 |
| -dbg | Switch to generate debug info. and diagnostic files from *script-specific functions* | Null | 0 | §5.2.3 |
| -sdbg | Switch to also generate debug info and diagnostic files from *C-module executions* | Null | 0 | §5.2.1 |
| -v | Switch to increase verbosity to stdout from *script-specific functions* | Null | 0 | generic |
| -sv | Switch to include more verbosity to stdout from *C-module executions* | Null | 0 | generic |

**Table 1: Command-line inputs and options**

### 3.2 Assumptions, Parameter Interplay, and Processing Details

Below we list the assumptions pertaining to the input images and parameters. Recommendations on parameter values are also given. Many of these are checked during early execution of *icore*. If not satisfied, the program aborts with a message and non-zero exit status written to standard error.

- The input lists of FITS intensity images (–imglist), masks (–msklist), and uncertainty images (–unclist) must all have the same number of filenames listed in one-to-one correspondence. Each FITS image may be supplied as a gzipped file.
- All input frames are expected to overlap with some predefined footprint on the sky with WCS/dimensions/pixel scale defined by –ra, –dec, –rot, –sizeX, –sizeY, and –pa_coad. This footprint also defines the dimensions of the output co-add products. Note the output interpolation grid for outlier detection uses the same WCS/dimensions but uses a pixel scale specified by –pa_odet. If the bulk of your input frames do not overlap with (or touch) this footprint, you'll be wasting precious memory.
- All image inputs must be in FITS format.
- All input intensity, mask and uncertainty images must have the same native pixel scale (scales along *x* and *y* can differ); the same projection type (CTYPE header keywords); the same NAXIS1, NAXIS2 values (with NAXIS1 ≠ NAXIS2 allowed); and the same EQUINOX.
- If FOV-distortion information is available, this must be represented in the FITS headers of the intensity images using the Simple Imaging Polynomial (SIP) convention. The WCS keywords CDELT1, CDELT2, CROTA2 may be encoded in CD matrix format.
- It is recommended that all image pixel scales (either from CDELT or CD keywords) be represented in the FITS headers to at least 8 significant figures.
- The only projection types recognized by the software are: TAN, SIN, ZEA, STG and ARC. These are specific to the fast plane-to-plane reprojection algorithm.
- For PRF-weighted co-adds (–n_coad = 1) or HiRes'ing (–n_coad > 1), a list of PRFs must be specified. This can be either one PRF applicable to the entire focal plane of the detector frames, or, multiple PRFs if the PRF is non-isoplanatic. The PRFs can be specified explicitly using an input list (–psflist), or by providing the path to their location and a PRF base-filename: –psfdir and –basepsf respectively. The script first checks if –psfdir and –basepsf were specified. If not, it checks for –psflist. If –psflist is not specified, the processing diverts to making a simple co-add using overlap-area weighting (i.e., –sc_coad is forced "on").
- If multiple PRF images are to be specified via –psflist (i.e., to handle a focal-plane dependent PRF), they must all be the same size (NAXIS1, NAXIS2 values the same) and have the same pixel scale (CDELT1, CDELT2). An example of a minimal FITS header for an input PRF is shown below, where the CDELT* and FPALOC* values are not realistic. See further below on how to derive the FPALOC* keywords.

```
SIMPLE  =                    T / file does conform to FITS standard
BITPIX  =                  -32 / number of bits per data pixel
NAXIS   =                    2 / number of data axes
```

```
NAXIS1  =                 <Num> / length of data axis 1
NAXIS2  =                 <Num> / length of data axis 2
CRPIX1  =       <0.5*(Num + 1)> / reference pixel for axis 1
CRPIX2  =       <0.5*(Num + 1)> / reference pixel for axis 2
CTYPE1  =            'RA---SIN' / projection type for axis 1
CTYPE2  =            'DEC--SIN' / projection type for axis 2
CDELT1  =        -0.00019097222 / axis 1 scale [deg/pix]
CDELT2  =         0.00019097222 / axis 2 scale [deg/pix]
FPALOCX =                 508.0 / center x coord of grid square
FPALOCY =                 711.2 / center y coord of grid square
```

- The values of the CTYPE1, CTYPE2 keywords in the PRF headers must be the same as those in the headers of the input intensity images.
- The PRF CDELT1, CDELT2 (pixel scale) values must be within some tolerance "–ct_coad" (default=0.0001 arcsec) of the internal co-add cell-grid scale (=pa_coad*pc_coad). For either the *X* or *Y* (CDELT) PRF pixel scale, these must satisfy: `|pa_coad * pc_coad - prf_CDELT| ≤ ct_coad (in arcsec).` Therefore, depending on the desired accuracy, it is wise to first pick an *absolute* output mosaic pixel scale (–pa_coad) and a cell size factor (–pc_coad) before deriving the PRF(s).
- The FPALOCX and FPALOCY keywords in the PRF headers (see example header above) are only necessary if *multiple* PRFs are specified in –psflist. These specify the location in the native *X, Y* coordinate system of an input frame at which the PRF applies. This is for cases where the PRF is non-isoplanatic (i.e., varies over the FPA). The values of these keywords refer to the center coordinates of a square over which the PRF applies. The specific square regions are defined beforehand by partitioning the FPA into an $n \times n$ grid, with a PRF derived in each. The total number of input PRFs is therefore $n^2$. For a grid square labeled by integer coordinates $(i, j)$ where $1 \le i \le n$ and $1 \le j \le n$ (see Figure 2), its center coordinates will be:

$$\mathrm{FPALOCX} = \left(\frac{\mathrm{NAXIS1_f}}{n}\right)\left(i - \frac{1}{2}\right);$$
$$\mathrm{FPALOCY} = \left(\frac{\mathrm{NAXIS2_f}}{n}\right)\left(j - \frac{1}{2}\right),$$

  where ($\mathrm{NAXIS1_f}$, $\mathrm{NAXIS2_f}$) refer to the dimensions of an input image frame. So for example, a grid square located at partition $(i, j)$ = (03, 04) in a 5 × 5 grid (see Figure 2) will have in the PRF header: FPALOCX = (1016/5)*(3 – 0.5) = 508.0, and FPALOCY = (1016/5)*(4 – 0.5) = 711.2. These keywords are used by ICORE to match a specific PRF from the input list to the pixel being processed.
- The $n \times n$ PRF images are then supplied as an input list to ICORE (parameter –psflist). Note that only square grids are allowed, i.e., the number of input PRF images must be a perfect square: 1, 4, 9, 16, 25 etc. Multiple input PRFs are usually only needed if one is after accurate resolution enhancement (input parameter –n_coad > 1). For PRF-weighted co-addition (–n_coad = 1), it is sufficient to use one (averaged) PRF for the entire array.

The author has yet come across a detector whose PRF is highly variable to warrant using several or more PRFs for co-addition.

- The PRFs must be volume-normalized to unity. This is internally checked. The sum of all PRF values in an input PRF image must not differ from unity by more than 1.0E-06. This tolerance is hard-coded.
- The internal *cell* to output mosaic pixel size ratio (command-line parameter –pc_coad) must be expressible in the form 1/*integer*, where *integer* = 1, 2, 3…, and lie within the range: 0.2 ≤ pc_coad ≤ 1.0. The lower limit is hard-coded.
- The input value for –pa_coad must satisfy: 0.1**minrawscale* ≤ pa_coad ≤ *minrawscale*, where *minrawscale* `= min[input_image_CDELT1, input_image_CDELT2]` and the `input_image_CDELTs` are input image pixel scales.
- The output grid pixel scale for outlier detection, –pa_odet <*in arcsec*> must satisfy: *minscale* ≤ pa_odet ≤ *maxscale*, where *minscale* `= sqrt[MINAREAR*inp_image_CDELT1*inp_image_CDELT2]`; *maxscale* = `sqrt[inp_image_CDELT1*inp_image_CDELT2]`, and `MINAREAR = 0.1`. The `inp_image_CDELTs` are input image pixel scales. Therefore, the grid pixel scale is constrained by the ratio of output/input pixel area.
- The input pixel mask FITS images if specified, are expected to have a BITPIX=32 (i.e., 32-bit signed integer format). However, only the first 31 bits (excluding the sign bit) are used in processing. Masks with BITPIX=16 or 8 or even -32 (floating point) can still be stored. Only the integer part of the float will be stored for input masks with BITPIX=-32.
- The input fatal bit-string template specification for co-addition or HiRes'ing: `–m_coad` allows one to flag pixels according to certain conditions/criteria. The maximum value allowed is $2^{31}$ = 2147483647. If this value is specified, then all pixels with values ≥$2^0$ in the input masks will be omitted from co-addition or HiRes'ing.
- There is another input bit-string for tracking saturated pixels in input frames: `–ms_coad` [$0 \leq$ value $\leq 2^{31}$]. This requires knowledge of which bits define saturation in the input masks. This information is used to tag saturated locations with value "100" in an output 8-bit mask with filename specified by `–om_coad`. Bad pixels specified by the `–m_coad` bit-string (see above) are also tagged in this mask and assigned value "1". For a stack of frames, non-zero values in the `–om_coad` mask imply that a saturated or bad pixel occurred at least once at that sky location in the stack. This mask is only generated for PRF-interpolated co-adds (–n_coad 1). Furthermore, if the `–ms_coad` saturation bit-string is specified, this must be included in the overall bad-pixel bit-string specified by: `–m_coad`.
- The area-overlap PRF-weighted interpolation method (command-line option –if_coad = 1) can only be invoked when (i) generating a PRF-weighted (–n_coad = 1) co-add, **and** (ii) when *no rotation* of the input PRF during re-projection is desired (command line option –rf_coad = 0). The interpolation method will default to the "nearest neighbor" method (–if_coad = 0) if either rotation of the PRF is specified (–rf_coad = 1), or, resolution enhancement is desired (–n_coad > 1).
- It is recommended that the "nearest neighbor" method with rotation included (–if_coad = 0 and –rf_coad = 1) be used when performing resolution enhancement.

- If “nearest neighbor” interpolation is used and the output mosaic pixel scale (from –pa_coad) is *not* a fraction expressible as (1/*integer*)***input image pixel scale* where *integer* = 1, 2, 3…, then systematic patterns in the output depth-of-coverage map may result. These are normalized out of the intensity images since there is an implicit division by the coverage at every location (see formalism in §6). However, to *minimize* systematic variations in the coverage map when nearest neighbor interpolation is used, it is advised that: (i) a new output mosaic pixel scale be chosen that satisfies the (1/*integer*)***input image pixel scale* criterion, and (ii) that the input PRFs are sampled to a pixel scale ≤ 0.25 × input image pixel scale. If you are adamant on using a specific output mosaic pixel scale, then the more robust (but slower) area-overlap weighting PRF-interpolation method (–if_coad 1) can be used.
- The pixel units in the output intensity co-add (–o1_coad) and corresponding uncertainty images (–o3_coad or –o4_coad or –o6_coad) reflect the input image units and can be scaled using the –sf_coad <scale> parameter. For example, if the input image units are in Data Number (DN or counts), one may want to re-scale these according to pixel area so that the total counts in a photometric aperture are conserved.
- Any of the five major processing steps can be turned off/on using command line switches: background (offset) matching: –bmatch; throughput (gain) matching: –tmatch; outlier detection: –odet; co-addition or HiRes’ing: –coad; and quality assurance: –qa.
- The –flxbias switch turns on ringing suppression in HiRes (§8.5). This requires the –bmatch (background matching) switch to be specified in order to trigger the extended structure detection algorithm (see §4.2). This special processing allows the frame SVB to be replaced by a constant equal to the minimum median background over all –svbgrid × –svbgrid partitions of a frame if extended structure is detected through the –ratmax parameter.
- The –arbconst input parameter allows one to specify a constant signal to add to all input image pixels to ensure positivity for HiRes processing (§8.5). This is *only operative* if the –tmatch switch is also set. The latter also performs throughput (gain) matching of the input images if photometric zero-point information is present in the input image headers (§4.1). If no zero-point information is available, throughput matching is not performed.
- The –mcmprod switch is used to generate HiRes ancillary products. If specified (§8.1, §8.3), you will also need to supply the –o5_coad and –o6_coad output filenames.

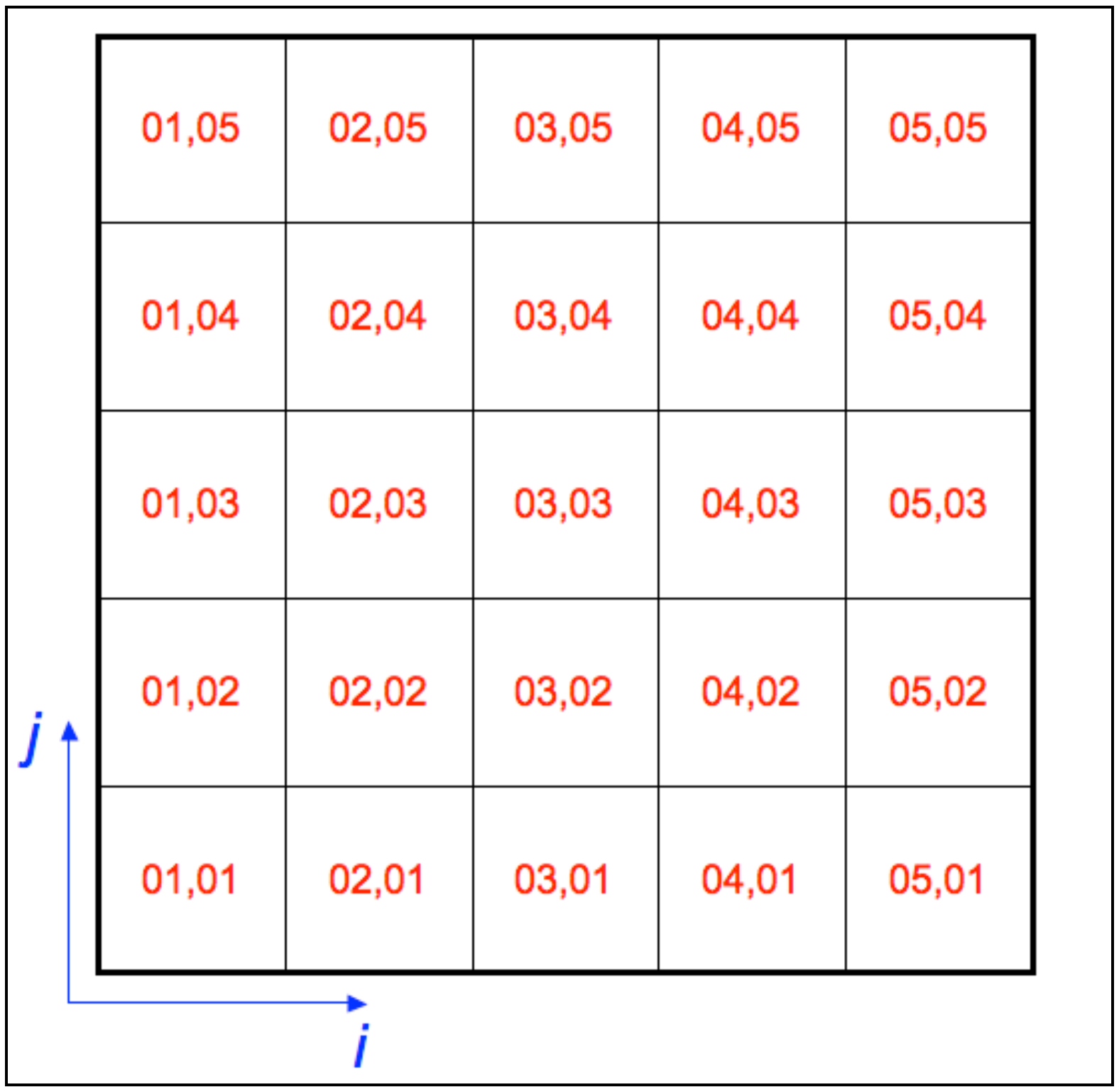

**Figure 2: Example of a 5 × 5 partition of an input frame for defining FPA position-dependent PRFs with an indexing scheme for computing the FPALOCX and FPALOCY keywords (see §3.2).**

# 4 PREPARATORY STEPS

## 4.1 Throughput (Gain) Matching

The first (optional) step is to scale the frame pixel values to a common photometric zero-point using calibration zero-point information in the input FITS image headers. This step is only necessary if the calibration zero-point (or conversion factor between instrumental DN to absolute flux units) varies amongst the input frames. This step can be controlled using the –tmatch switch in *icore*.

If the –tmatch switch is set, the software reads the zero-points from the input FITS image headers with keyword name specified by –magzpk (default = *MAGZP*). If throughput matching is desired and this keyword is not present, it can be derived (in magnitude units) using:

$$\begin{aligned} MAGZP &= M_{true} - M_{inst} \\ &= -2.5\log_{10}\left[\frac{f_{true}}{f_0}\right] + 2.5\log_{10}\left[DN\right] \\ &= 2.5\log_{10}\left[\frac{f_0}{c}\right], \end{aligned}$$

where $M_{true}$ and $M_{inst}$ are the *true* and *instrumental* magnitudes of a calibrator source, *DN* (Data Number) are the raw image pixel counts of the calibrator source from photometry (with background subtraction), $f_{true}$ is its "true" flux density (e.g., Jy), $f_0$ is the flux corresponding to magnitude zero, and $c = f_{true}/DN$ = the "DN-to-[Jy]" conversion factor (or whatever absolute units are involved). Some instruments (e.g., on *Spitzer*) only report values for *c*, in which case it would need to be converted to some corresponding *relative MAGZP* as above.

Let's label the *MAGZP* value in an input FITS image header as $magzp_i$. Given a common (or target) zero-point $magzp_c$ specified by –magzp, the pixels ($p_{old}$) in frame *i* are rescaled according to:

$$p_{new} = p_{old}10^{0.4\left(magzp_c - magzp_i\right)}.$$

The same operation is performed on the input uncertainty frames if specified.

Since the input frame pixel values are modified (likewise for the background level-matching step described below), local copies of the frames (intensities and uncertainties) are made to avoid overwriting the originals. The modified frames are created under the directory specified by –outdir.

The target zero point ($magzp_c$) is then written to the FITS headers of co-add image products using the same keyword specified by –magzpk. This enables the calibration of photometric measurements on a co-add. Note that no zero-point information is written to the headers of output image products if *any* input image header is missing a zero-point or has the wrong format. If this occurs, a warning is written to standard output to indicate that the output pixel values may be erroneous due to incomplete throughput matching.

The FITS keyword name for the accompanying zero-point uncertainty can be specified by –magzpuk (default = *MAGZPUNC*). If present in the input FITS image headers, these uncertainties are median-combined to derive an effective zero-point uncertainty (with output keyword name from –magzpuk) to accompany the co-add *MAGZP* value. If these uncertainties are absent or null ("-999"), the value "-999" is assigned to the zero-point uncertainty in co-add image products.

## 4.2 Frame Background/Level Matching

Frame exposures taken at different times usually show variations in background levels due to for example: instrumental transients, changing environments, scattered and stray light. The goal is to obtain seamless (and/or smooth) transitions between frames near their overlap regions prior to co-addition. We will want to equalize background levels from frame to frame, but at the same time preserve natural background variations and structures as much as possible. This step is only performed if the –bmatch switch is specified. *icore* supports three methods:

### 4.2.1 Method 1: 'robust' tilted plane-fitting

This method can be invoked by specifying "–bmeth 0". Here are the steps.

1. Fit a robust plane to each frame. By "robust", we mean relatively immune to the presence of bright sources and extended structure. Our goal is to capture the global underlying background level in a frame. There are of course cases where structure may span over most of a frame, and hence the background will be over-represented. The planar fit is parameterized by $z = f(x, y)$, where $z$ is the background level and $x$, $y$ are frame-pixel coordinates. We need at least three $(x, y, z)$ values per frame to (exactly) fit a plane. We choose the $x, y$ to be the centers of square partitions 72 x 72 pixels in size and $z$ is the median in each. Seven different partition configurations (of three $x,y,z$ points) are used per frame. The coefficients across all seven fits are then medianed to further ensure robustness of the planar fit.

2. The robust planar fits are subtracted from each respective frame. This effectively flattens the frames and places them on a zero-baseline.

3. Compute a global median $M$ of all frame pixels contributing to the co-add footprint.

4. Add this global median $M$ (a constant) to each of the "zero-level" frames (from 2). The frames have now been matched to a common background level. This will be more-or-less representative of the natural background over the co-add footprint region.

The above method also includes a method to ameliorate biases from the possible presence of bright extended structure. Presence of extended structure over a frame is searched for by thresholding on the ratio of quantile differences in the pixel distribution of each input frame, e.g.,

$$Q_d = \frac{q_{0.84} - q_{0.5}}{q_{0.5} - q_{0.16}}.$$

The minimum threshold for this ratio above which bright extended structure is suspected in a frame can be specified via –ratmax. Testing has revealed that values of $Q_d$ >~ 2 [= default] usually indicate a highly skewed distribution and hence contamination from bright extended structure. If extended structure is detected in a frame, the frame is partitioned into –bgrid × –bgrid squares (default=8 × 8) and the square with the *lowest* median background value is used to represent the underlying frame background. The allows the software to pick a region with minimal bright extended structure. One can therefore force this mode (over planar fitting) by simply setting a small a small $Q_d$ (–ratmax ) value, e.g., 0.5. The –bgrid value can be tuned according to prior knowledge of the spatial scale of expected background variations. This method still has its limitations (e.g., if structure covers an entire frame and there are few pixels to work with to define a background), but it extends the robustness of the algorithm.

### 4.2.2 Method 2: generic surface fitting with outlier rejection

This method can be invoked by specifying "–bmeth 1" and is generally more powerful than method 1 if tuned correctly. Here are the steps.

1. Regularize frame intensity pixels by replacing values > *clsig**σ + mode with upper threshold value: "*clsig**σ + mode". Here *clsig* is the –clsig input parameter and σ is a robust measure of the frame sigma using the lower tail: *median – 16^th^ percentile*. This replacement is known as *winsorisation* in the statistics literature. It reduces the impact of bright sources and structure (outliers) on the fit.

2. Partition the regularized frame into –bfgrid x –bfgrid squares (default = 9 x 9).

3. Compute median ($z$) of pixels in each square partition with grid centers: $x$, $y$.

4. Fit a 2D polynomial: $z = f(x,y)$ of order "–order" (default = 2) to regularized and median-partitioned frame. In general, a 2D polynomial of order $N$ can be written:

   $$z = \sum_{m=0}^{N}\sum_{n=0}^{m} a_{mn} x^{m-n} y^{n},$$

   where $a_{mn}$ are the fit coefficients. The number of coefficients is $(N + 1)(N + 2)/2$.

5. Generate an image of the polynomial fit and subtract from original input frame (after any

throughput matching of course – see §4.1)

6. Compute global median $M$ of all frame pixels contributing to the co-add footprint and add to the background-subtracted frame in 5.

7. Repeat steps 1 – 6 on all input frames. The frames have now been matched to a common background level $M$. This will be more or less representative of the natural background level over the co-add footprint region.

Note that like method 1, no attempt is made to preserve background variations (gradients) spanning the full coadd footprint. Also, it's important to exercise caution on using a too high order fit. The higher the order, the more likely real astrophysical structure at higher spatial frequencies will be removed, especially if the outlier threshold (–clsig) is not properly tuned. We recommend not going above order = 3. Order = 2 is safest if you'd like to retain contiguous, bright extended structure (e.g., a galaxy) spanning <~ 40% of a frame by area, but only if the outlier threshold –clsig is low enough to clip the structure of interest before fitting.

### 4.2.3 Method 3: global minimization of relative offsets in frame overlaps

This method is the most powerful of all and can be invoked by specifying "–bmeth 2". The above methods will usually ensure continuity of the background across a co-add footprint, but they are prone to removing astrophysical structure, especially if it spans a large fraction of an input frame. One can avoid this by matching backgrounds *only* in the common overlap regions between a set of frames. For simplicity, we assume the difference in levels in the overlap region between any two frames is spatially constant. This is adequate for most scenarios. We solve for all overlap offsets (a single number per frame) using a global linear least squares method with no priors or regularizing constraints. The offsets are computed relative to one of the input frames (see below) since the purpose here is only to ensure a continuous, seamless background.

For $N$ input frames with pairs of overlapping frames indexed $m,n$, our cost function to minimize is defined as

$$L = \sum_{m=1}^{N} \sum_{n>m}^{N} \left[p_m + \varepsilon_m - p_n - \varepsilon_n\right]^2, \qquad (4.2.1)$$

where $p_m$, $p_n$ are the "matching" pixel values from overlapping frames $m$ and $n$ respectively, i.e., that "see" the same point on the sky. Since we assume the difference $p_m - p_n$ is spatially constant across an overlap region, this difference can be taken to be the median difference in pixel values, or some other robust estimate. $\varepsilon_m$, $\varepsilon_n$ are the offsets we wish to derive and apply to all pixels of frames $m$ and $n$ respectively so their background levels will match. The sums are over all the input frames that mutually overlap with any other frame. $L$ will be globally minimized at a specific set of values $\varepsilon_i$ (the solution set) where its partial derivatives vanish:

$$\left.\frac{\partial L}{\partial \varepsilon_m}\right|_{\varepsilon_m=\varepsilon_i} = 2\sum_{m=1}^{N}\sum_{n>m}^{N}\delta_{mi}\left[p_m - p_n + \varepsilon_m - \varepsilon_n\right] = 0, \qquad (4.2.2)$$

where

$$\delta_{mi} = \begin{cases} 1 \text{ for } m = i = 1,\ 2,\ 3\ldots\ N-1 \\ 0 \text{ for } m \neq i \end{cases}$$

This is with respect to $N - 1$ values of $\varepsilon_i$ since $N$ includes a "reference" image frame (index $r$) whose offset $\varepsilon_r$ is defined to be exactly zero. I.e., all other frame offsets $\varepsilon_i$ will be relative to this reference frame (see below for how the reference frame is selected). Rearranging Eq. 4.2.2, this leads to a set of $N - 1$ simultaneous equations in $N - 1$ unknowns $\varepsilon_i$ generated from the relation:

$$\sum_{n=1(\neq i,=r)}^{N}\varepsilon_i - \sum_{n=1(\neq i,\neq r)}^{N}\varepsilon_n = -\sum_{n=1(\neq i,=r)}^{N}\left(p_i - p_n\right) \qquad (4.2.3)$$

for frames indexed by $i = 1, 2, 3, \ldots N - 1 \neq r$, i.e., excluding the reference frame $r$ which is implicitly assigned $\varepsilon_r = 0$. For a given frame $i$, the summations in Eq. 4.2.3 are evaluated over all other frames $n$ which *overlap* with $i$.

We define the reference frame to be a new "dummy" image constructed using (i) a WCS cloned from the input image closest to the geometric center of the requested footprint geometry, and (ii) with a constant level equal to the global median of all the input image pixels (after any optional throughput/gain matching). This dummy image is appended to the end of the input image array initially containing $N - 1$ images. We could have chosen any input image as the reference, but this choice avoids picking an image with an abnormal or negative background level (see below for a further measure to avoid this). It also allows all image levels to be corrected more-or-less towards the global median level after the offset corrections are applied.

We can recast the set of simultaneous equations formed by Eq. 4.2.3 as a matrix equation **A.X** = **B** with 'unknown' column vector **X** = $(\varepsilon_1\ \varepsilon_2\ \varepsilon_3\ \ldots\ \varepsilon_{N-1})^{\mathrm{T}}$ and matrix elements for **A** and **B** as follows:

$$A_{ii} = \sum_{n=1(\neq i,=r)}^{N}\delta_{n\cap i}\ ,$$

$$A_{ni} = A_{in} = -\delta_{n\cap i(\neq r)}$$

$$B_i = -\sum_{n=1(\neq i,=r)}^{N}\delta_{n\cap i}\left(p_i - p_n\right)$$

where

$$\delta_{n\cap i} = \begin{cases} 1 \text{ if } n\cap i \neq 0 \text{ (frames } n \text{ and } i \text{ overlap)} \\ 0 \text{ if } n\cap i = 0 \text{ (frames } n \text{ and } i \text{ do not overlap)} \end{cases}$$

The solution to **X** then involves inverting the matrix **A**. After the offset solutions are obtained, they are checked for plausibility. If any abnormally deviate from the overall core-distribution in offset solutions, they are readjusted. This regularization process entails replacing any outlying offset solution $\varepsilon_i$ satisfying $|\varepsilon_i - median\{\varepsilon_i\}| > offtol*\sigma$ with $median\{\varepsilon_i\}$, where σ is a robust estimate of the overall RMS in offsets: $\sigma \approx 0.5*(84\%tile - 16\%tile)$. *offtol* is an input parameter specified by –offtol. Outlying offset solutions should rarely happen, but numerical instabilities in the minimization process could lead to such cases. During processing, we advise checking for output warnings of the type (for example):

```
=== Warning: offset solution 3 deviates from median by > 1000-sigma;
replacing 3000 with 110...\n";
```

If you find this warning happening often, we suggest increasing the input tolerance value supplied to –offtol.

If too many frames end up with many negative pixels after the offset corrections are applied, there will be consequences for downstream processing. First, this can cause problems for the outlier rejection step (if run in integer processing mode with –ip_odet=1) and second, this will cause problems for HiRes which requires positivity. This can happen with input frames having pixel values close to zero and spanning a small dynamic range (e.g., sometimes encountered with *Spitzer*-IRAC data). To avoid or at least mitigate this, we further regularize the final frame levels by adding a global constant to ensure positivity across all pixels. This constant is computed dynamically and is equal to: *globalmed – globalmedmin*, where *globalmed* is the global median of all input frame pixels *before* any background-level matching and *globalmedmin* is the minimum frame median (across all frames) *after* the offset corrections are applied. The latter ensures that frame levels are brought to > zero values (in case any were abnormally negative) before adding the initial global median. However, if you repeatedly find too many negative pixels after background-level matching with –bmeth=2, we suggest rerunning with –bmeth=1, or even turning off background-matching entirely (–bmatch 0) if you think it's unnecessary.

After applying the (possibly regularized) offset solutions to the respective frame pixels, the background matched frames are written to the output working directory specified by –outdir. To speed up processing under this method, the frames may be optionally rebinned (down-sampled) internally using the –reimg switch with a resampling factor of –refac *per axis*. Furthermore, when computing mutual overlap pixel differences $p_i - p_n$, we only use the frame edges and not all the pixels that may be common between two overlapping frames. The number of edge pixels used (inwards from an edge and after any internal rebinning) can be specified by the –edgw parameter. The larger this value, the slower the process. But if –edgw is too small, pixel differences and background solutions will be inaccurate. The current defaults should be adequate for most applications.

We advise that this method be used whenever possible since it does not suffer from the biases that can affect methods 1 and 2. However, there are two crucial requirements for this global minimization method to work: (i) *there must be three or more input frames*, and (ii) *every frame* *must overlap with at least one other frame in the input list by at least one pixel*. If there are any disjoint frames or breaks between clusters of overlapping frames, the matrix **A** will be singular and no solution is possible. You will then need to resort to either method 1 or 2. There will be a fix to circumvent singular cases in a future version.

# 5 OUTLIER DETECTION AND MASKING

## 5.1 Overview

The goal of outlier detection is to identify frame pixel measurements of the same location on the sky which appear inconsistent with the (bulk) remainder of the sample at that location. This assumes multiple frame exposures of the same region of sky are available. Potential outliers include cosmic rays, latents (image persistence), instrumental artifacts (including bad pixels), poorly registered frames from gross pointing errors, supernovae, asteroids, and basically anything that has moved or varied appreciably with respect to the inertial sky over the observation span of a set of overlapping frames. Outlier detection and flagging has been implemented in the AWOD module (A WISE Outlier Detector), and is executed by the *icore* script if the –odet switch has been set.

In summary, the method involves first projecting and interpolating each input frame onto a common grid with user-specified pixel scale optimized for the detector's Point Spread Function (PSF) size. The interpolation is performed using the overlap-area weighting method (analogous to using a top hat kernel). This accentuates and localizes the outliers for optimal detection (e.g., cosmic ray spikes). When all frames have been interpolated, robust estimates of the first and second moments are computed for each interpolated pixel stack $j$. We adopt the sample median (*med*), and the Median Absolute Deviation (MAD) as a proxy for the dispersion:

$$\sigma_j \approx 1.4826\ med\{|p_i - med\{p_i\}|\}, \tag{1}$$

where $p_i$ is the value of the $i^{th}$ interpolated pixel within stack $j$. The factor of 1.4826 is the correction necessary for consistency with the standard deviation of a Normal distribution in the large sample limit. The MAD estimator is relatively immune to the presence of outliers where it exhibits a breakdown point of 50%, i.e., more than half the measurements in a sample will need to be declared outliers before the MAD gives an arbitrarily large error.

The final step involves re-projecting and re-interpolating each input pixel again, but now testing each for outlier status against other values in its stack using the pre-computed robust metrics. A pixel with value $p_i$ is declared an outlier if for given “upper” ($u_{thres}$) and “lower” ($l_{thres}$) tail thresholds, either of the following is satisfied:

$$\begin{aligned} p_i &> med\{p_i\} + u_{thres}\sigma_j \\ p_i &< med\{p_i\} - l_{thres}\sigma_j \end{aligned} \tag{2}$$

The $u_{thres}$ and $l_{thres}$ thresholds are specified by the –tu_odet and –tl_odet command-line inputs to *icore* respectively. If declared an outlier, a bit is set (value specified by –m_odet) in the accompanying frame mask (listed in –msklist) for use downstream. The algorithm also includes an adaptive thresholding method in that if a pixel is likely to contain “real” signal (e.g., from a source), the upper threshold is automatically inflated by a specified amount (–r_odet) to reduce the incidence of outlier flagging at that location. To distinguish between what's real or not, we generate a background subtracted *median*-SNR co-add using all the input pixels. The background and local noise are computed using spatial median filtering and quantile differencing: $\sigma \approx q_{0.5} - q_{0.16}$ respectively. The idea here is that since these metrics are relatively outlier resistant, a large median pixel value in the co-add (or SNR derived therefrom) is likely to contain signal associated with a source. Therefore, when flagging outliers using Eq. 2, we also threshold on the co-add SNR (with minimum tolerable value –ts_odet) to determine if $u_{thres}$ should be inflated by the factor –r_odet. Details are given in §5.2.2.

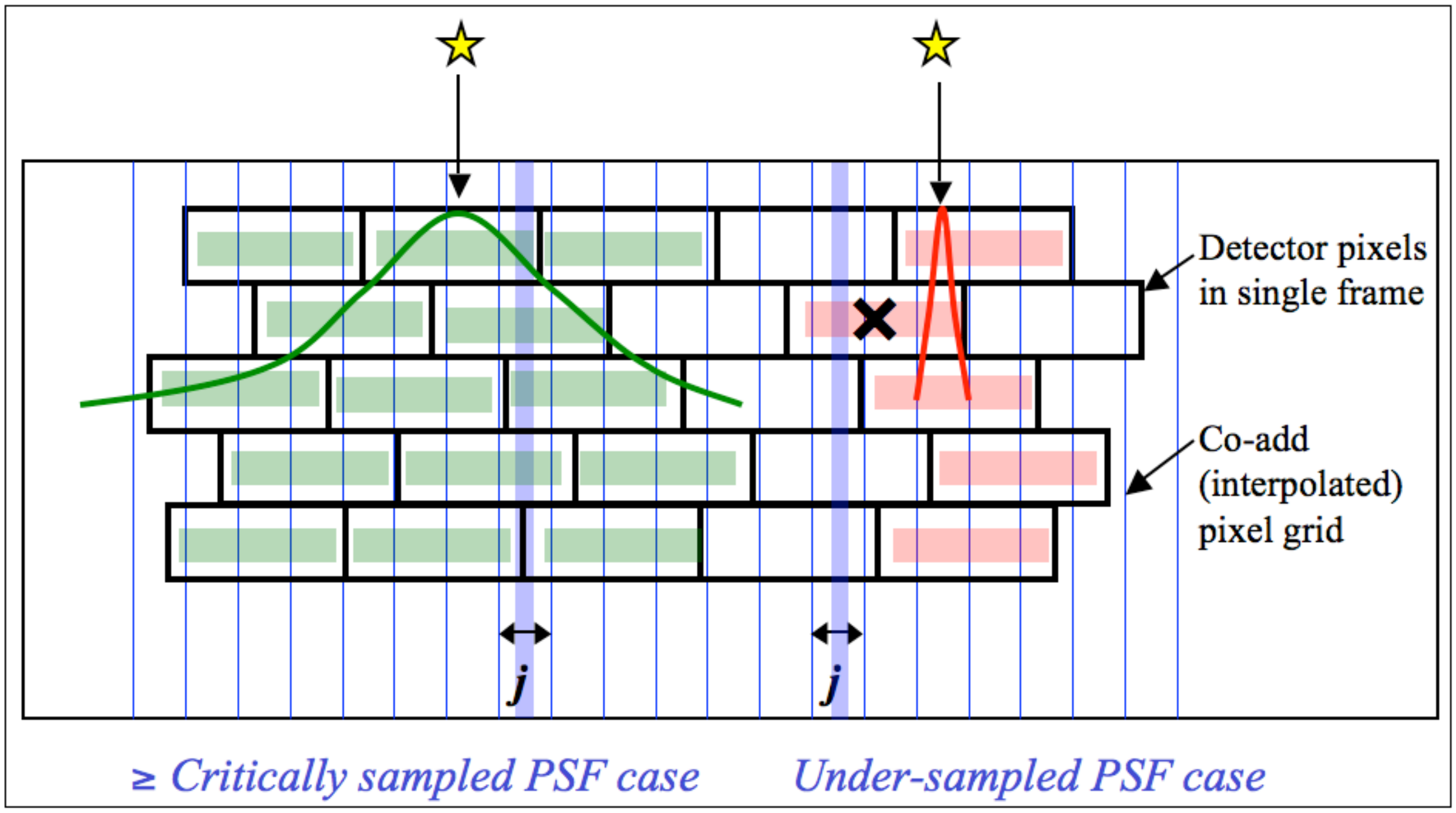

**Figure 4: one-dimensional schematic of stacking method for detecting outliers for well sampled (*left*) and under-sampled (*right*) cases. The input pixel marked "×" contains signal from a source and is in danger of being flagged when outlier detection is performed at location *j* in the output grid.**

We require typically at least five samples (overlapping pixels) in a stack for the above method to be reliable. This is because the MAD measure for $\sigma$, even though robust, can itself be noisy when the sample size is small. Simulations show that for the MAD to achieve the same accuracy as the most optimal estimator of $\sigma$ for a normal distribution (i.e., the sample standard deviation), the sample size needs be ≈2.74× larger. A noisy $\sigma$ will adversely affect the ability to perform reliable outlier detection. The minimum number of samples (or depth-of-coverage) above which AWOD will attempt to test for outliers can be specified by the –ns_odet parameter (default=5). This can be set to 4 as the absolute minimum, but in general, we don't recommend going below 5 since the above method becomes severely unreliable.

Another requirement to ensure good reliability is to have good sampling of the instrumental PSF, i.e., at the Nyquist rate or better. When well sampled, more detector pixels in a stack can be made to align within the span of the PSF, and any pixel variations due to PSF shape are minimized. On the other hand, a PSF which is grossly under sampled can artificially increase the scatter in a stack, with the consequence of erroneously flagging pixels containing true source signal. Figure 4 illustrates these concepts.

Figure 5 shows the Reliability and Completeness of outlier detection using stacks of simulated frames. The simulation contains point sources with PSF sampled at the Nyquist rate, Poisson noise, and single cosmic-ray hits (outliers). For depths-of-coverage ≥ 8 for example, we expect to detect outliers to completeness and reliability levels of >~80% for a nominal threshold of ~5$\sigma$. Note that the only source of "unreliability" in this simulation are noise spikes. The purpose here was to check that the algorithm performs as expected for different thresholds and frame depths.

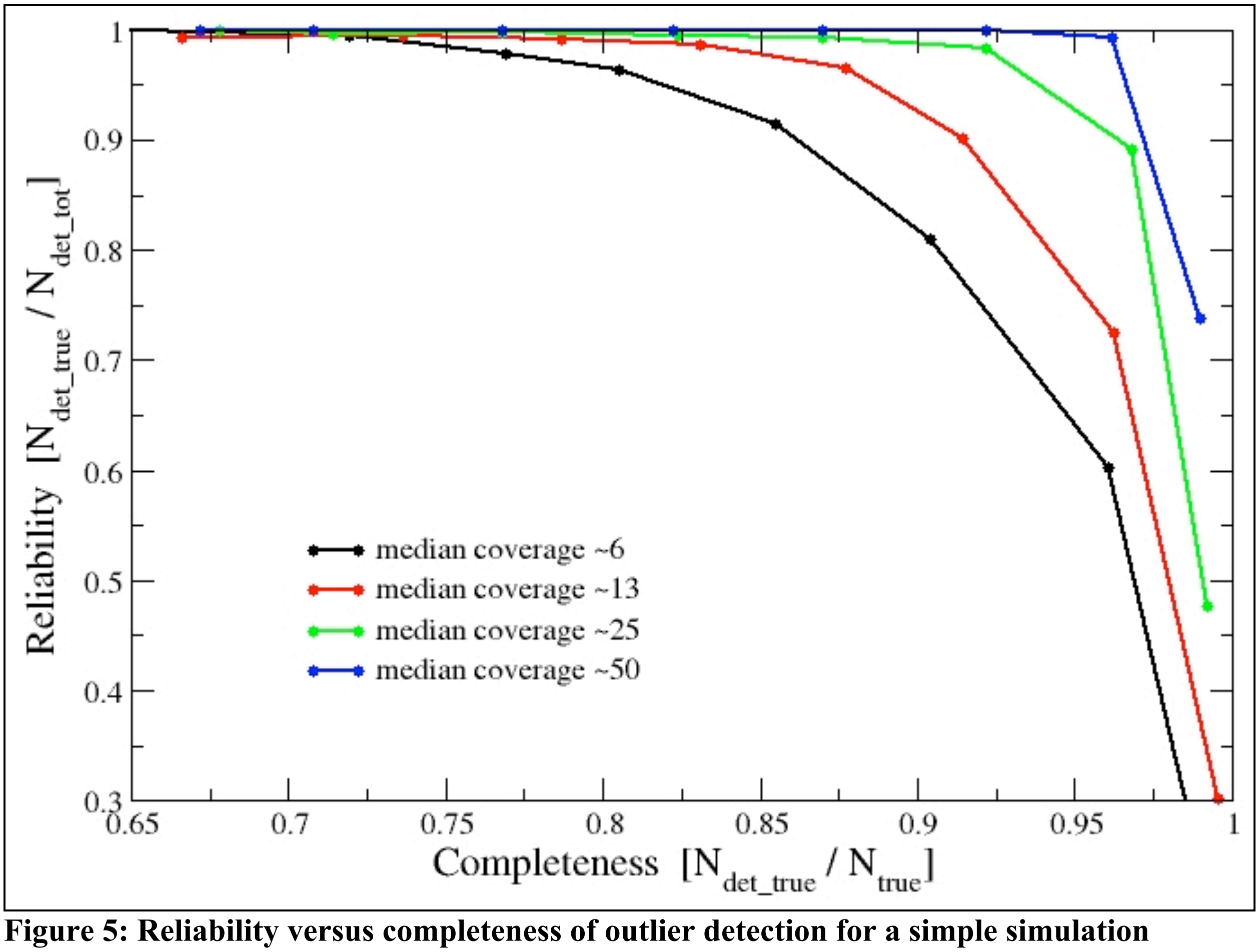

**Figure 5: Reliability versus completeness of outlier detection for a simple simulation containing critically-sampled point sources and Poisson noise. Results are shown for four median depths-of-coverage. The dots on each curve going from right to left correspond to detection thresholds of 3, 4, 5, 6, 7, 8, 9 and 10σ.**

## 5.2 Outlier Detection Implementation Details

Above we gave a broad overview of the outlier detection step. Here we expand on some of the algorithmic details. The parameters pertaining to outlier detection (execution of the AWOD module) are suffixed by "_odet" in the *icore* script. The processing flow is given in Figure 6.

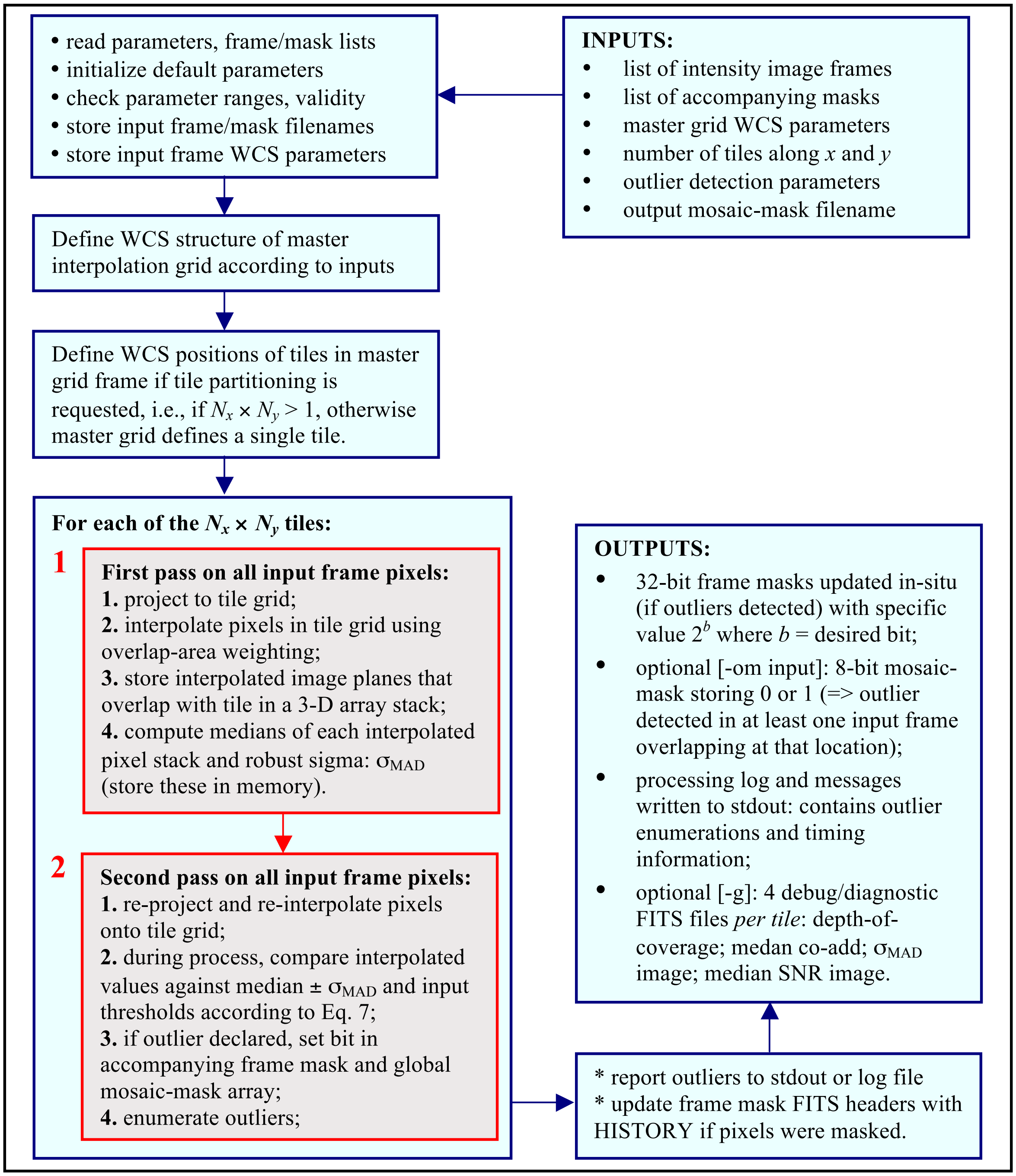

**Figure 6: Processing flow in AWOD. Red boxes represent the main computational steps.**

First, the AWOD program sets up the WCS of the master interpolation grid based on the input parameters: -sizeX, -sizeY, -ra, -dec, -rot, -pa_odet, -pb_odet and other generic WCS keywords from the first listed input frame. In absolute units (projected on the sky), the master grid for outlier detection is slightly larger than the final co-add footprint by the amount "pb_odet", where –pb_odet is an input parameter in *arcsec* (default=0). This parameter is additional padding for the standard footprint defined by -sizeX, -sizeY, and is recommended to be set to the PRF linear size if a PRF-interpolated co-add (or HiRes image) is desired downstream. This is because the PRF-interpolation method for co-addition will attempt to map information from outside the standard footprint from a distance equal to half the PRF size. The "–pb_odet" padding ensures that possible outliers lying just outside the desired footprint (but within the detector's PRF size) will not introduce artifacts on the co-add (or HiRes'd image) near the boundary of the footprint.

The master grid can be further partitioned into $N_x \times N_y$ tiles (via the –nx_odet and –ny_odet inputs). This is to assist with memory management when storing 3-D cubes of interpolated image frames (see §11.2 for computing memory). If tiling is desired, then each tile is assigned its own WCS with unique R.A., Dec, reference pixel coordinates, and rotation (CROTA2) equal to that of the master grid. Each tile will define a "common" interpolation grid for all input frames that overlap with it. If there is enough system memory and partitioning is not desired (i.e., $N_x \times N_y = 1$), then the "common" tile grid is the master grid itself.

For each tile (or master grid for no tiling) there are two passes through all the frame pixels [see two red boxes embedded in Figure 5]. The goal of the first pass is to compute robust (≈ outlier resistant) measures of location and spread in all interpolated pixel stacks (e.g., the median and $\sigma_{MAD}$). These products will be used in the second processing pass to actually detect and flag outliers. We outline the specifics of each step.

### 5.2.1 First Pass Computations

The first pass projects frame pixels using a fast pixel-to-pixel WCS transformation library onto the interpolation grid. This transformation also corrects for FOV distortion using the SIP FITS representation. The interpolation uses the overlap-area weighting method to compute the flux in the output grid pixels. In essence, this method involves projecting the four corners of an input pixel directly onto the output grid (with rotation included). Input/output pixel overlap areas are then computed using a textbook algorithm for the area of a polygon, and then used as weights when summing the contribution from the pixels of an input frame with signals $D_i$. The signal in grid pixel $j$ from frame $k$ is given by:

$$f_{jk} = \frac{\sum_i a_{ikj} D_{ik}}{\sum_i a_{ikj}}, \qquad (3)$$

where $a_{ikj}$ is the overlap area between pixel $i$ from input frame $k$ with output grid pixel $j$ (see right schematic in Figure 7).

The projection and interpolation is only performed for those frames that overlap with the tile grid of interest. This overlap pre-filtering is performed using a coarse method whereby an overlap is declared if the separation between the centers of a tile and input frame is less than the sum of the radii of circles that circumscribe the tile and frame. Note that this frame-to-tile overlap test is only performed if "–d_odet 1" is specified (the default). This should be set to 0 if co-adding in the rest frame of a moving object (see §8.9), indicating not to perform this test and not to omit frames that don't overlap. This is because the WCS in each frame is readjusted to force the frames to overlap and "fool" the co-adder downstream. The interpolated planes for each overlapping frame are stored in a 3-D cube, whose memory allocation is described in §11.2.

The median is then computed for each interpolated pixel stack *j* by first sorting the pixel samples using the *heapsort* algorithm. This uses the *gsl_sort_float*() routine from the GNU Scientific Library. The median is computed as the 50$^{th}$ percentile of the sorted stack, and we account for even numbered samples. We also compute a robust measure of the dispersion. For speed and efficiency, we use an approximation to the actual $\sigma_{MAD}$ defined by Eq. 1 and call it the pseudo-MAD. This approximation works directly off a sorted array by selecting the appropriate indices. We define the pseudo-MAD as follows, where $k = 0, 1, 2, 3, \ldots N$ is the sample number in a sorted pixel stack, *stack*[*k*], at location *j* in the interpolation grid.

$$\sigma_{MADj} \approx 1.482602 \times \frac{1}{4} \times \left\{ \left( stack\left[\frac{3N}{4}\right] + stack\left[\frac{3N}{4} - 1\right] \right) - \left( stack\left[\frac{N}{4}\right] + stack\left[\frac{N}{4} + 1\right] \right) \right\}, \tag{4}$$

where the factor 1.482… is the correction to ensure Normality in the large sample limit. The array indices are rounded down to the nearest integer. The depth-of-coverage $N$ (number of samples in a stack) must satisfy $N \geq$ ns_odet (default=5). If $N <$ ns_odet, the $\sigma_{MADj}$ value is set to the hard-coded "flag" value -10000 so it can be detected and avoided downstream.

There is an option to smooth the $\sigma_{MADj}$ image values using spatial median filtering. This is only performed if "–b_odet 1" is specified on the command-line. The default is to perform no median filtering. If requested, the window side-length of the median filter (in odd number of grid pixels) can be specified by –w_odet (default=3). The reason for smoothing the $\sigma_{MADj}$ image is to avoid using erroneous (usually underestimated) values of $\sigma_{MADj}$ when the depth-of-coverage is low, i.e., $N <\sim 10$. As mentioned above, $\sigma_{MADj}$ itself is not an efficient estimator of scale for small samples. Smoothing replaces each value with a median of the neighborhood, and this is expected to be more-or-less representative of the "true dispersion" in the stack at each location. If for example, the $\sigma_{MADj}$ values happen to be underestimated in some regions (due to their noisy nature), this will bias flux thresholds ($med_j \pm t\sigma_{MADj}$ ) towards low values and hence inadvertently flag "good samples" as outliers. There is also an option to scale the *post-filtered* $\sigma_{MADj}$ values with a factor specified by –s_odet (default = 1.5). This allows the user to scale the $\sigma_{MADj}$ values for each stack for consistency with expected RMS fluctuations in an image (Gaussian or otherwise). This enables one to set thresholds according to a prior measured image

RMS, rather than the pseudo-MAD in pixel stacks that may be subject to under-sampling and other systematics.

AWOD includes an adaptive thresholding method in that if it encounters a pixel that's likely to be associated with a "real" signal (e.g., a source), then the upper and lower thresholds are automatically inflated by a specified amount to avoid (or reduce the incidence of) flagging such sources as outliers in the second pass (see §5.2.2). To distinguish between what's real or not, we compute a "median SNR" image corresponding to the interpolated pixels $j$. This is computed using:

$$SNR_j \approx \frac{median_j - bckgnd}{RMS_j}, \quad (5)$$

$$\begin{aligned}
&\text{where:}\\
&bckgnd = median\left(median_j;\ \text{for all}\ j \in \left\{median\ N_j\right\}\right);\\
&RMS_j \approx RMS\left(median_j;\ \text{for all}\ j \in \left\{median\ N_j\right\}\right) \times \sqrt{\frac{median\ N_j}{N_j}},\\
&\text{and the base RMS:}\\
&RMS\left(median_j;\ \text{for all}\ j \in \left\{median\ N_j\right\}\right) \approx bckgnd - 16^{th}\,\text{ptile}.
\end{aligned}$$

The first term in the numerator is the median signal for pixel stack $j$, and the second term is a *global* background measure. This background is computed as the median signal over all stack medians such that their depth-of-coverage $N_j$ is equal to the *median depth* over the entire tile grid. The denominator is an estimate of the spatial RMS fluctuation for pixel stacks at the median depth-of-coverage, relative to their median signal (the base RMS), and then appropriately rescaled to represent the pseudo-local RMS at any depth $N_j$ (third line in Eq. 5). More precisely, the base RMS is computed using the lower tail values of the pixel distribution to avoid being biased by sources and other "spatial outliers". The 16$^{th}$ percentile (or more exactly, the quantile corresponding to a lower-tail probability of 0.1586) is used, analogous to the standard deviation of a Normal distribution, which can also be derived from quantiles: $\sigma = \mu - q_{0.1586} = 0.5*(q_{0.8413} - q_{0.1586})$. The local SNR computed this way is an approximation and assumes the background does not vary wildly. The SNR image (Eq. 5) is used for regularizing the $\sigma_{MADj}$ image (see next paraegraph) and as a proxy to determine if a location contains real source signal to assist the outlier flagging process (see 2$^{nd}$ pass below).

AWOD has an option to regularize (or homogenize) the $\sigma_{MADj}$ values where they are too low relative to their overall neighboring values, or too high in regions below some minimum SNR (using the median SNR image described above). This regularization occurs *before* any *local* smoothing of the $\sigma_{MADj}$ values through the –b_odet parameter described above. It ensures global robustness and stability of $\sigma_{MADj}$ values against wild frame backgrounds (e.g., that may have not been properly background matched) or noisy regions where the depth-of-coverage (stack sample size) is low, causing the $\sigma_{MADj}$ themselves to be noisy and unreliable for outlier detection. This operation is only performed if "–odet_h 1" is specified. In general, the homogenization process involves replacing $\sigma_{MADj}$ values with their global median over the tile (or footprint) if the following criteria are satisfied:

$$\sigma_{MADj} > median[\sigma_{MADj}] + n_{thres}\sigma_{spatial}[\sigma_{MADj}] \text{ and } SNR_j \leq SNR_{\min}$$

or

$$\sigma_{MADj} < median[\sigma_{MADj}] - n_{thres}\sigma_{spatial}[\sigma_{MADj}] \text{ regardless of } SNR_j$$

where the parameters $n_{thres}$ and $SNR_{min}$ are the command-line inputs: –q_odet and –k_odet respectively. $\sigma_{spatial}$ is a robust estimator of the global (spatial) variation in $\sigma_{MADj}$: *median – 16$^{th}$ percentile*.

At the end of first pass computations, we have three image products stored in memory for a tile grid: median, pseudo-MAD ($\sigma_{MAD}$), and a SNR image. These are needed for the second pass. If the debug switch (–sdbg) is set, these three images and the depth-of-coverage map are saved to disk as FITS images in the directory specified by –outdir.

### 5.2.2 Second Pass Computations

Given the robust metrics from the first pass, we now re-project (with distortion correction) and re-interpolate all the input frame pixels onto the tile grid again. The only difference here is that as each pixel from the $k^{\text{th}}$ input frame is projected, it is compared to the *existing* median, $\sigma_{MAD}$ and SNR values at the same pixel location $j$ in the tile grid to determine if it is an outlier. In general, an interpolated pixel from frame $k$, $f_{jk}$ (Eq. 3) is declared an outlier with respect to other pixels in stack $j$ if its value satisfies:

$$f_{jk} > median_j + u_{thres}\sigma_{MADj} \text{ and } SNR_j \leq SNR_{\min} \quad (6)$$

or

$$f_{jk} > median_j + r * u_{thres}\sigma_{MADj} \text{ and } SNR_j > SNR_{\min}$$

or

$$f_{jk} < median_j - l_{thres}\sigma_{MADj} \text{ and } SNR_j \leq SNR_{\min}$$

or

$$f_{jk} < median_j - r * l_{thres}\sigma_{MADj} \text{ and } SNR_j > SNR_{\min}$$

where the parameters $u_{thres}$, $l_{thres}$, $r$, and $SNR_{min}$ correspond to the command-line inputs: –tu_odet, –tl_odet, –r_odet, and –ts_odet respectively. This is a refinement to Eq. 2. The addition here is that the upper-tail threshold, $u_{thres}$ is inflated by the factor $r$ if the *median SNR* in pixel $j$ exceeds some $SNR_{min}$ of interest. The median is relatively immune to outliers (as long as they don't contaminate ≥50% of the sample), therefore, if the median is relatively large, it's likely that the pixel contains "real" source signal. The inflation factor $r$ (–r_odet) can be set to some arbitrarily large value to avoid masking at locations where $SNR > SNR_{min}$ (–ts _odet). An example where one may want to invoke this is with undersampled data (see Figure 4).

If an input pixel from frame *k* leads to an outlier in the interpolated space of some tile, then a bit-value is set in it's accompanying mask image. The mask value is specified by the command-line input: –m_odet <$2^b$ *value*>, where *b* is the required bit. The default "–m_odet 0" means *no* mask updating. This could be useful if running a test to get an idea of the outlier count for threshold tuning purposes. To avoid updating (and possibly corrupting) the original input masks with outlier information, it is strongly recommended that you specify the –cpmsk command-line switch so that the original masks are copied into the local processing directory –outdir before "destructively" updating them.

In addition to updating the input masks, a value "1" is also written to an output mask array at the appropriate WCS location with the same dimensions as the master interpolation grid. This is only performed if an output FITS filename: –om_odet <*outfname*> was specified. This product is also referred to as an *outlier map* in this document. This map indicates that an outlier was detected in *at least one* of the frame pixels in the stack at that location on the sky.

Another detail in second pass processing is estimating the interpolated pixel flux $f_{jk}$. This can be either done "exactly" using the area-overlap method (e.g., Eq. 3), or approximately using nearest-neighbor interpolation. The approximate method was implemented for speed, and its accuracy depends on the ratio of output-to-input pixel area, $a_j / a_i$. The value at which the nearest-neighbor method kicks in can be controlled by the –ta_odet <*max out/in*> parameter. This threshold represents the maximum ratio of output-to-input pixel area below which the nearest neighbor method is used. Above this threshold, the exact polygonal area-overlap method is triggered (Eq. 3). The default value for –ta_odet is 0.25, i.e., if an output pixel occupies a quarter of an input pixel or less by area (i.e., $a_j / a_i \leq 0.25$), then the overlap will start to be appreciable (left schematic in Figure 7). The smaller this ratio, more output pixels will overlap *entirely* with an input pixel. In this case, the output pixel that's nearest to the projected input pixel's location will be assigned a signal $f_j \approx (a_j / a_i)D_i$ in interpolated space.

If however $a_j / a_i > 0.25$, it is *less likely* that an input pixel will entirely overlap an output pixel (right schematic in Figure 7). In this case, all four corners of the input pixel need to be projected to compute the exact overlap area enclosed by a polygon. This is more time-consuming, but it uses less system memory in the end. If the ratio of output-to-input pixel area is >~ 0.25, then the exact overlap-area method is strongly recommended to ensure the best interpolation accuracy and representation of input pixels. Unless execution time is critical, we advise keeping "–ta_odet 0.25" (the default). If the nearest neighbor method is forced when the output/input pixel areas are really > 0.25, the variance in interpolated pixel stacks will be inflated. To compensate, you'll need to inflate the thresholds for outlier detection (–tl_odet and –tu_odet), otherwise you may end up flagging many genuine input pixels as outliers.

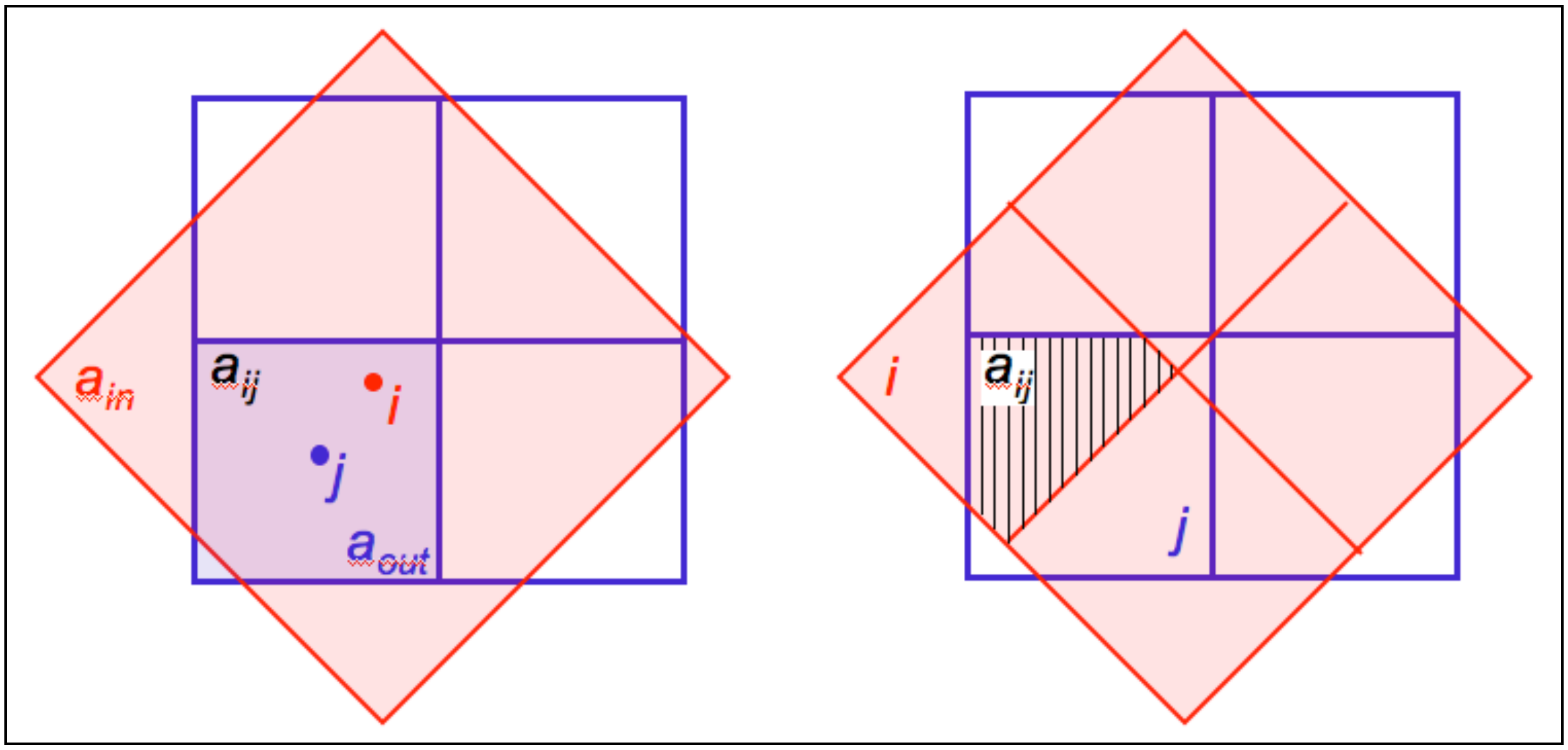

**Figure 7: Left: schematic where $a_{out}/a_{in}$ <~ 0.25 with overlap area $a_{ij} \approx a_{out}$. Here the nearest neighbor to output pixel *j* is input pixel *i*. Right: schematic where $a_{out}/a_{in} = 1$ and the overlap area $a_{ij}$ (shaded polygon) is formally computed.**

At the end of second pass processing, outlier pixels are enumerated *per frame* and reported to standard output. It's important to note that these enumerations may contain duplicates if partitioning of the master grid into tiles was specified (i.e., with –nx_odet > 1 and/or –ny_odet > 1). This is because each tile is padded with additional pixels to avoid incomplete representation of input pixels near tile boundaries after projection. Duplicates in the reported outlier enumerations will never occur in single-tile mode (–nx 1 and –ny 1), assuming one has sufficient memory to run in this mode. In the end, the total number of outliers detected per frame can be inferred by counting the number of pixels in the mask images updated with the outlier-bit. Average and median statistics on the number of outliers detected per frame are also reported in the output QA meta-data table (–qameta *<outfilename>*).

### 5.2.3 Optimization Options

The *icore* script includes an option to manage memory usage for outlier detection. This can be exercised by setting the "–partition" switch. Here the input image (and accompanying mask) list is partitioned if the number of input images exceeds a maximum specified by –nmaxodet. The number of tiles along *x* and *y* (–nx_odet and –ny_odet) are then reset to 1. In brief, the input image (and mask) files are first randomized to ensure the final partitioned lists contain images that more-or-less randomly (and uniformly) sample the footprint depth-of-coverage. This is because the input lists could already be ordered according to sky location for example and hence a serial partition thereof will not sample the whole footprint. Next, the optimum number of images per partition $N_{opt}$ is computed by recursively halving the number of inputs $N_{tot}$ until $floor(N_{opt}) \leq N_{max}$ (where –nmaxodet $< N_{max} >$) is satisfied. The input images are then segregated into "$floor[(N_{tot}/N_{opt}) - 1]$" sublists, each containing $N_{opt}$ images. The remaining "$N_{tot} - N_{opt}$

*floor*[($N_{tot}$/$N_{opt}$) – 1]" images are then added to another sublist. So for example, if $N_{tot}$ = 161 and $N_{max}$ = 150, there will be two sublists, each containing 80 and 81 images. AWOD is then executed on each sublist ***k*** and outlier image maps are generated for each run in the format: <*basename*_sublist***k***.fits> where *basename* is read from the input specification: –om_odet <*basename*>. All "sublist" outlier maps are combined at the end to generate a single overall map with filename specified by –om_odet <*basename*>. See §11.2 for details on how to compute the expected memory allocation under "–partition" processing.

Another feature in the outlier detection step is an option to "expand" outlier-masked pixels initially found above to ensure *more* complete masking of partially masked regions. This may occur from latent (image persistence) artifacts, moving objects, and/or other outlier "fuzzies" whose cores may only have been completely flagged. This post-processing is only triggered if the –expodet switch is set. The masks (residing in –outdir if –cpmsk was specified) are updated with additional outliers. The number of additional outliers is also written to standard output. The input parameters controlling this step are: –nei_odet, –nsz_odet, and –exp_odet. In brief, if a frame pixel was tagged as an outlier upstream and if a region of size "nsz_odet × nsz_odet" centered on it contains > nei_odet additional outliers (where nei_odet ≤ nsz_odet × nsz_odet), outlier expansion will be triggered around all outliers in this region. This entails blanketing "exp_odet × exp_odet" pixels around each outlier into more outliers (with bit specified by –m_odet in the mask). Figure 8 illustrates the mechanism. If the –dbg switch is set, residual FITS image masks of "new (outlier expanded) mask – input mask" are generated.

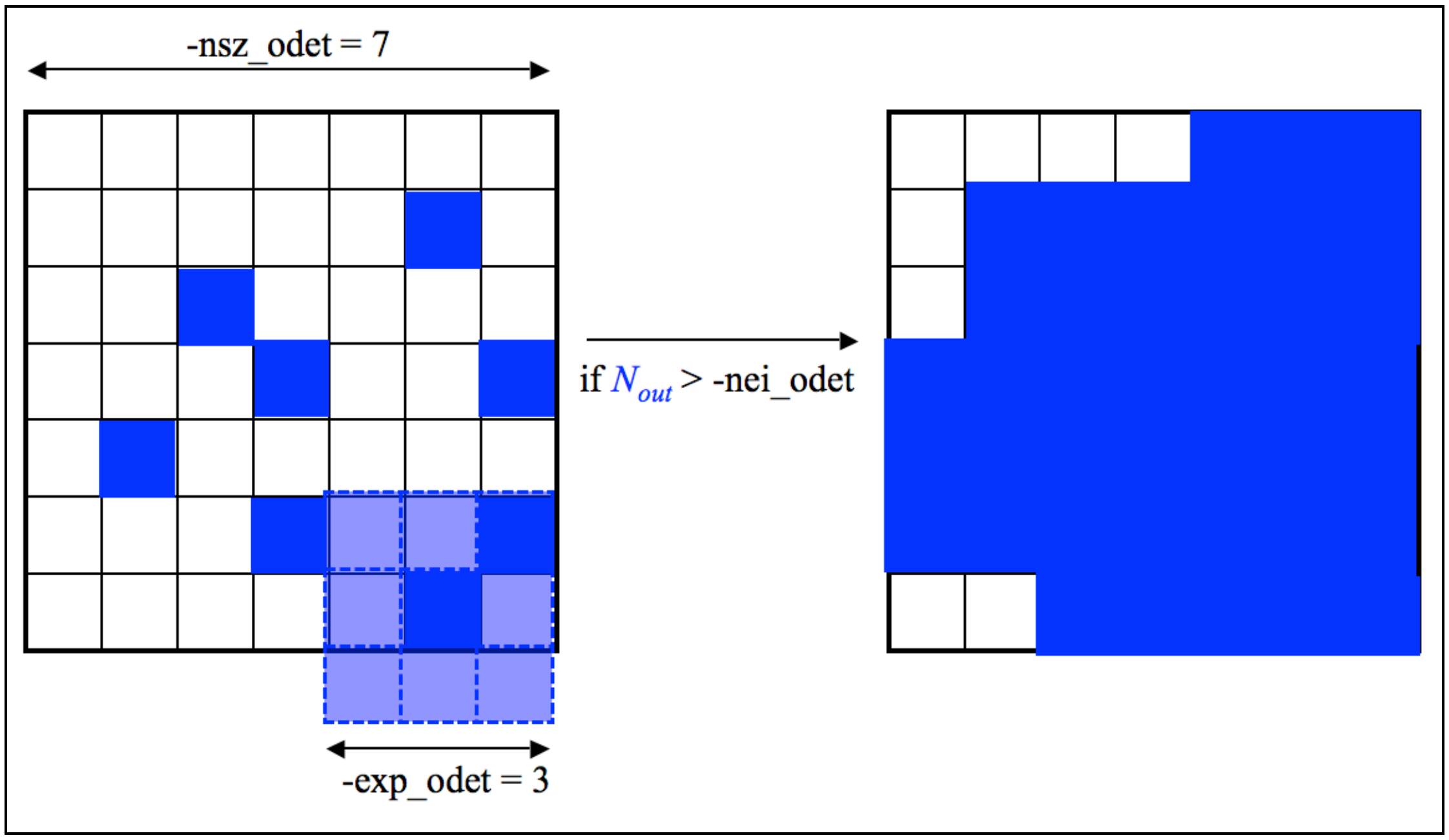

**Figure 8: Schematic of outlier expansion method described in §5.2.3. $N_{out}$ is the number of outliers initially detected within the region "nsz_odet × nsz_odet". If this number exceeds the user-specified threshold "-nei_odet", an additional "exp_odet × exp_odet – 1" outliers will be tagged around each pre-existing outlier. On the right is the final blanketed region for this example.**

Since outlier detection is the most memory intensive task, memory usage can be further reduced by invoking 2-byte integer storage and arithmetic when computing the stack estimators. This can be invoked by specifying: "–ip_odet 1" (where default = 0 ⇒ use 4-byte floating point). If specified, the floating-point values of *interpolated* pixels $p_f$ from each 2-D frame are transformed into the 2-byte dynamic range: $-2^{15} \leq p_i \leq 2^{15} - 1$ ($-32768 \leq p_i \leq 32767$) using a scaled log transformation:

$$p_i = scale * \log_e[p_f],$$

where $p_f > 0$ and *scale* is a scaling factor specified by the –is_odet input parameter. This parameter has default value 2000.0 and is tuned for WISE data. For the general case, the scale factor can be tuned such that it approximately satisfies:

$$scale < \frac{32767}{\log_e\left[\max(p_f)\right]},$$

where $\max(p_f)$ is the expected (approximate) maximum value of an *interpolated* pixel (or some percentile in the high-tail of the interpolated pixel distribution). This can be estimated from the typical maximum value of an input frame pixel multiplied by the interpolation factor: *output_pixel_area*/*input_pixel_area* ~ (*pa_odet***pa_odet*)/(*x_scal***y_scal*) where *pa_odet* is the pixel scale of the output interpolation grid (parameter: –pa_odet) and *x_scal***y_scal* is the input frame pixel area. If finding a suitable scale factor proves difficult, we suggest using the default floating-point processing (–is_odet 0).

### 5.2.4 Frame-based outlier rejection

To further ensure coadd quality and optimize S/N, *icore* includes an option to reject entire frames from co-addition based on the number of "bad" pixels tagged upstream. The various flavors of bad pixels enumerated per frame, examples of possible causes, their default maximum tolerable numbers above which a frame will be rejected if any are exceeded, and input parameters defining these are summarized in Table 2. Note that the maximum possible thresholds are the number of pixels in each frame. The default thresholds are set intentionally high (a giga-pixel) so this functionality is not triggered in the general case, however, some detectors are now starting to exceed a giga-pixel and the thresholds should be reset accordingly.

| Bad pixel flavor | Examples | Parameter threshold | Default threshold [#pixels] |
|---|---|---|---|
| Temporal outliers using pixel-stack statistics from AWOD. Includes spatially expanded outliers as described in §5.2.3. Note: there is no temporal outlier detection for depths-of-coverage ≤ 4 | Moon glow; moving objects and heavy streaking from satellite trails; high cosmic ray count from SAA; other glitches; bad pointing solution; latents and diffraction spike smearing at high ecliptic latitudes | **-tn_odet**: threshold for max tolerable number of temporal outliers per frame above which entire frame will be omitted from co-addition | 1.0E+09 |
| Spike or hard-edged pixel glitches/outliers detected using median filtering in ICal pipeline. Includes hi and lo spikes. Mask bit-string param. for specifying glitch pixels is –mg_odet | Heavy cosmic-ray hit rate, e.g., from SAA; solar storms; excessive popcorn-like noise | **-tg_odet**: threshold for max tolerable number of spike-glitch pixels per frame above which entire frame will be omitted from co-addition | 1.0E+09 |
| Saturated pixels from saturation tagging in ICal pipeline (in any SUTR). Mask bit-string param. for specifying saturated pixels is –ms_coad | Warm transient behavior; anneals (if not explicitly filtered upstream); super-saturating sources | **-tsat_odet**: Threshold for max tolerable number of saturated pixels per frame above which entire frame will be omitted from co-addition | 1.0E+09 |

**Table 2: Frame-outlier rejection criteria**

The filenames of all rejected frames from the above operation are written to standard output (if –v was specified) and listed in a text file suffixed with *_bad.txt* under –outdir.

## 6 CO-ADDITION USING PRF INTERPOLATION

One of the interpolation methods in ICORE involves using the detector's Point Response Function (PRF) as the interpolation kernel. The PRF is simply the instrumental PSF convolved with the pixel response. When knowledge of the intra-pixel responsivity is absent, the pixel response is assumed to be uniform, i.e., a top hat. The PRF is what one usually measures off an image using the profiles of point sources. Each pixel can be thought as collecting light from its vicinity with an efficiency described by the PRF.

The PRF can be used to estimate the flux at any point in space as follows. In general, the flux in an output pixel $j$ is estimated by combining the input detector pixel measurements $D_i$ using PRF-weighted averaging:

$$f_j = \frac{\sum_i \left( r_{ij} / \sigma_i^2 \right) D_i}{\sum_i r_{ij} / \sigma_i^2}, \qquad (7)$$

where $r_{ij}$ is the value of the PRF from input pixel $i$ at the location of output pixel $j$. The $r_{ij}$ are volume normalized to unity, i.e., for each $i$, $\sum_j r_{ij} = 1$. This will ensure flux is conserved. The inverse-variance weights ($1/\sigma_i^2$) are optional and default to 1 (parameter –wf_coad). The $\sigma_i$ can be fed into ICORE as 1-$\sigma$ uncertainty frames, e.g., as propagated from a prior noise model. The sums in Eq. 7 are over all input pixels in all input frames. Eq. 7 represents the intensity image of a "PRF-interpolated" co-add (invoked with "–n_coad 1" and written to output: –o1_coad). With multiple overlapping input frames, this will result in a co-add. The 1-$\sigma$ uncertainty in the co-add pixel flux $f_j$, as derived from Eq. 7 is given by

$$\sigma_j = \left[ \sum_i w_{ij}^2 \sigma_i^2 \right]^{1/2}, \qquad (8)$$

where $w_{ij} = (r_{ij}/\sigma_i^2) / \sum_i r_{ij}/\sigma_i^2$. Equation 8 assumes the measurement errors (in the $D_i$) are uncorrelated. Note that this represents the co-add flux uncertainty based on priors (output is specified by –o3_coad). With $N_f$ overlapping input frames and assuming $\sigma_i$ = constant throughout, it's not difficult to show that Eq. 8 scales as: $\sigma_j \sim \sigma_i / \sqrt{(N_f P_j)}$, where $P_j = 1 / \sum_i r_{ij}^2$ is a characteristic of the detector's PRF, usually referred to as the effective number of "noise pixels". This scaling also assumes that the PRF is isoplanatic (has fixed shape over the focal plane) so that $P_j$ = constant. Furthermore, the depth-of-coverage at co-add pixel $j$ (output –o2_coad) is given by the sum of all overlapping PRF contributions at that location: $N_j = \sum_i r_{ij}$. This effectively indicates how many times a point on the sky was visited by the PRF of a "good" detector pixel $i$, i.e., not rejected due to prior-masking. If no input pixels were masked, this reduces to the number of frame overlaps, $N_f$.

In general, the PRF is usually non-isoplanatic, especially for large detector arrays. ICORE allows for a list of spatially varying PRFs to be specified (via –psflist), where each PRF corresponds to some pre-determined region (e.g., a partition of a square grid) on the detector focal plane.

Figure 9 shows a schematic of a detector PRF mapped onto the co-add output grid. The PRF boundary is shown as the dashed circle and is centered on the detector pixel. To ensure accurate mapping of PRF pixels and interpolation onto the co-add grid, a finer cell-grid composed of "pixel cells" is set up internally. The cell size can be selected according to the accuracy to which the PRF can be positioned. The PRF is subject to thermal fluctuations in the optical system as well as pointing errors if multiple frames are being combined. Therefore, it does not make sense to have a cell-grid finer than the measured positional accuracy of the PRF. The cell pixel linear size is given by the parameter product: "pc_coad*pa_coad" (arcsec) and is constructed such that it also equals the input PRF pixel size to within some tolerance specified by –ct_coad (default = 0.0001 arcsec; see also §3.2 for specification details).

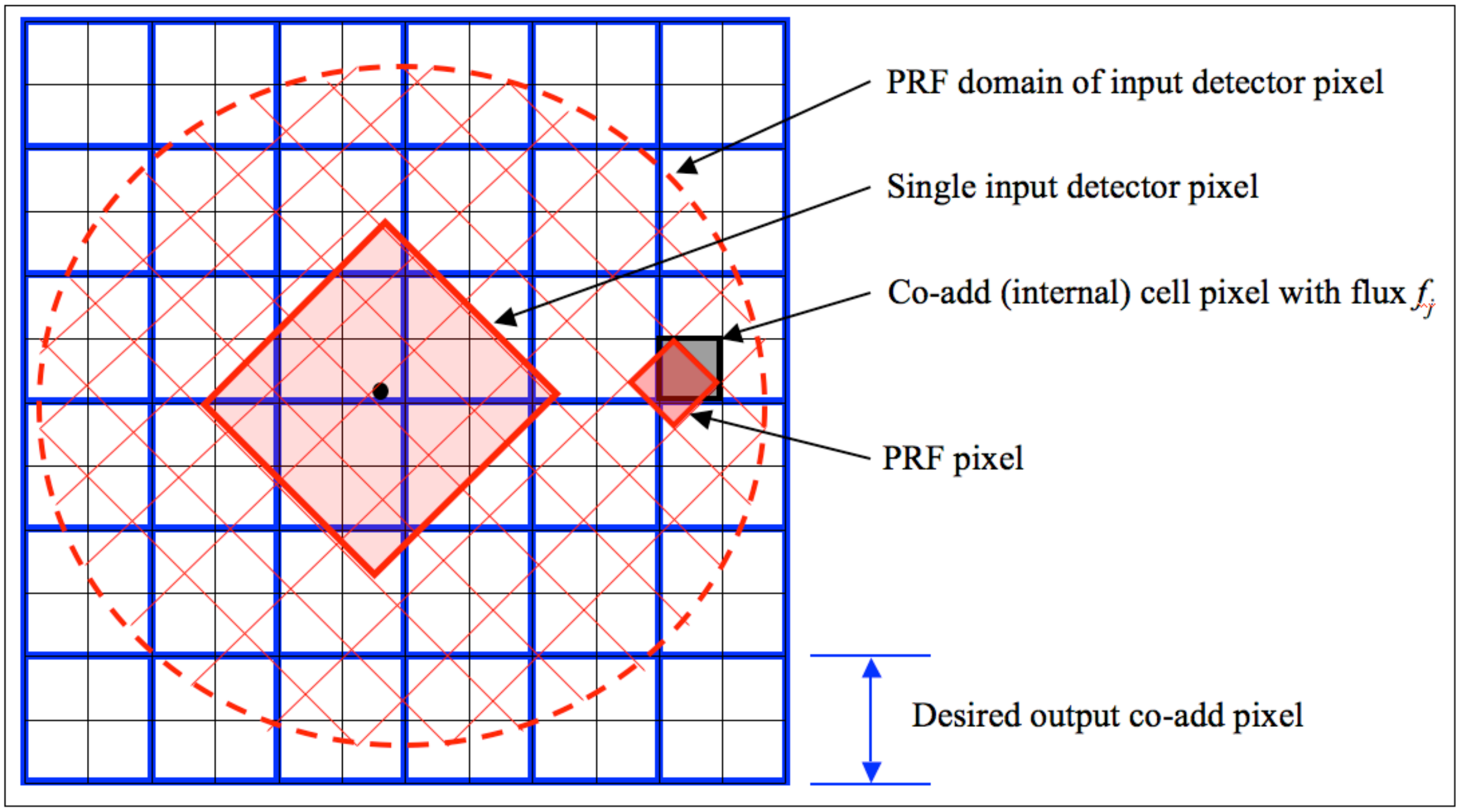

**Figure 9: Schematic of PRF interpolation for a single input pixel.**

The PRF pixels are mapped onto the cell-grid frame by first projecting the center of the detector pixel with distortion correction if necessary, and then using a local transformation with rotation to determine the positions of the PRF pixels in the cell-grid:

$$\begin{pmatrix} C_x \\ C_y \end{pmatrix} = \begin{pmatrix} D_x \\ D_y \end{pmatrix} + \begin{pmatrix} \cos\theta & -\sin\theta \\ \sin\theta & \cos\theta \end{pmatrix} \begin{pmatrix} P_x - P_x^0 \\ P_y - P_y^0 \end{pmatrix},$$

where ($P_x$, $P_y$) is a PRF pixel coordinate, ($P^0{}_x$, $P^0{}_y$) center-pixel coordinates of the PRF, ($D_x$, $D_y$) are the detector pixel coordinates in the cell-grid frame, and the outputs ($C_x$, $C_y$) are the coordinates of the PRF pixel in the cell-grid frame. $\theta$ is the rotation of the input image with respect to the co-add (footprint) frame. Rotation of the input PRFs can be turned on/off using the –rf_coad option, where "–rf_coad 0" implies no rotation and we explicity set $\theta = 0$. Turning this off can significantly speed up processing and should suffice for PRF-interpolated co-adds. For HiRes'd products however (which use the same interpolation scheme), we recommend that PRF rotation be invoked (with "–rf_coad 1"), unless of coarse the input PRFs are sufficiently axisymmetric. "Nearest-neighbor" interpolation then entails rounding the output coordinates ($C_x$, $C_y$) to the nearest integer pixel. Note that if the input PRFs are too coarsely sampled, then systematic variations may result in the output products (see §3.2 for details on how to alleviate this).

The value of a PRF-weighted detector pixel flux $r_{ij}D_i$ in a co-add cell pixel $j$ can then computed using either a nearest-neighbor match (–if _coad 0), or, an overlap-area weighting method *with no rotation* of input detector PRF-to-cell grid (–if _coad 1). The latter is more accurate but slower. The nearest neighbor method should suffice for most purposes. Under the overlap-area PRF-interpolation scheme (–if _coad 1), the contribution from an input PRF pixel (for a given detector pixel) is apportioned, by area, to its four neighboring cell pixels. A schematic is shown in Figure 10. Given overlap areas $a_{ijk}$ between a PRF pixel $j$ (from detector pixel $i$) and four neighboring cell pixels $k$ of the *same size*, Eq. 7 for output cell pixel flux is replaced by:

$$f_k = \frac{\sum_i \sum_j^{N_{PRF}} \frac{a_{ijk} r_{ij} D_i}{\sigma_i^2}}{\sum_i \sum_j^{N_{PRF}} \frac{a_{ijk} r_{ij}}{\sigma_i^2}} \qquad (9)$$

It's important to note that the PRF transformed in this manner (regardless of interpolation method) does not use the full *non-linear* WCS transformation from input image to co-add frame. For large co-add footprints (linear extent >~ 16°), there could be an inaccurate representation of the PRF flux *distribution* towards the footprint edges - a consequence of projection off a sphere. After all the input pixels with their PRFs have been mapped, the internal co-add cells are down-sampled to the desired output co-add pixel sizes (–pa_coad).

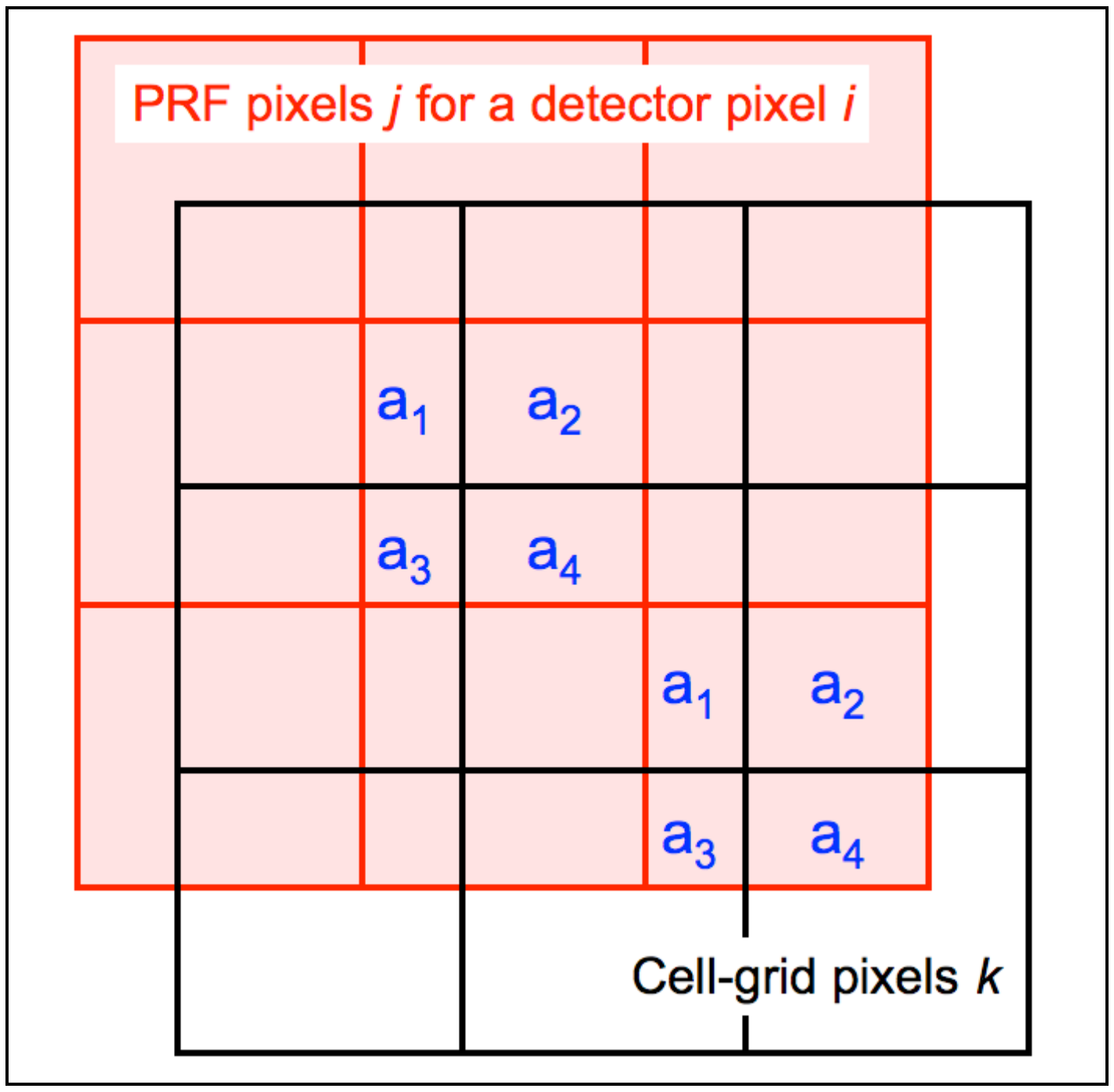

**Figure 10: Schematic of PRF pixel-to-cell grid interpolation using area-overlap weighting. Four neighboring cell pixels are incremented in proportion to the PRF pixel overlap areas.**

### 6.1 Advantages and Pitfalls of PRF-Interpolation

There are three advantages to using the PRF as an interpolation kernel. First, it reduces the impact of masked (missing) input pixels if the data are well sampled, even close to Nyquist. This is because the PRF tails of neighboring "good" pixels can overlap and stretch into the bad pixel locations to effectively give a non-zero coverage and signal there in the co-add. Second, Eqs 7 and 8 can be used to define a linear matched filter optimized for point source detection. This effectively represents a cross-covariance of a point source template (the PRF) with the input data. It leads to smoothing of the high-frequency noise without affecting the point source signal sought. In other words, the SNR per pixel in the co-add is maximized for detecting point source peaks. The inclusion of inverse-variance weighting further ensures that the SNR is maximized since it implies the co-add fluxes will be maximum likelihood estimates for normally distributed data. The third advantage is that the PRF allows for resolution enhancement by "deconvolving" its effects from the input data (see §8).

Use of the PRF as an interpolation kernel also has its pitfalls, at least for the process of co-add generation. The operation defined by Eq. 7 leads to a "smoothing" of the input data in the co-add grid. This smoothing is minimized for a top-hat PRF spanning one detector pixel (equivalent to overlap-area weighting). This leads to smearing of the input pixel signals and one consequence is that cosmic rays can masquerade as point sources (albeit with narrower width) if not properly masked. For point sources with Gaussian profiles, their effective width will increase by a factor of $\sim\sqrt{2}$. Furthermore, a broad kernel will cause the noise to be spatially correlated in the co-add, typically on scales (correlation lengths) approaching the PRF size. Correlations are minimized for top-hat kernels. Both the effects of flux smearing and correlated noise must be accounted for in photometric measurements off the co-add, both in profile fitting and aperture photometry. The compensation for flux smearing can be handled through an appropriate aperture correction. Ignorance of correlated noise will cause photometric uncertainties to be underestimated. Methods on how to account for correlated noise in photometry are discussed in §13.

## 7 CO-ADDITION USING OVERLAP-AREA WEIGHTING AND DRIZZLE

Equation 7 can be compared to the popular pixel overlap-area weighting method, e.g., as implemented in the *Montage*[2] and *MOPEX*[3] tools. In fact if the PSF is grossly under-sampled, then the PRF is effectively a top hat spanning one detector pixel. The interpolation as described above then reduces to overlap-area weighted averaging where the interpolation weights $r_{ij}$ become the input($i$)-to-output($j$) pixel overlap areas $a_{ij}$. ICORE also implements the overlap-area weighting method for co-addition, in case detector PRFs are not available. This can be invoked by specifying "–sc_coad 1".

In essence, this method involves projecting the four corners of an input frame pixel directly onto the output co-add grid using the full WCS transformation with distortion correction if applicable. Input/output pixel overlap areas are computed using a formula for the area of a polygon and then used as weights when summing the contribution from all input detector pixels with signals $D_i$. The signal in co-add pixel $j$ (written to output: –o1_coad) is given by:

$$f_j = \frac{\sum_i \frac{a_{ij}}{\sigma_i^2} D_i}{\sum_i \frac{a_{ij}}{\sigma_i^2}}, \qquad (10)$$

where the inverse variance weighting ($1/\sigma_i^2$) is optional and can be turned on/off using the –wf_coad parameter. Uncertainties based on priors, analogous to Eq. 8 are also generated (output: –o3_coad).

An additional product that can only be generated for overlap-area weighting is an image of the standard-deviation of the input pixel stacks divided by the square-root of the depth-of-coverage at each location. This represents an alternative measure of the uncertainty in a co-add pixel and is generated if the –o4_coad output filename is supplied. In general, the variance in a pixel stack at location $j$ can be written:

$$\mathrm{var}_j = \frac{N_j}{N_j - 1}\left[\left\langle D_i^2 \right\rangle_j - \left\langle D_i \right\rangle_j^2\right],$$

where $N_j$ is the depth-of-coverage and the pre-factor $N_j$ / ($N_j$ - 1) is the correction to convert the sample variance to an unbiased estimate of the population variance. The angled brackets denote area-weighted averages of all the input detector pixel signals overlapping with co-add pixel $j$. The uncertainty in the co-add signal $f_j$ is then given by

[2] http://montage.ipac.caltech.edu/
[3] http://ssc.spitzer.caltech.edu/dataanalysistools/tools/mopex/

$$\sigma_{SDj} = \sqrt{\frac{\text{var}_j}{N_j}}.$$

Using the above expressions, where $<D_i>_j \equiv f_j$ is given by Eq. 10, the uncertainty can be written:

$$\sigma_{SDj} = \frac{1}{\sqrt{N_j - 1}} \left[ \frac{\sum_i \frac{a_{ij}}{\sigma_i^2} D_i^2}{\sum_i \frac{a_{ij}}{\sigma_i^2}} - f_j^2 \right]^{1/2}. \quad (11)$$

This is only computed for pixels where $N_j > 1$, otherwise $\sigma_{SDj} = 0$ is written to the output (–o4_coad) image. In general, estimates from Eq. 11 are expected to be larger than those estimated using priors (analogous to Eq. 8). This is because many more sources of error can contribute to the dispersion in a pixel stack, e.g., under-sampling of PSFs (even close to critical), pointing and registration errors, and temporal variations in detector/instrumental signatures.

An added feature tied of the overlap-area interpolation method is a "drizzle" option, which has its roots in construction of the Hubble Deep Field images from HST[4]. It has the advantage of minimizing the degree of correlated noise in a co-add and can also improve the spatial resolution if the depth-of-coverage is appreciable, especially for under-sampled data. Drizzling can be invoked by specifying a value $0 \leq d < 1$ for the *linear* drizzle factor parameter: –d_coad <*d*> where *d = new pixel scale (or drop size) / input (native) pixel scale*. I.e., the input pixel area is shrunk by a factor of $d^2$ before proceeding with overlap-area interpolation onto the output grid (see Figure 11). This parameter has a default of 1 and implies the entire input pixels are mapped onto the output grid with no shrinkage.

### 7.1 Rule-of-Thumb for Drizzle

Note that if the drizzle factor *d* is too small, one may end up with holes (gaps) in the output co-add if there is insufficient depth-of-coverage. We recommend choosing a pixel drop size (*d*input native scale*) that is small enough to minimize degradation of the output image, but large enough such that after all frames have been projected, the coverage is still fairly uniform. For *randomly dithered* input frames, a conservative approach is to attempt drizzling only if the overall depth-of-coverage is $N_f$ >~ 10 and setting $d \approx 2/\sqrt{N_f}$ with an output co-add pixel size (parameter –pa_coad) of >~ *d*input native scale*. Again, this rule-of-thumb assumes the frame dithering is random with non-integer pixel steps. For WISE frames for example with native pixel scale ~2.75 arcsec and covering a region to uniform depth of ~16, $d = 2/\sqrt{16} = 0.5$ and pa_coad = 1.375 arcsec. In general however, we found that pa_coad for WISE can be globally fixed at 1 arcsec/pixel to support large projects that sample a range in depths-of-coverage over the sky.

[4] http://www-int.stsci.edu/~fruchter/dither/dither.html

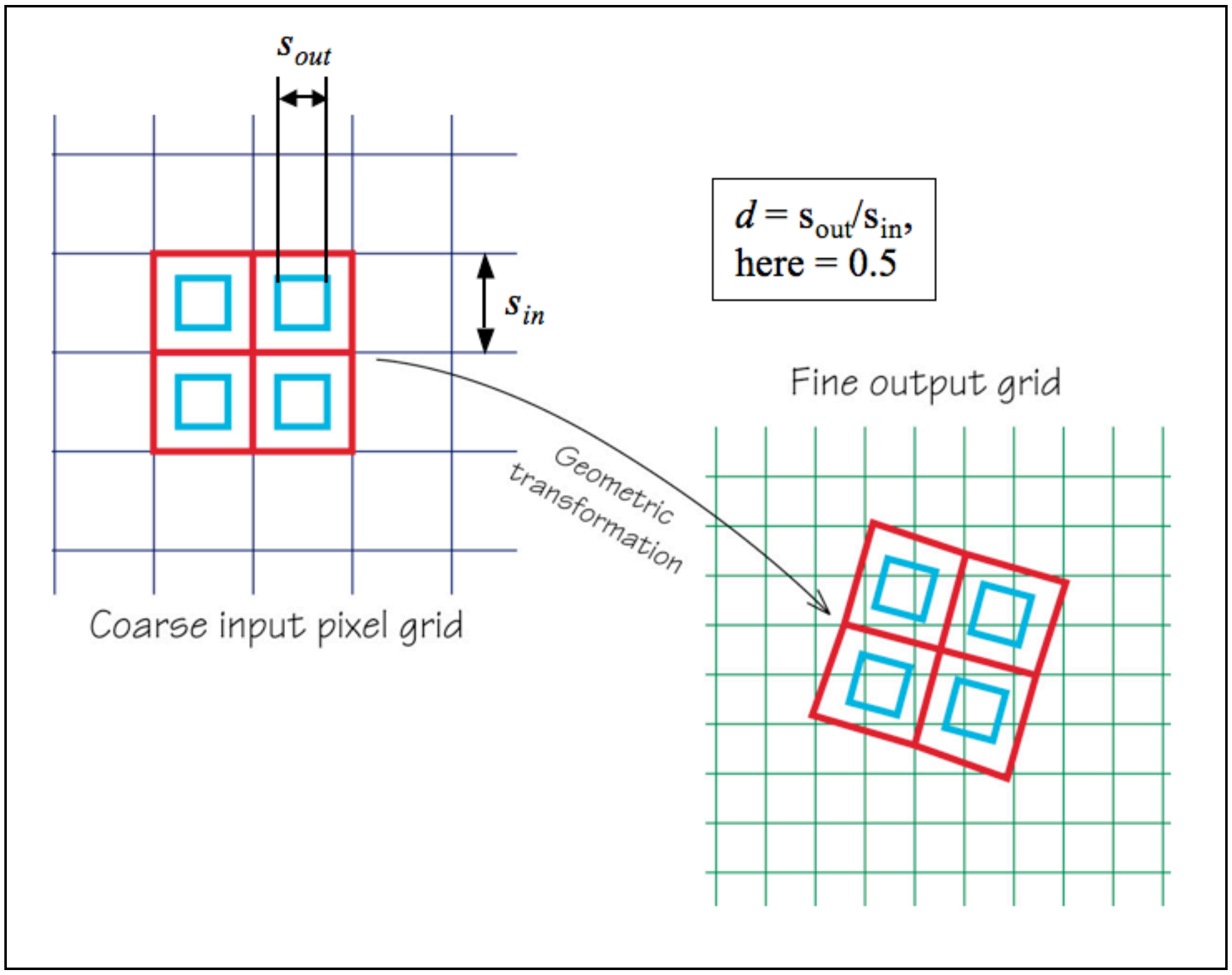

**Figure 11: Schematic showing definition of linear drizzle factor *d* (*icore* parameter - d_coad). Base figure adapted from: http://www-int.stsci.edu/~fruchter/dither/drizzle.html**

## 8 EXTENSION TO RESOLUTION ENHANCEMENT

We now describe a generic framework for co-addition that includes resolution enhancement (HiRes). Above we referred to the concept of combining frames to create a co-add. The HiRes problem asks the reverse: what model or representation of the sky propagates through the measurement process to yield the observations within measurement error? As a reminder, the measurement process is a filtering operation performed by the instrument's PRF:

$$\text{sky (truth)} \otimes \text{PSF} \otimes \prod \otimes \text{sampling} \rightarrow \text{noisy measurements}, \qquad (12)$$

where

$$\text{PSF} \otimes \prod \equiv \text{PRF}.$$

The Π symbol represents an individual pixel's response, usually assumed to be uniform (top-hat). Our goal is to infer a plausible model of the sky or "truth" given the instrumental effects.

### 8.1 The Maximum Correlation Method

The HiRes algorithm in ICORE is based on the Maximum Correlation Method (MCM). This was originally implemented to boost the scientific return of data from *IRAS* approximately 20 years ago (Aumann et al. 1990; Fowler & Aumann 1994), and is still provided as an online service to users. We have now implemented MCM in a form which is suitable for use on any imaging data that are compatible with the FITS and WCS standards, and the SIP convention for distortion. The versatility of MCM is that it implicitly generates, as its very first step (or first iteration), a PRF-interpolated co-add as described in §6. The algorithm is as follows.

1. First we begin with a flat model image of ones, i.e., a "maximally correlated" image:

   $$f_j^{n=0} = 1 \;\; \forall \; j, \qquad (13)$$

   where the subscript $j$ refers to a pixel in the upsampled output grid, and $n$ refers to the iteration number (input parameter –n_coad). This starting image is a first guess at the "truth" that we plan to reconstruct. Obviously this is a bad approximation, since it represents what we know without any measurements having been used yet. We could instead have used prior information as the starting model if it was available.

2. Next, we use the detector PRF(s) to "observe" this model image, or predict the input detector measurements. Starting with $n = 1$, the predicted flux in each detector pixel $i$ is obtained by a "convolution":

   $$F_i^n = \sum_j r_{ij} f_j^{n-1}, \qquad (14)$$

   where $r_{ij}$ is the response (PRF value) of pixel $i$ at the location of output model pixel $j$. Eq.

14 is a tensor inner product of the model image with the flipped PRF (see below for why we need to flip the PRF). It may not be a true convolution since the kernel $r_{ij}$ may be non-isoplanatic.

3. Correction factors are computed for each detector pixel $i$ by dividing their measured flux, $D_i$, by those predicted from the model (Eq. 14):

$$K_i^n = \frac{D_i}{F_i^n}. \tag{15}$$

4. For each model pixel $j$, all "contributing" correction factors, i.e., contributed by the overlapping PRFs of all neighboring detector pixels $i$ are averaged using response-weighted averaging (with optional $1/\sigma_i^2$ weighting):

$$C_j^n = \frac{\sum_i \left(r_{ij}/\sigma_i^2\right) K_i^n}{\sum_i r_{ij}/\sigma_i^2}. \tag{16}$$

5. Finally, the model image pixels are multiplied by their respective averaged correction factors (Eq. 16) to obtain new refined estimates of the model fluxes:

$$f_j^n = f_j^{n-1} C_j^n. \tag{17}$$

If we are after a simple PRF-interpolated co-add, we terminate the process at step 5 (with output written to –o1_coad). In fact, Eq. 16 is analogous to the co-addition equation (Eq. 7) in that a starting model image with $f_j^0 = 1$ implies a correction factor $K_i^1 \equiv D_i$ since a PRF volume-normalized to unity predicts $F_i^1 = 1$ (Eq. 14). Therefore after the first ($n = 1$) iteration of MCM, *co-add* fluxes will automatically result: $f_j^1 = f_j^0 C_j^1 = f_j$.

If we desire resolution enhancement, the above process is iterated, where the model image from step 5 is used to re-predict the measurements in step 2. This process of iteratively refining the model continues until the model reproduces the measurements to within the noise, i.e., the predictions from Eq. 14 are consistent with the measurements $D_i$. If input prior uncertainties ($\sigma_i$) are available, this convergence can be formally checked using a global $\chi^2$ test that uses all the input detector pixels:

$$\chi_n^2 = \sum_{i=1}^{N} \frac{\left(D_i - F_i^n\right)^2}{\sigma_i^2}, \tag{18}$$

where we expect $\chi_n^2 \approx N$ (the number of degrees of freedom, = the number of unmasked input pixels). If input uncertainties are provided (–unclist), the reduced chi-square, $\chi_n^2/N$, on a single frame basis is written to standard output following every iteration $n$. This is expected is converge to ≈1 as iterations increase, provided input uncertainties have been validated and adequately represent noise fluctuations in the input data. Alternatively, convergence can also be checked by examining the correction factors for each detector pixel (Eq. 15), where we expect $K_i^n \approx 1$ within the noise, or, via the averaged correction factors (Eq. 16), where $C_j^n \rightarrow 1$ after many iterations. Iterating much further beyond the initial signs of convergence has the potential of introducing unnecessary (and usually unaesthetic) detail in the model. This is important to ensure a parsimonious HiRes solution.

An image of the $C_j^n$ in the internal *cell*-grid frame (defined in §6) can be generated at each iteration *n* if –mcmprod is specified. This product is generically named "mosaic-cellcor.fits_iter<*n*>" and written to the directory specified by –outdir. Furthermore, the final correction-factors in the *downsampled* output grid (at ending iteration specified by –n_coad [*n* > 1]) can be written to an image specified by –o5_coad. The final HiRes'd intensity image is written to output specified by –o1_coad. For comparison/diagnostic purposes, an intensity image from the *first* iteration (≡PRF-interpolated co-add) in the *downsampled* grid can be written to output specified by –of_coad. Furthermore, if –mcmprod was set, intensity images in the internal *cell*-grid frame at each iteration are generated under –outdir with generic name: "mosaic-cell.fits_iter<*n*>".

Therefore, it is an algorithmic property of MCM that it only modifies (or de-correlates) a *flat* starting model image to the extent necessary to make it reproduce the measurements within the noise. A PRF-interpolated co-add (from the first MCM iteration) will generally not satisfy the measurements after it is convolved with the detector PRFs, i.e., when subject to the measurement process (Eq. 12). Figure 12 shows a schematic of the MCM process.

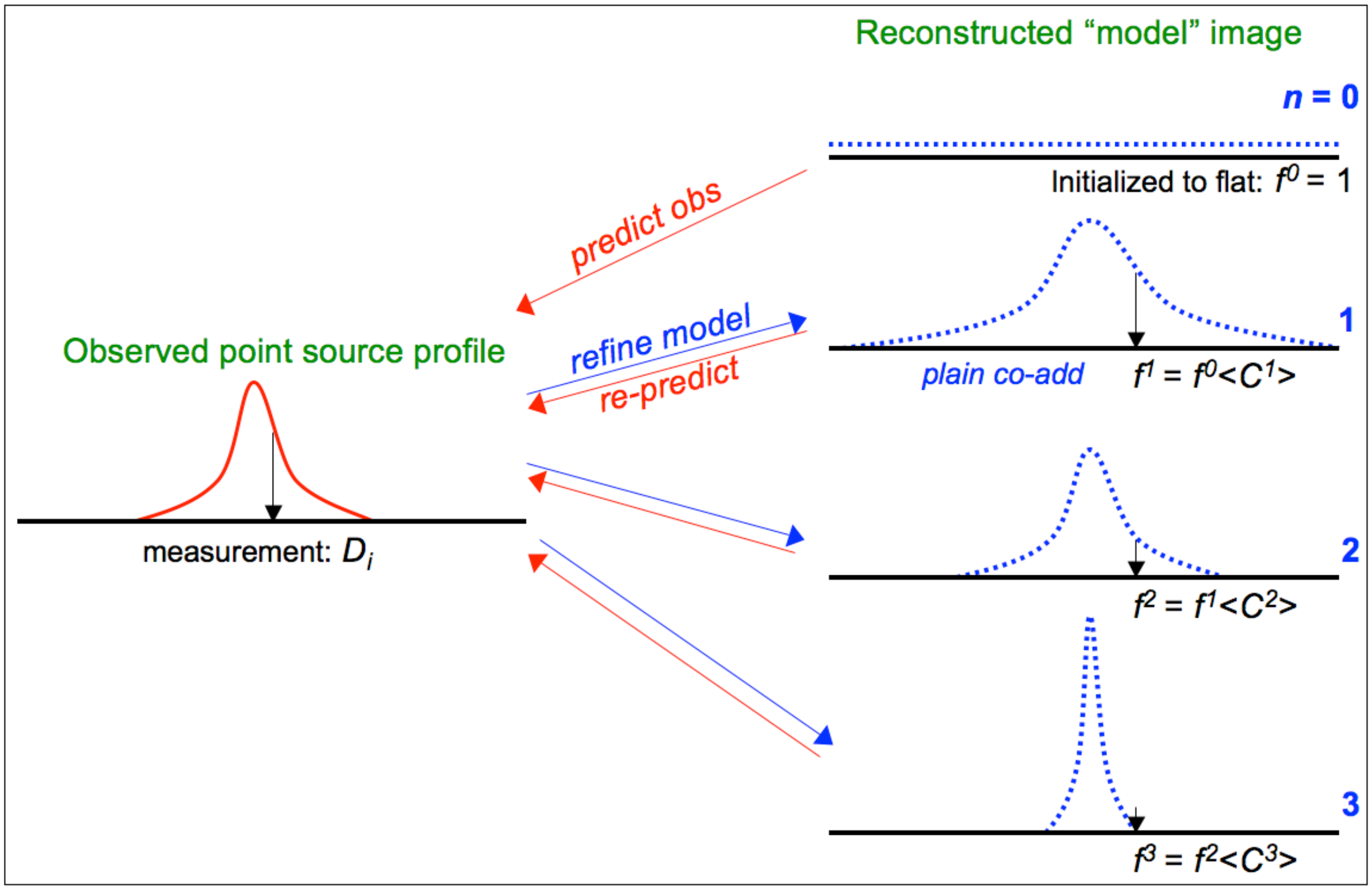

**Figure 12: Cartoon showing reconstruction of a "point source" using MCM. The process consists of iteratively refining the model using spatially dependent correction factors *C* (Eq. 16) such that the "convolved model" reproduces the measurements within the noise (equivalent to the *C* converging to unity).**

As a detail, the input PRFs are first flipped in $x$ and $y$ (or equivalently rotated by 180°) when HiRes'ing is performed ($n > 1$). This is to conform to the usual rules of convolution and assumes the input PRFs were made by combining images of point sources observed with the same detector in the same native $x$-$y$ pixel frame. For PRF-interpolated co-adds however (that terminate at $n = 1$), the PRFs are not flipped since a cross-covariance with the input data is instead needed. The PRFs here are used as matched filters to generate products optimized for point source detection (see §6).

It is also worth noting that MCM reduces to the classic Richardson-Lucy (RL) method if the following are assumed: (i) the PRF is isoplanatic so that a constant kernel allows for Fourier-based deconvolution methods to be used; (ii) the inverse-variance weighting of measurement correction factors is disabled from the PRF-weighted averaging (Eq. 16), or equivalently if all the input variances $\sigma_i^2$ are assumed equal. This implies the solution will converge to the maximum likelihood estimate for data that are Poisson distributed. With inverse-variance weighting included, the solution converges to the maximum likelihood estimate for Gaussian distributed data. This is usually always satisfied for astronomical image data in the limit of high photon counts; (iii) there is no explicit testing for global convergence at each iteration by checking, for example, that the solution reproduces the data within measurement error (Eqs 14 and 18). This criterion was indeed suggested by Lucy (1974), although it is seldom used in modern implementations of the RL method.

In the absence of prior information for the starting model, MCM implicitly gives a solution which is the "smoothest" possible, i.e., has maximal entropy. This should be compared to maximum entropy methods (e.g., Cornwell & Evans 1985) which attempt to minimize the $\chi^2$ of the differences between the data and the convolved model, with an additional constraint imposing smoothness of the solution. MCM requires no explicit smoothness constraint. MCM can indeed use a regularizing constraint in the form of non-flat starting model, (e.g., an image of the sky from another detector or wavelength), but this jettisons the idea of an image with maximally correlated pixels, and the refined model image will not be the smoothest possible. Smoothness is important because it can be used to convey fidelity in a model. In general, the solution to a deconvolution problem is not unique, especially in the presence of noise. Many models can be made to fit the data, and many methods invoke regularization techniques in order to select a plausible solution. A consequence is that some methods give more structure or detail than necessary to satisfy the data, and there is no guarantee that this structure is genuine. MCM adopts the Occam's razor approach. Given no prior constraints (apart from satisfying the input data), MCM will always converge on the simplest solution. This will be the smoothest possible.

The default prior (starting) model for ICORE is a flat image of ones, i.e., a "maximally correlated" image (Eq. 13). However, there is also an option to internally create an overlap area-weighted co-add from the input data (using the method described in §7 with *no* drizzling) and use it as the starting model (instead of Eq. 13). This can be invoked by specifying "–fp_coad 1". This gives the iterations a "head-start" in reconstructing the HiRes solution since the first iteration co-add created with a flat prior (default: –fp_coad 0), i.e., the PRF-interpolated co-add, is smeared (with more power at low-frequencies) compared to an area-weighted co-add. As described above, a non-flat prior will not lead to the smoothest (and simplest) solution. Artifacts and glitches present in the non-flat prior (area-weighted co-add) may have adverse effects on the converged HiRes solution.

### 8.2 Specifying a Non-flat Image Prior

Aside from defaulting to a "flat" image prior, or creating an internal one from an overlap-area weighted co-add of the input data (see above), there is also an option to specify your own image prior. This could be a higher resolution image from another instrument or even perhaps a different wavelength. If another wavelength is used, it's important to note that the physics of emission could be vastly different from the wavelength at which the input images were observed. The HiRes solution must be interpreted with caution since the MCM process may introduce "features" in the output product carried over from the assumed prior that may only be unique to the wavelength at which the prior was observed.

The filename of a specific image prior can be supplied via the –pri_coad option. It must be in FITS format and contain a valid WCS (with distortion in SIP format if applicable). This image will be automatically used if specified and only if the number of MCM iterations (–n_iter) is > 1. As a detail, this image prior is only used in the *first* MCM processing pass if the ringing suppression switch (–flxbias; §8.6) is set. Note that the target object or region of interest in the prior must fall in the footprint defined by the *icore* input parameters: –sizeX, –sizeY, –ra, –dec, and –rot, i.e., the image prior need not completely overlap with the desired output footprint geometry, nor does it need to have the same pixel scale. This is because the input image prior is resampled and reprojected onto the desired output footprint (in the internal cell-grid format as needed by MCM) before use. The interpolation method used here is the simple overlap-area weighting method (§7).

Following resampling onto the cell-grid footprint, the prior image is further reformatted and regularized to conform to the requirements of MCM. This procedure consists of subtracting a robust background level (the mode of all valid pixels); replacing pixel signals $\leq N\sigma$ with zero where $N$ = –clipt input parameter and $\sigma$ is a robust global pixel sigma based on percentiles; and then adding one to *all* pixels, including any blank space. Furthermore, if the –isolate switch is set, more regularization is performed to "isolate" the target region of interest in the prior image. This consists of forcing all pixels *outside* a radius of –isorad from the target position –rat, –dect to have value one.

The reprojected and regularized image prior is written to the generic name: –outdir/cellgrid-prior-int.fits. This product is then used as input for MCM in place of the default flat prior defined by Eq. 13.

### 8.3 The CFV Diagnostic

A powerful diagnostic from MCM is the Correction Factor Variance (CFV). This represents the variance about the PRF-weighted average correction factor (Eq. 16) at a location in the output grid for iteration $n$: $V_j^n = \langle K_i^2\rangle_j - \langle K_i\rangle_j^2$, or

$$V_j^n = \sum_i w_{ij}\left[K_i^n\right]^2 - \left[\sum_i w_{ij}K_i^n\right]^2, \qquad (19)$$

where $w_{ij} = (r_{ij}/\sigma_i^2) \,/\, \sum_i r_{ij}/\sigma_i^2$, and the detector-pixel correction factors $K_i^n$ were defined in Eq. 15. At early iterations, the CFV is generally high everywhere because spatial structure has not yet been resolved, and the model contradicts the measurements when subject to the measurement process. If after convergence, all the detector-pixel measurements contributing a non-zero response at some location $j$ agreed exactly with their predicted fluxes (Eq. 14), then all the $K_i^n$ would be ≈ 1 and the CFV ($V_j^n$) at that location would be zero. Areas with a relatively large CFV indicate the presence of input pixel measurements which do not agree with the majority of the other measurements (e.g., outliers). It could also indicate noisy data, saturated data, regions where the PRF is not a good match (e.g., erroneously broad), or that a field has not yet converged and would benefit from further iteration. By thresholding the CFV, one can therefore create a mask for a HiRes image to assist in photometry, e.g., to avoid outliers and unreliable detections from amplified noise fluctuations (see below). An image of the CFV in the internal *cell*-grid frame (defined in §6) can be generated at each iteration *n* if –mcmprod is specified. This product is generically named “mosaic-cellcfv.fits_iter<*n*>” and written to the directory specified by –outdir.

Figure 13 shows example products from ICORE at three iteration levels of the spiral galaxy M51a (this is discussed further in §8.6). Corresponding CFVs are shown in Figure 14. To illustrate the concept of the CFV, outlying input measurements were not masked in the *left* and *middle* images of Figure 14. These refer to iteration levels $n = 1$ and $n = 40$ respectively, with the latter corresponding to convergence. The smooth CFV image on the *far right* in Figure 14 was created from data with outliers detected and masked *a priori* using the algorithm described in §5. The uniformity of this CFV image indicates that the bulk of outlying measurements were indeed masked. Figure 15 compares a PRF-interpolated co-add (HiRes after 1 iteration), an overlap area-weighted co-add (§7), and a HiRes run to 50 iterations for globular cluster 47 Tucanae.

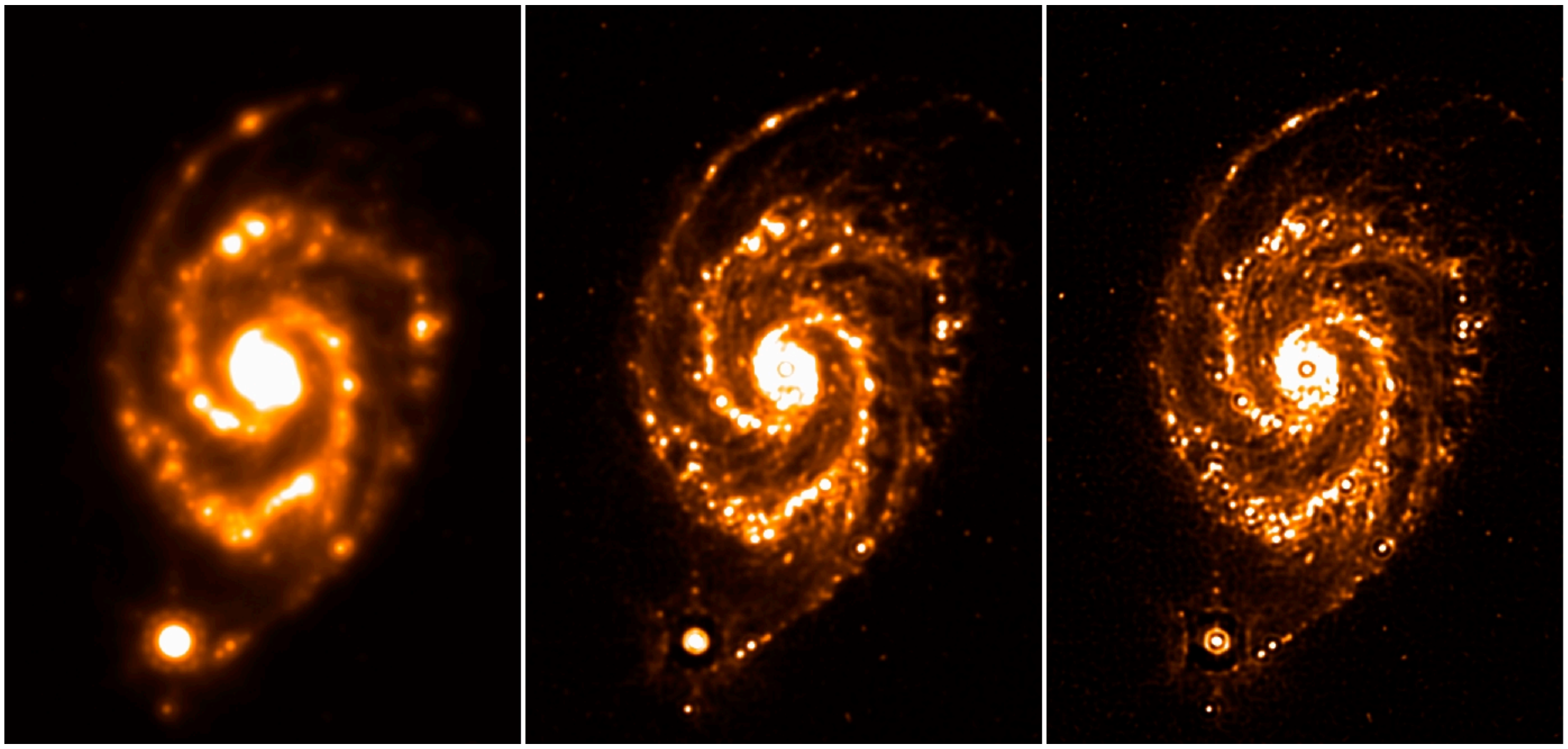

**Figure 13: M51 (The Whirlpool Galaxy) from *Spitzer*-MIPS 24µm observations. *Left*: co-add (after 1 iteration); *Middle*: HiRes after 10 iterations; *Right*: HiRes after 40 iterations. Each run combined 216 frames to give a median depth-of-coverage of ~45. The fields span ~7′ × 11′.**

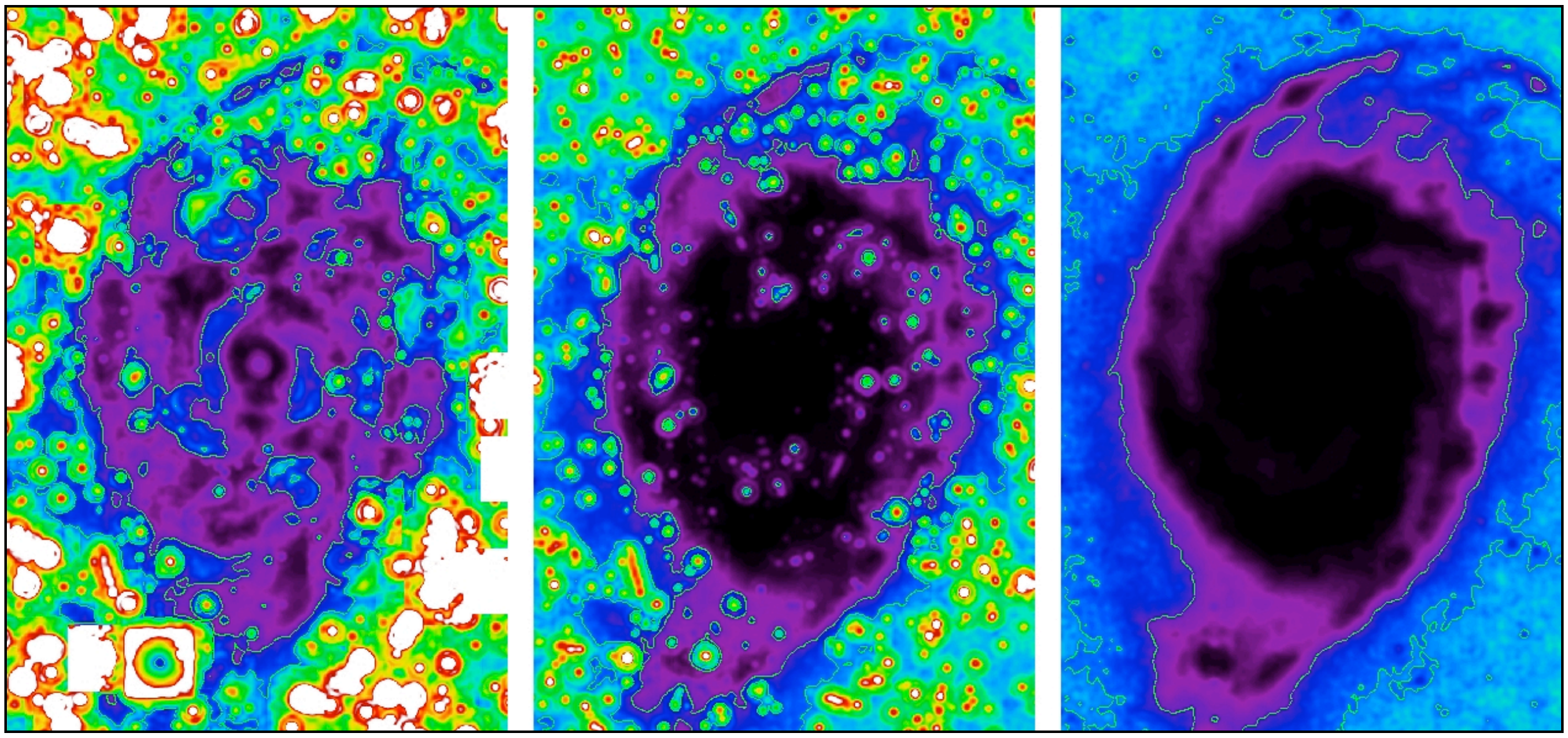

**Figure 14: Correction Factor Variance (CFV) images for M51 whose intensity images were shown in Fig. 13. For a description of the CFV, see §8.3. *Left*: CFV after 1 iteration with outlying measurements purposefully retained; *Middle*: CFV after 40 iterations with outlying measurements also purposefully retained; *Right*: CFV after 40 iterations but with outliers masked (omitted) prior to HiRes'ing. Darkest regions correspond to lowest values of the CFV ($V_j$ <~ 0.1), and the brightest to highest values ($V_j$ >~ 100).**

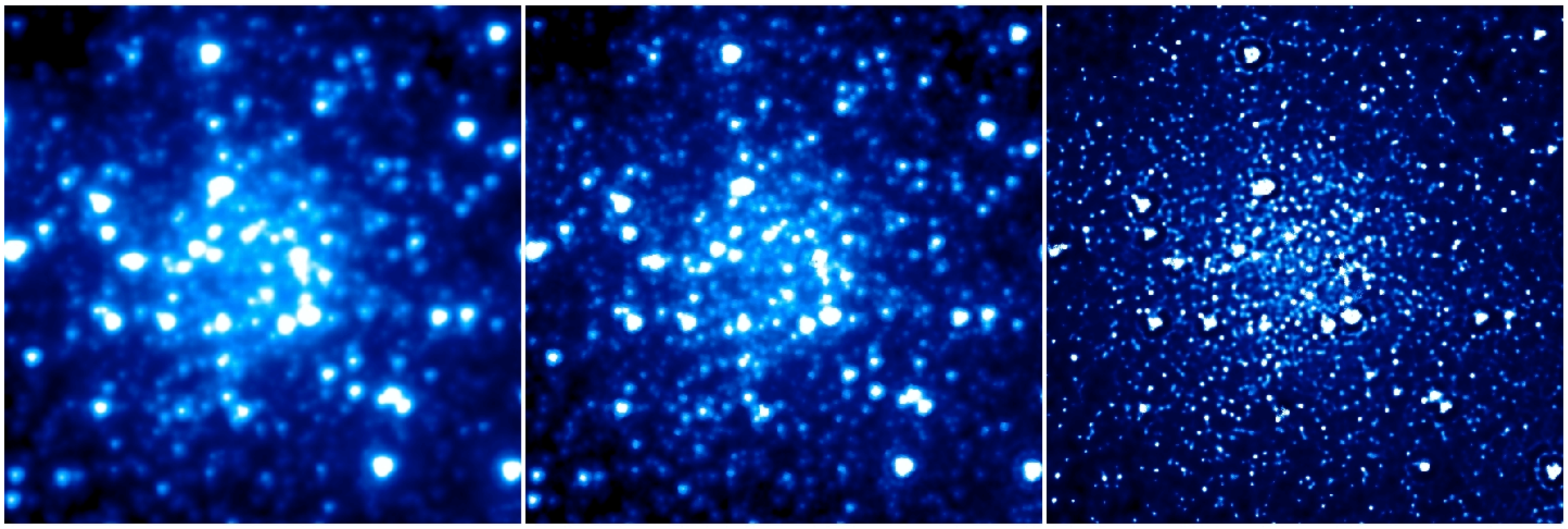

**Figure 15: *Spitzer*-IRAC 4.5μm observations of globular cluster 47 Tucanae. *Left*: PRF-interpolated co-add (= HiRes after 1 iteration); *Middle*: co-add from overlap area-weighting method with drizzle factor = 1 (see §7 for details); *Right*: HiRes after 50 iterations. The fields span 2.4′ × 2.4′ and the median depth-of-coverage is ~37.**

## 8.4 HiRes Uncertainties

### 8.4.1 *a posteriori* (data-derived) estimates

Apart from providing a qualitative diagnostic, the CFV can also be used to compute *a posteriori* (data-derived) uncertainties in the pixel fluxes $f_j^n$ in a HiRes image. These are written to the output image filename specified by –o6_coad. In general, the 1-$\sigma$ uncertainty at iteration $n$ can be written in terms of the CFV as:

$$\sigma_j^n = c^n f_j^n \sqrt{\frac{V_j^n}{\sum_i r_{ij}}}, \tag{20}$$

where $V_j^n$ is the CFV defined by Eq. 19 and the sum is over the responses from all measurements $i$ at output pixel $j$, i.e., the effective depth-of-coverage. $c^n$ is a correction factor to account for re distribution of noise power across spatial frequencies from one iteration to the next. At low iterations, power is relatively high at low frequencies, i.e., the noise is correlated across pixels. As iterations increase, noise is de-correlated and power migrates to high frequencies. The spectrum approaches that of white noise, provided the input measurement noise was spectrally white. Figure 16 shows the evolution of the pixel noise distribution with iteration number $n$ using a simulation. Gaussian white noise was assumed as input. The first iteration represents the PRF-interpolated co-add where noise is maximally correlated.

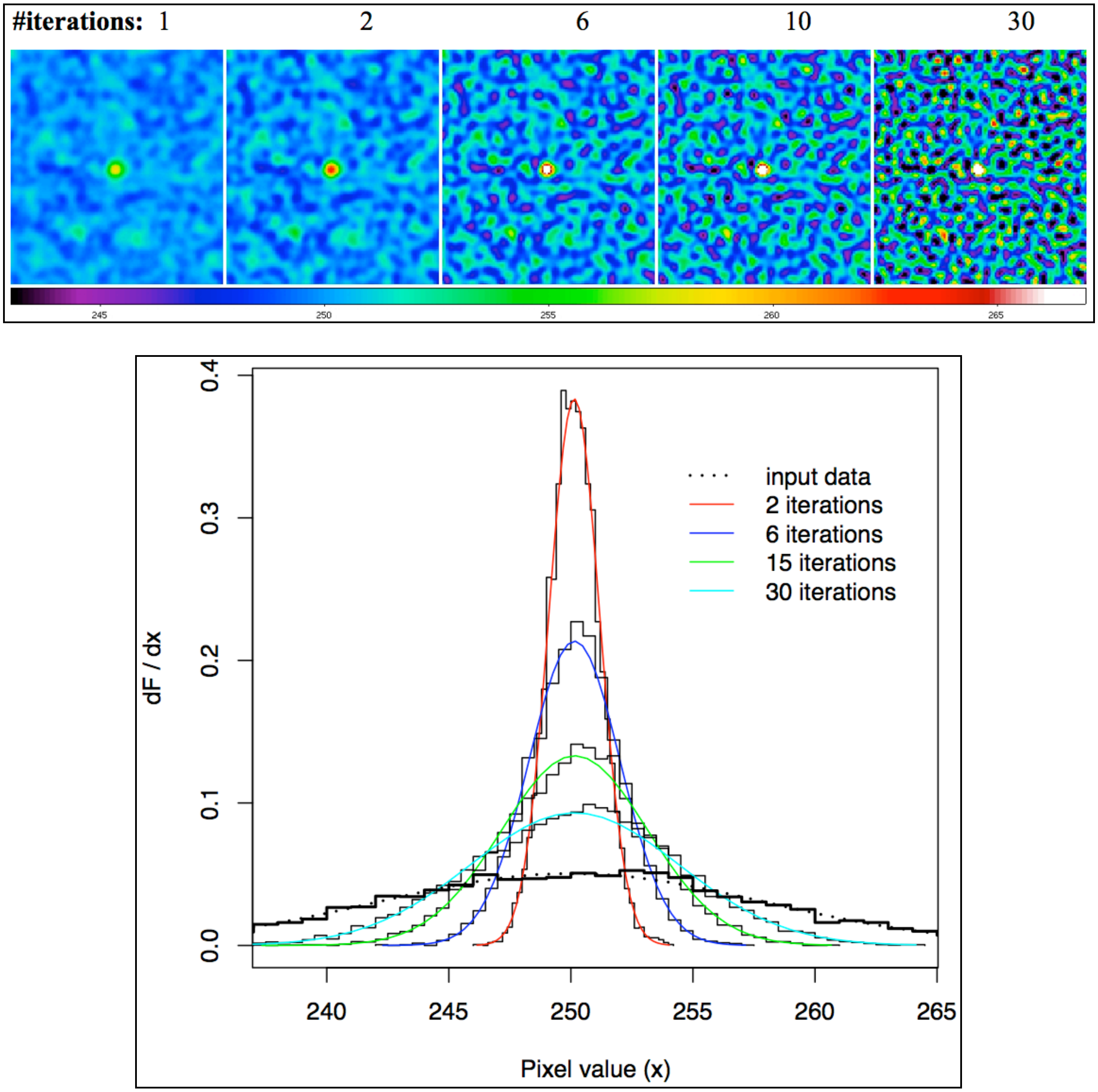

**Figure 16:** ***Above*: series of HiRes simulation images at different iteration numbers assuming as input Gaussian white noise and single point source in the middle of the frame. *Below*: normalized distributions of the background noise fluctuations at several iterations.**

For $n = 1$ (giving a co-add), it can be shown that $c^1 \equiv 1/\sqrt{P_j}$, where $P_j$ is the effective number of noise pixels defined in §6. With $c^1$ written this way, Eq. 20 becomes equivalent to the co-add pixel uncertainty defined in Eq. 8. In general, the $c^n$ at any iteration $n \geq 1$ can be approximated from the output image products as:

$$c^n \approx \frac{\sigma_{RMS}\left[f_j^n\right]}{\left\langle \sigma_j^n [c^n = 1] \right\rangle}, \qquad (21)$$

where $\sigma_{RMS}$ is the standard-deviation (or some robust equivalent) of the pixel noise fluctuations within a "source-free" stationary background region with ≈uniform depth-of-coverage in the $f_j^n$ image. The denominator is the mean (or some robust equivalent) of Eq. 20 with $c^n = 1$ in the same region.

In ICORE, the region for computing $c^n$ is chosen by partitioning the HiRes'd intensity image ($f_j^n$) into a grid of –siggrid × –siggrid squares (default = 8 × 8) and then selecting the region with the *smallest* robust spatial RMS defined by percentiles in the pixel distribution:

$$\sigma_{RMS} \approx 0.5\left[p(84\%tile) - p(16\%tile)\right].$$

This estimate is used in the numerator of Eq. 21. It minimizes biases to the local RMS from fast-varying backgrounds, high frequency structure, and regions with high source-confusion over the footprint. If this "minimum RMS" region is still contaminated, the $c^n$ scaling factor may be overestimated (depending on how the denominator of Eq. 21 responds), leading to overestimated HiRes uncertainties from Eq. 20. This is not catastrophic since having uncertainties slightly overestimated errs on the conservative side. Furthermore, if source confusion is high everywhere, this will implicitly include confusion noise in the HiRes uncertainties. The location of the "optimal" region is written to standard output during processing and therefore we recommend visually checking the intensity image to ensure it does not contain significant structure at that location. At the time of writing, the $c^n$ uncertainty-scaling factor is always computed and applied for HiRes cases (–n_coad > 1). If you think it's grossly wrong, then a manual rescaling of the uncertainty images (–o6_coad for CFV-derived, or –o3_coad for prior-derived; see §8.4.2 for latter) will be necessary.

It's important to note that the CFV-derived uncertainties are most useful when (i) there are multiple overlapping input images (say >~ 10) and (ii) when the background levels are close to those representative of the true sky (and/or telescope) emission. These backgrounds need not be in absolute calibrated units, but in detector counts (DN). This is because the CFV-derived pixel uncertainty as defined by Eq. 20 depends on the pixel flux and effective depth-of-coverage contributed by all overlapping PRFs. Any arbitrary/artificial level that's added to the input images via the –arbconst parameter (§8.5) can lead to biased estimates in this product. Nonetheless, the CFV-derived uncertainties can still serve as a diagnostic when HiRes'ing a single input image with or without a "true" representative background, for example, to assess the relative level of convergence between iterations.

The above rescaling method gives pixel uncertainties which are more or less statistically compatible with *background* noise fluctuations in the HiRes'd image and must be interpreted with caution. This is because the background fluctuations may not be representative of the Poisson-noise where source signal is high, e.g., if the detector with which the frames were acquired is read-noise limited, it is dangerous to assume that the scaling computed above also applies at the location of sources for the purpose of estimating photometric uncertainties. The uncertainties could be slightly underestimated where the Poisson contribution is high. This may cancel out in the end if the background uncertainties were overestimated in the first place as described above. However, if the input frames are background-photon dominated (typical for mid/far IR data), the HiRes uncertainties will be appropriate for all fluxes in the HiRes'd image, including at the location of sources. This will allow one to estimate uncertainties in source photometry. Note that correlated noise is also expected to be minimal in a converged HiRes image, and even less if products were created with ringing suppression turned on (see below).

### 8.4.2 Uncertainties from propagating priors

For low depths-of-coverage, estimates of the data-derived uncertainty using the CFV (Eqs 20 and 21, output: –o6_coad) may not be accurate. Even though rescaling is still involved to satisfy local RMS spatial fluctuations, there's no guarantee that uncertainties at the location of sources will be accurate enough for photometry. An alternative estimate is written to the product specified by –o3_coad. This estimate uses the first iteration prediction using priors (Eq. 8, as applies to a PRF-interpolated co-add), and then it is rescaled using the same method described in §8.4.1 to account for the migration of noise power with increasing iteration. This assumes prior uncertainties were provided on input (–unclist) and were validated before use. In effect, the –o3_coad output product is a "pseudo-prior" uncertainty image since it uses input prior uncertainties mapped into the PRF-interpolated co-add and then rescaled to satisfy background fluctuations in the data. This product is also subject to the cautionary notes described in §8.4.1.

### 8.4.3 Signal-to-Noise Ratio Images

There are two flavors of SNR images computed. One computed using CFV-derived 1-$\sigma$ uncertainties (§8.4.1), output: –snc_coad, and another computed using pseudo-prior uncertainties (§8.4.2), output: –snu_coad. The –snu_coad can be generated for all iteration products $n \geq 1$ (including the first iteration PRF-interpolated co-add), while the –snc_coad is only generated for HiRes products ($n > 1$). In either case, the SNR for output pixel $j$ is computed using:

$$SNR_j = \frac{f_j - SVB_j}{\sigma_j}, \qquad (22)$$

where $f_j$ is the output HiRes or co-add image signal and $SVB_j$ is an estimate of the Slowly Varying Background at the same location. The *SVB* image is computed as follows: each input frame is partitioned into a grid of –svbgrid × –svbgrid squares (default = 3 × 3); each square is replaced by the median (or optionally the mode if –modfilt is set) of all pixels therein; the block-filtered frames are then Gaussian-smoothed using a 2-D Gaussian kernel with parameters: linear extent –gausize and sigma –gausigm, both in units of native frame pixels. Internal re-binning (down-sampling) of frames is also performed to speed up the convolution. The *SVB* frames are written to the working directory specified by –outdir and then co-added internally using PRF-interpolation (with masking if frame masks were specified). The output *SVB* co-add image is named "mosaic-int-bckgnd.fits" and generated under the –outdir directory. At the end of processing, this *SVB* image is used to compute the SNR images defined by Eq. 22, outputs: –snu_coad and –snc_coad for each of the $\sigma$ measures described above.

As a detail during the single frame *SVB* computations described above, if "–bmatch 1" is also specified, a search for bright/extended structure is also performed using the –ratmax ($Q_d$) parameter described in 4.2. If detected, the frame *SVB* is set to a constant equal to the *minimum* block-median (or optionally block-mode) value over all –svbgrid × –svbgrid partitions. This avoids any significant structure from biasing the underlying frame background.

The HiRes SNR images generated here will be the maximum possible since MCM would have converged to the maximum likelihood estimate for data that were Poisson or Gaussian distributed. The latter is usually always satisfied for astronomical image data in the limit of high photon counts. In particular, the SNR image for a PRF-interpolated co-add (from output –snu_coad using "–n_coad 1") defines a linear-matched filter optimized for point-source detection (§6.1).

## 8.5 Ringing Suppression

Like most deconvolution methods, MCM can lead to ringing artifacts in the model image. This limits super-resolution, i.e., when attempting to go well beyond the diffraction limit of an imaging system. In general, ringing occurs because the reconstruction process tries to make the model image agree with the "true" scene with access to only the low spatial frequency components comprising the data. The input data are usually band limited, and information beyond some high spatial frequency cutoff can never be recovered. The best we can ever reconstruct is a "low-pass filtered" version of the truth, with the filter determined by the maximum spatial frequency the observations provide. This includes the finite sampling by pixels. A hard high frequency cutoff will lead to sinc-like oscillations in real image space. The magnitude of the ringing depends on the strength of a source relative to the local background intensity level.

It is no accident that a solution with ringing is the smoothest (and simplest) solution possible with MCM. Anything smoother (with more low frequency power) will not satisfy the measurements when subject to the measurement process (Eq. 12). However, since a large number of less-smooth solutions can reproduce the observations, those without ringing are generally more desirable. Therefore, we relax our request for the smoothest image and use *prior* knowledge that the background and (desired) source fluxes are physically distinct and separable. There have been numerous approaches that have used this philosophy (e.g., Lucy 1994). A flowchart of the ringing suppression algorithm in ICORE is shown in Figure 17. Below we expand on some of the details. Ringing suppression is only executed as part of HiRes'ing (–n_coad > 1) if the –flxbias switch was specified. See below for further tuning details.

1. We first generate an image of the slowly varying background for each input frame on some specific scale using block-median or mode filtering (if –modfilt is set) over an *N* x *N* grid (–svbgrid <*N*>). This includes a Gaussian convolution to smooth out any block discontinuities.
2. This is then subtracted from the respective input frames to create the "source" images; negative noise fluctuations are set to zero, and a tiny positive offset added;
3. MCM is then run on the background-subtracted images until convergence (with –n_coad <iterations>). This operation enforces a positivity constraint for reconstruction of the source signals. It ensures that source flux won't ring against an essentially zero background level so power can be forced into high spatial frequencies;
4. The background images are co-added and then optionally added to the HiRes'd source-image product. This is only performed if the –addbck switch is specified. The background intensity co-add in the final *downsampled* grid is generically named: "mosaic-int-bckgnd.fits" and written to the working directory specified by –outdir. A background intensity co-add in the internal *cell*-grid frame (defined in §6) is also generated and named "mosaic-cell-bckgnd.fits".
5. After the background has been added to the HiRes'd source-only product, MCM is re-executed for several iterations (or specifically –n_coadn <iterations>) using this as the starting model image and the original frames as input. This step re-adjusts the solution and attempts to restore the intrinsic noise properties of the HiRes process, i.e., what one would have obtained if no background were removed or positivity constraint enforced. It ensures that photometric uncertainties don't become biased and the final solution adequately reproduces the measurements within the noise.

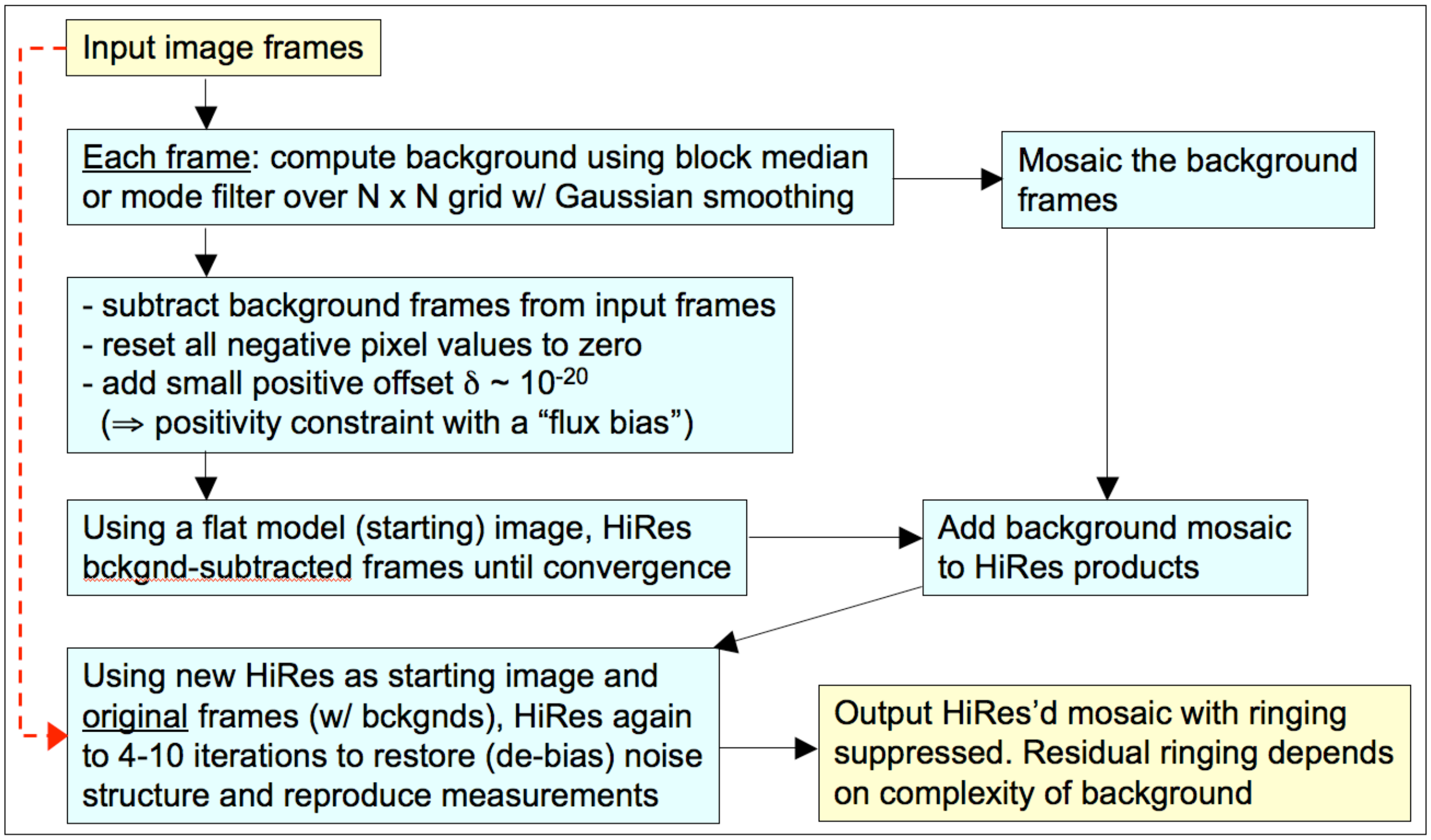

**Figure 17: ringing suppression algorithm to support HiRes**

Note that ringing suppression is probably the most difficult of all components to tune in HiRes, in particular when one has extended structure covering a footprint. Here, the frame background estimates could be over-estimated, resulting in a loss of flux (and information in general) when subtracted to create the "pure-signal", positively-constrained frames. The lost information will not participate in HiRes'ing, although it is (optionally) added back to the final HiRes'd image. This ensures flux is conserved. The goal then is to minimize SVB estimates from being contaminated from extended (and interesting) structure. This can be done by choosing a –svbgrid size such that >~50% of the pixels in each partition capture the background signal (or uninteresting parts in the frames). Some contamination is inevitable, especially for partitions which fall on top of extended structures or objects covering most of a frame's FOV. Therefore it is advised that the SVB intensity co-add (generically named –outdir/"mosaic-int-bckgnd.fits") be examined to assess whether the partition grid size needs to be made larger (–svbgrid value smaller) or smaller (–svbgrid value larger). An example where it may need to be smaller is if one is interested in capturing the background on smaller scales so ringing can be mitigated around sources superimposed on uninteresting "fast-varying" backgrounds or structures. One can also examine the individual SVB frames corresponding to each input frame under –outdir. These have suffix: "_svb.fits".

If tweaking –svbgrid doesn't help, there is one more functionality that can rescue the day. Along with the –flxbias switch (that triggers ringing suppression in the first place), it is recommended that the –bmatch (background matching) switch also be specified. This will assist in the explicit detection of extended structure over each frame (see §4.2) using the –ratmax threshold. If detected, it allows the frame background to default to a constant equal to the *minimum* median value over all –svbgrid × –svbgrid partitions of a frame (as opposed to a block-median filtered SVB). The goal here is to select the partition "most representative" of the underlying background. Depending on your extended structure or object, this may require decreasing the –ratmax threshold so that detection is triggered for your input frames (i.e., for a majority of them if you like). For super-extended structures that cover most of your input frames (or even a single frame), you may need to reduce the partition grid size (larger –svbgrid value) to try and tease out those "minimum background" regions.

The requirement of positive fluxes for the input image pixels when HiRes'ing can also be ensured by specifying –arbconst <*const signal*> and the –tmatch switch, even if the –flxbias switch was *not* set. This is an arbitrary global constant added to all the input pixels and is only operative if the –tmatch switch was also set. As a detail, –tmatch also performs throughput (gain) matching if photometric zero-point information was available in the input image headers (§4.1), otherwise, throughput matching is bypassed. The constant specified by –arbconst is later removed from the HiRes intensity product to ensure conservation of flux or surface brightness. We advise using this parameter only if input image levels are close to zero and/or many negative pixel values are present. We also recommend setting –arbconst to the *smallest* possible value that ensures global positivity in the input pixels. Large values can lead to significant (but small) deviations in the expected flux or surface brightness in HiRes'd products.

## 8.6 Noise Suppression Algorithm

A consequence of HiRes'ing is that noise-spikes at high spatial frequencies can be amplified with increasing iteration. The standard MCM algorithm does not distinguish between noise, residual outliers, and real-source information. All components are reconstructed by the deconvolution process, but not necessarily at the same rate. "Fitting of the noise" is a feature of RL-like algorithms since noise is usually present at frequencies exceeding the maximum (band-limit) of the input PRF. If a HiRes solution is to satisfy the input data on convolution with the PRF at all locations and scales, high frequency noise must persist in the HiRes image, usually with greater amplitude. One can reduce noise in the output by having a high depth-of-coverage of input frames, but high enough depths may not always be available. It is expected that the largest noise-spikes are outliers in the data and can be detected and masked by the outlier algorithm (§5) during preprocessing. This depends on the assumed outlier thresholds of course. If one is not after a science quality product (i.e., in the quantitative sense with data-compatible uncertainties), one can live dangerously and drive the outlier detection thresholds (–tl_odet and –tu_odet) really low and attempt to mask spikes well into the noise. For reliable detection, this requires a moderately high depth-of-coverage (>~15) in the first place. Again, we don't advise attempting this unless the algorithm below cannot be tuned for your data.

The noise suppression algorithm in ICORE is still very experimental so please proceed with caution. It has been tested on the M51 case shown in Figure 13 and gives reasonable results. This makes use of the CFV diagnostic described in §8.3, in particular its change from one iteration ($n$) to the next ($n + 1$): $\Delta CFV$. In regions of the output footprint devoid of sources and structure, and dominated by high-frequency noise, the CFV is seen to drop at a faster rate with increasing iteration than in regions with *unresolved* structure that require more iterations for convergence. A CFV which exhibits little change from one iteration to the next (within the noise) is an indication of convergence. Therefore, the rate of convergence for MCM is position dependent. One can use the $\Delta CFV$ diagnostic by not HiRes'ing the "noisy/void" regions to the same ending iteration number (–n_coad [> 1]) desired to resolve the interesting structure.

In summary, the HiRes'ing for output pixel $j$ is terminated at iteration $n$ if the absolute %-change in the CFV satisfies:

$$\%\Delta CFV = 100\left|\frac{CFV_j^n - CFV_j^{n+1}}{CFV_j^n}\right| < thres, \quad (23)$$

where *thres* is specified by the input parameter –h_coad with default=0 implying this processing is not activated, there is no noise suppression, and all regions are HiRes'd to the same number of iterations (–n_coad). Iterations are terminated internally by forcing the averaged correction factor $C_j^n$ to 1 (Eq. 16 in §8.1) for pixel $j$, implying no further resolution enhancement for this pixel. The HiRes flux $f_j^n$ is frozen at iteration $n$ and other regions may continue to be HiRes'd until Eq. 22 is satisfied, or the ending iteration specified by –n_coad is reached, whichever is earlier. If –h_coad > 0 [in %] is specified, output filenames for products –o5_coad, –o6_coad must also be specified. A related ancillary output product is an image of the ending iteration number (–oi_coad), showing where the HiRes'ing has terminated according to Eq. 23. This product can be used with the HiRes'd image (–o1_coad) to tune the threshold parameter –h_coad. The individual CFV images at each iteration (generated with –mcmprod) could also be useful (see §8.3). A further benefit of the noise suppression algorithm is that by focusing on those regions where resolution enhancement to maximal iteration is needed (which could comprise a small fraction of the footprint), one could reduce the overall run-time. In practice however, it is recommended that small footprints containing *just* the structure of interest be HiRes'd to high iteration, so noise suppression may not be useful in the end.

### 8.7 HiRes in Practice

Like most deconvolution methods, MCM does not alter the information content of the input image data. The signal and noise at a given frequency are scaled approximately together, keeping the SNR ~fixed. The process just re-emphasizes different parts of the frequency spectrum to make images more amenable to a certain kind of examination, e.g., for detecting previously unresolved objects and thereby increasing the completeness of a survey.

For optimal HiRes'ing, the input data will have to adequately sample the instrumental PSF to at least better than the Nyquist sampling frequency $2\nu_c$, where $\nu_c$ is the maximum frequency cutoff inherent in the PSF. For a *simple* diffraction-limited system with aperture diameter $D$, $\nu_c \propto D/\lambda$ and corresponds to the full width at half maximum (FWHM) of an Airy beam. Even if the detector pixels under-sample the PSF (below Nyquist), redundant coverage with $N_f$ randomly dithered frames can help recover the high spatial frequencies since the average sampling will scale as ≈ $1/\sqrt{N_f}$ of an input pixel. The better the sampling, the better the HiRes algorithm is at improving spatial resolution. For image data from the *Spitzer* IRAC and MIPS detectors with typically SNR >~ 5/pixel and 20 frame overlaps, the HiRes algorithm reduces the FWHM of the effective PRF to ≈ $0.33\lambda/D$ - a factor of 3 below the diffraction limit. This corresponds to almost an order of magnitude increase in flux per solid angle for a Gaussian profile. This enhancement assumes accurate knowledge of the PRF over the focal plane. As a rule of thumb, I've found that the attainable resolution from HiRes'ing is typically $R \sim 2*(native_pixel_scale)/\sqrt{N_f}$ assuming the $N_f$ overlapping frames were dithered in a way that sampled the detector pixels at the *sub-pixel level as uniformly as possible*. For example, to obtain a resolution of <0.2 arcsec from HiRes'ing IRAC band 1 frames (with pixel scale ~1.21 arcsec) would require <u>at least</u> $N_f = (2*1.21/0.2)^2 =$ 146 frames!

Example output from *icore* at three MCM iteration levels was shown in Figure 13. At high iterations, point source ringing starts to appear. The ringing around the satellite dwarf galaxy at the bottom is aggravated because the core is saturated in the data, and the PRF used for HiRes'ing (which is derived from unsaturated data) is not a good match. "Flat" core profiles in the data, due to either saturation or improperly corrected non-linearity, will contain relatively more power in the side-lobes than the actual PRF. When this PRF is used for HiRes'ing, these side-lobes will manifest as ringing artifacts in the HiRes image in order for it to reproduce the observations on convolution with the PRF. Even though ringing suppression was turned on in this example, ringing is still seen around some point sources. This is because these sources are superimposed on the extended structure of the galaxy. This structure acts like an elevated background against which point sources can ring. The ringing suppression algorithm relies on accurate estimation of the local background, and this can be difficult to achieve when complex structure is involved.

### 8.8 NaN'ing Low Depth-of-Coverage Regions

Due to the unreliability of temporal (stack) outlier detection at low depths-of-coverage and its complete absence for depths ≤ 4 (or more precisely ≤ value specified by parameter –ns_odet), the *icore* script has an option to replace pixels in all coadd and HiRes products with NaNs where the depth is less than some threshold (parameter: –cmin_coad; default = 4). This is only performed if "–crep_coad 1" is specified. The default is no replacement: "–crep_coad 0". If performed, products associated with intensity values, uncertainties (all flavors), and SNR maps (for either simple coaddition [§§6,7] or HiRes'ing [§8]), have their pixels replaced by NaNs.

### 8.9 Moving Object Co-addition and/or HiRes'ing

*icore* supports the ability to co-add images in the rest frame of a moving object (e.g., asteroids, comets, planets, artificial satellites etc..) and also perform HiRes'ing if desired. This is all driven by the "–mov" switch. If –mov is specified, the input list of intensity images (–imglist) must specify three fields: *image name*; *RA* [deg]; *Dec*[deg], where *RA*, *Dec* are equatorial coordinates of the moving target at each image epoch with corresponding *image name*. Example test scripts that co-add and HiRes a set of images of a known comet are named *icore_coaddmove_wise* and *icore_hiresmove_wise* respectively, and reside under the *testing* directory of the *icore* distribution.

In a nutshell, the software first transforms the WCS in the FITS headers of all input images to the first listed image. This first image will serve as the "reference" against which all other images (or rather the positions of the moving object therein) will be registered. This transformation assumes the astrometry (WCS) and distortion of all the input images is accurately calibrated. The reference pixel keywords (CRPIX1, CRPIX2) in the input image headers are then adjusted according to the observed *RA*, *Dec* of the object at each image epoch. This redefines the WCS coordinate system to be centered on the moving object so co-addition (and/or HiRes'ing) can proceed. An important assumption here is that we do not adjust for possible "moving rotation" (yet). Images are only registered in the moving-object frame using orthogonal translations. Compensation for rotation is important when co-adding resolved or extended objects (e.g., comets with extended tails and comae), and is especially important for maximizing gains from HiRes'ing.

To speed up later co-addition and/or HiRes'ing, FITS image stamps of size –stpX × –stpY and centered on the moving object in each image are cut-out and written to the output working directory: –outdir/stamps/. Furthermore, these image stamps are appended with a "S.fits". Uncertainty and mask-image equivalents (if specified via –unclist and –msklist respectively) are also generated under –outdir/stamps/. For stamp-cutouts near an edge of an input frame, the intensity and uncertainty stamps are appropriately padded with NaNs and the masks with zeroes. These stamps (with WCS redefined in the moving object frame) are then processed as normal using the same steps to produce co-adds and/or HiRes'd products. As a detail when generating PRF-interpolated co-adds or HiRes'd products from the stamps, only the first listed PRF image in the list specified by –psflist is used. Furthermore, owing to the stamp-image geometry, this PRF is reformatted (primarily the FPALOCX, FPALOCY keyword values are updated) and a new PRF image is written (with same filename) under –outdir. The input filename to –psflist is also updated and prepended with "New_" under –outdir.

Note: the output center RA, Dec of the intensity image product (co-add or HiRes image) will be forced to that of the first listed image in the file specified by –imglist. However, the footprint rotation and dimensions will be taken from the nominal user inputs: –rot, –sizeX, and –sizeY.

### 8.10 Photometric Performance and Expectations from MCM/HiRes

We have explored the impact of the MCM resolution-enhancement process on photometric flux and noise measurements made on HiRes'd images using a simulation. A simulation here is useful for two reasons. First, it enables us to validate the accuracy of outputs given knowledge of the truth, and second, it provides a method for unambiguously computing the output noise (in response to the input) by simulating repeated noise realizations sampled from an assumed prior noise-distribution model. The simulation used a single input image with input/output pixel sizes similar to that used in processing of WISE image data. We assumed a spatially flat background of 30 DN and added a point-source of total flux 3000 DN in the middle, then convolved with the WISE band 1 native PSF. We could have simulated another instrument or band, but any mock image suffices to illustrate the HiRe'sing performance in general. We then added Poisson noise to this image with variance $\sigma^2 = DN/g$, where we assumed an electronic gain factor of $g = 1$ for simplicity. Different input noise models and/or distributions don't change our general conclusions. 500 independent noise realizations (or trials) were simulated for the input image, and each trial was processed through the MCM-HiRes algorithm to seven different iterations: 1, 2, 4, 8, 16, 24, and 32. Output HiRes images at these iterations for a *single* simulation-trial are shown in Figure 18. At the top-left of this figure is the input raw simulated image for this trial. We used the same MCM processing parameters optimized for reconstructing WISE images in general. This includes ringing-suppression as described in §8.5.

One can see in Figure 18 that at low iterations, the noise becomes spatially correlated, then the positive noise spikes and the point-source at the center gradually "decorrelate" and sharpen with increasing iterations. Here the point-source FWHM went from ~ 5.8 arcsec (raw input simulated image) to ~ 1.75 arcsec (iteration 32), equivalent to an increase of 11× in the flux per solid angle at the point source position.

Figure 19 shows the dependence of the measured point-source flux (using standard aperture photometry; §13) on the number of MCM iterations from the single simulation-trial images and the average over all 500 independent trials at each iteration number. This shows that the measured fluxes are consistent with the true flux (within measurement error) or what one would measure in an "observed" noisy image. This is an important requirement for any image reconstruction process. Figure 20 shows the behavior of the photometric Signal-to-Noise (S/N) ratio and 1-σ uncertainty in the integrated point-source flux (i.e., the single simulation-trial error bars in Figure 19) as estimated from the standard-deviation of measurements over 500 independent trials. In general, the S/N in particular (as shown in Figure 20) is expected to decrease monotonically with iteration number. The hump in S/N at iterations 16 and 24 is an artifact of noise in the single-trial flux measurements at these iterations (Figure 19). This overall decrease in integrated S/N is a consequence of noise amplification in general with increasing iteration (red dashed curve in Figure 20). However, the drop in S/N is relatively small, measuring ≈ 6% over 20 MCM iterations.

Figures 21 and 22 illustrate the evolution of the peak pixel-flux of our simulated point-source, including its 1-σ uncertainty and signal-to-noise ratio as a function of MCM iteration number. All measurements pertain to single simulation trials. Figure 21 shows the effective rate at which the flux of a point-source is forced into its peak with increasing iteration (assuming the WISE pixel sampling). The increase in peak pixel-flux is a factor of ≈ 23 from iteration 1 to 32. Compared to the uncertainties in integrated flux measurements (Figures 19 and 20), the relative increase in the peak-flux uncertainty with iteration number (Figure 22) is appreciably greater. This implies local (pixel-scale) noise-fluctuations are more prone to amplification in the MCM process. Furthermore, we find the amount of amplification depends on the input noise level. For example, Figure 22 also shows the evolution of the background pixel spatial-RMS (dotted curve, ×10 for clarity). The input background (Poisson) noise simulated here is a factor of 8× (in σ) below that at the point-source peak position, and the background RMS varies by only a factor of ≈ 2 from iteration 1 to 32. Meanwhile, the peak pixel 1-σ flux uncertainty varies by a factor of ~ 60 over the same iteration interval. This shows that MCM is inherently a non-linear process in the reconstruction of signals in the presence of noise.

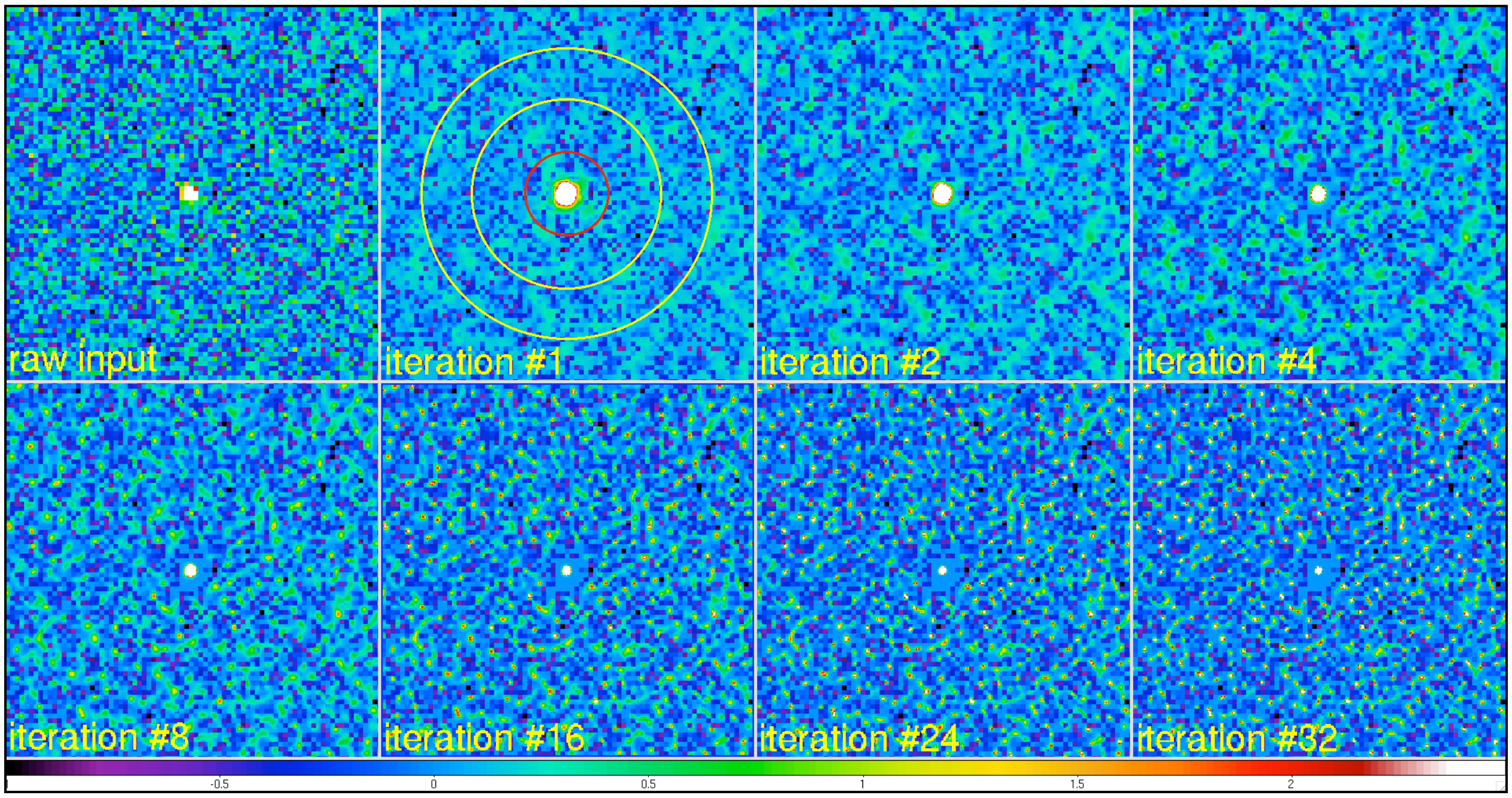

**Figure 18: Intensity images from the HiRes-MCM algorithm containing simulated Poission noise and a single point-source in the middle. The input simulated image is shown at top left. Outputs are shown for seven different MCM iterations (at the same image stretch). These illustrate the evolution of the noise structure. The "iteration 1" output image shows an overlay of the source aperture and background annulus used for the photometric analysis (see text and Figures 19-22 below).**

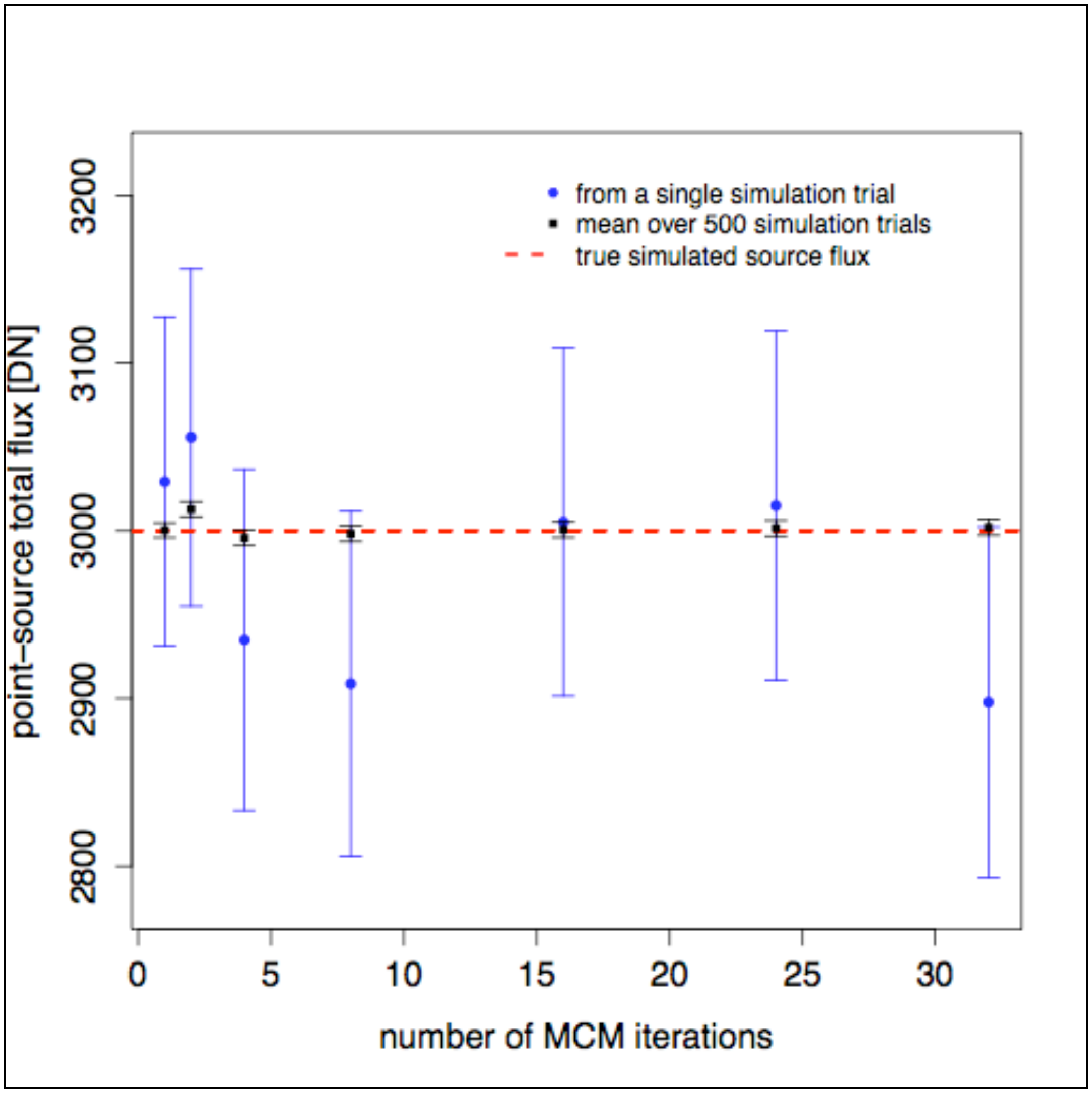

**Figure 19: Total flux of the central point-source shown in Figure 18 from aperture photometry as a function of iteration number. Measurements were made on image outputs from single simulation trials (*blue circles*) as well as an averaged stack of 500 trials at each iteration (*black squares*). The *red dashed line* shows the true input simulated flux (= 3000 DN). These results show that integrated flux measurements made on the HiRes'd images are consistent with the truth (within measurement errors) and hence unbiased.**

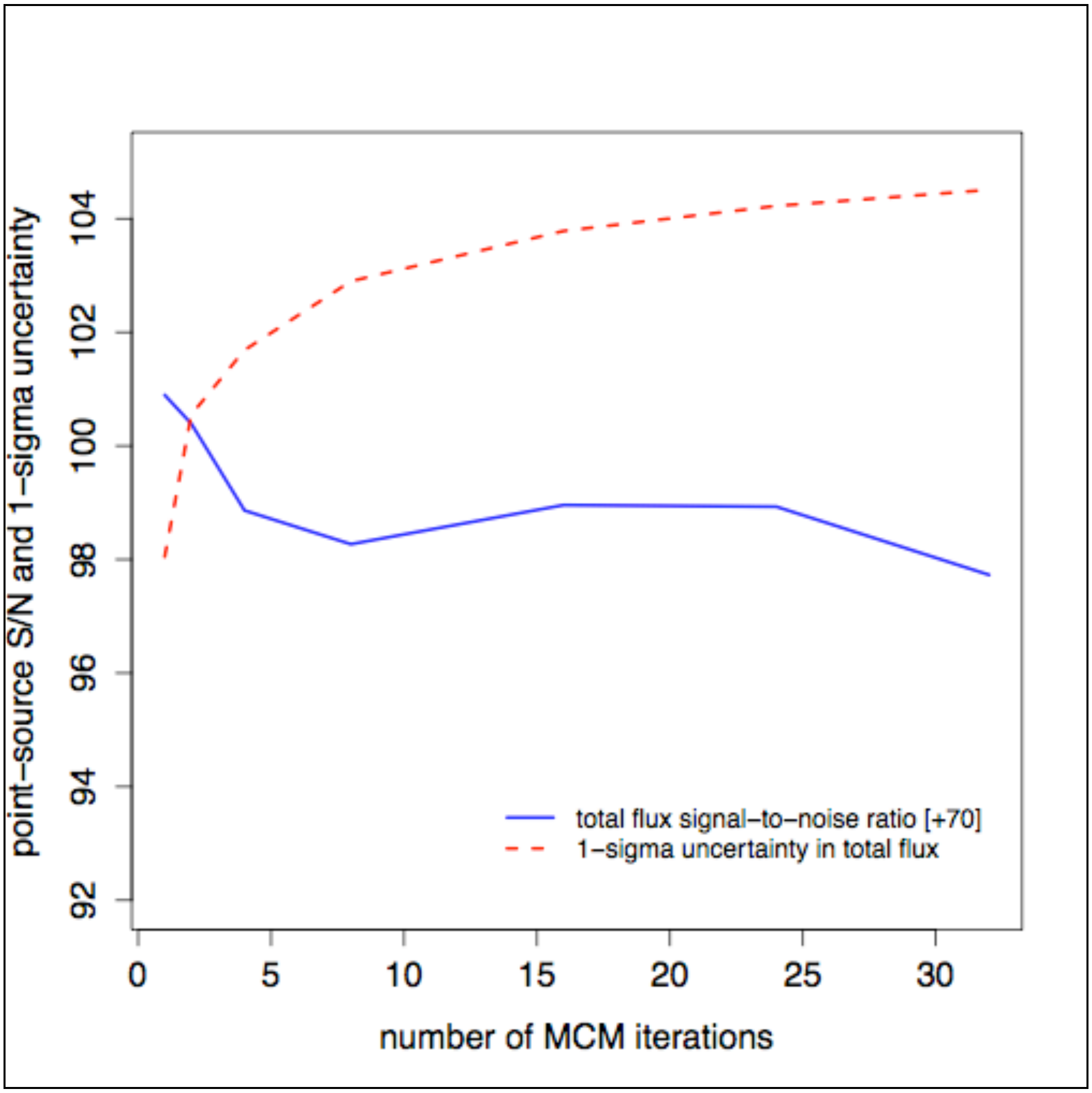

**Figure 20: Evolution of the Signal-to-Noise (S/N) ratio with MCM iteration number for the single simulation-trial measurements shown in Figure 19 (*blue solid curve*). An offset of 70 was added for display purposes. Also shown is the 1-σ uncertainty used for the S/N estimates (*red dashed curve*). This represents the standard-deviation in the measured point-source flux (using aperture photometry) over 500 independent simulation trials.**

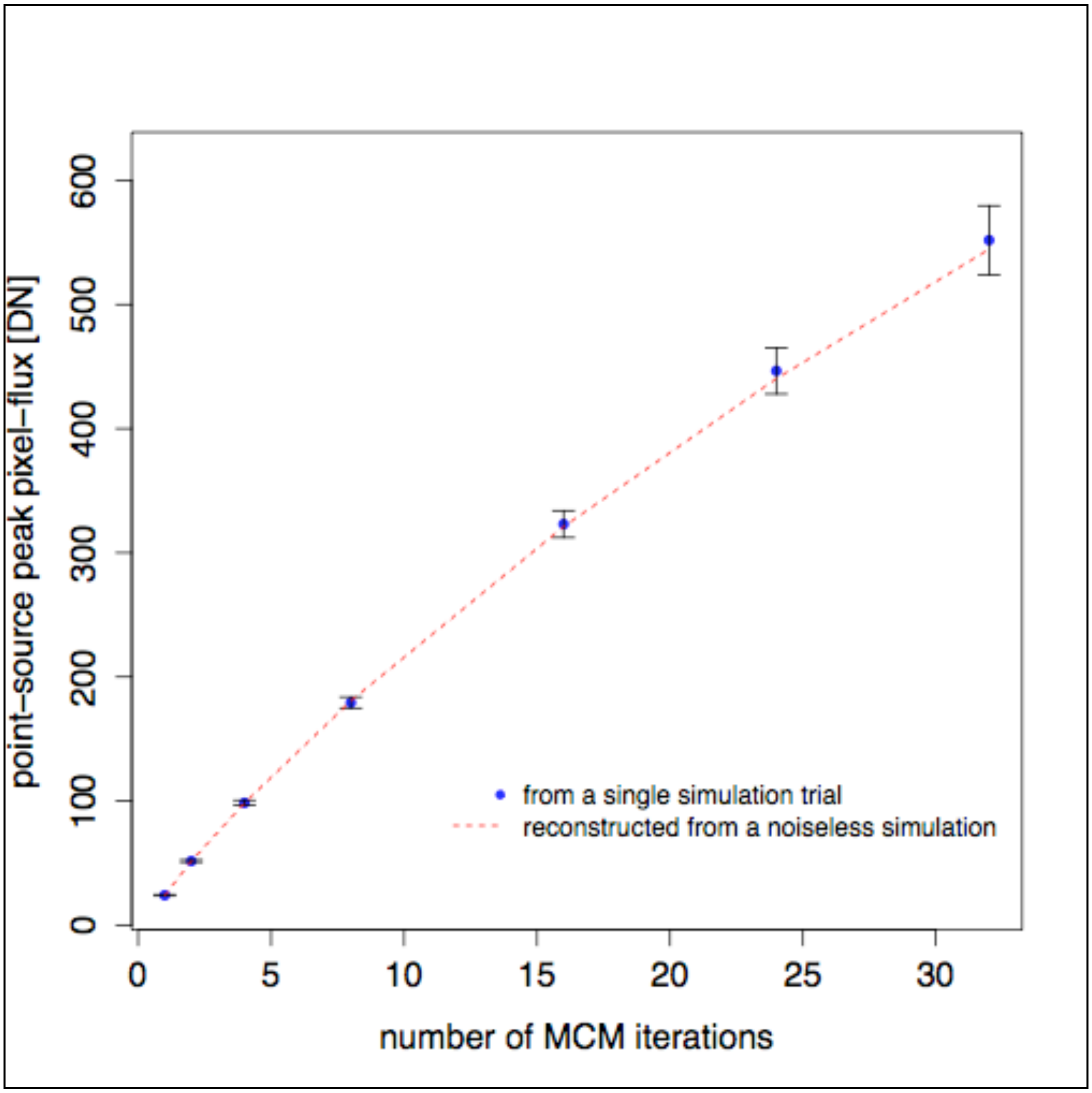

**Figure 21: Flux value in the peak pixel of the point-source shown in Figure 18 as a function of iteration number. Measurements were made on image outputs from single simulation trials (*blue circles*). The *red dashed line* shows the true reconstructed peak pixel flux one would obtain if noise were absent. In general, this plot shows the rate at which flux from the wings of a point-source is forced into its peak with increasing MCM iteration number.**

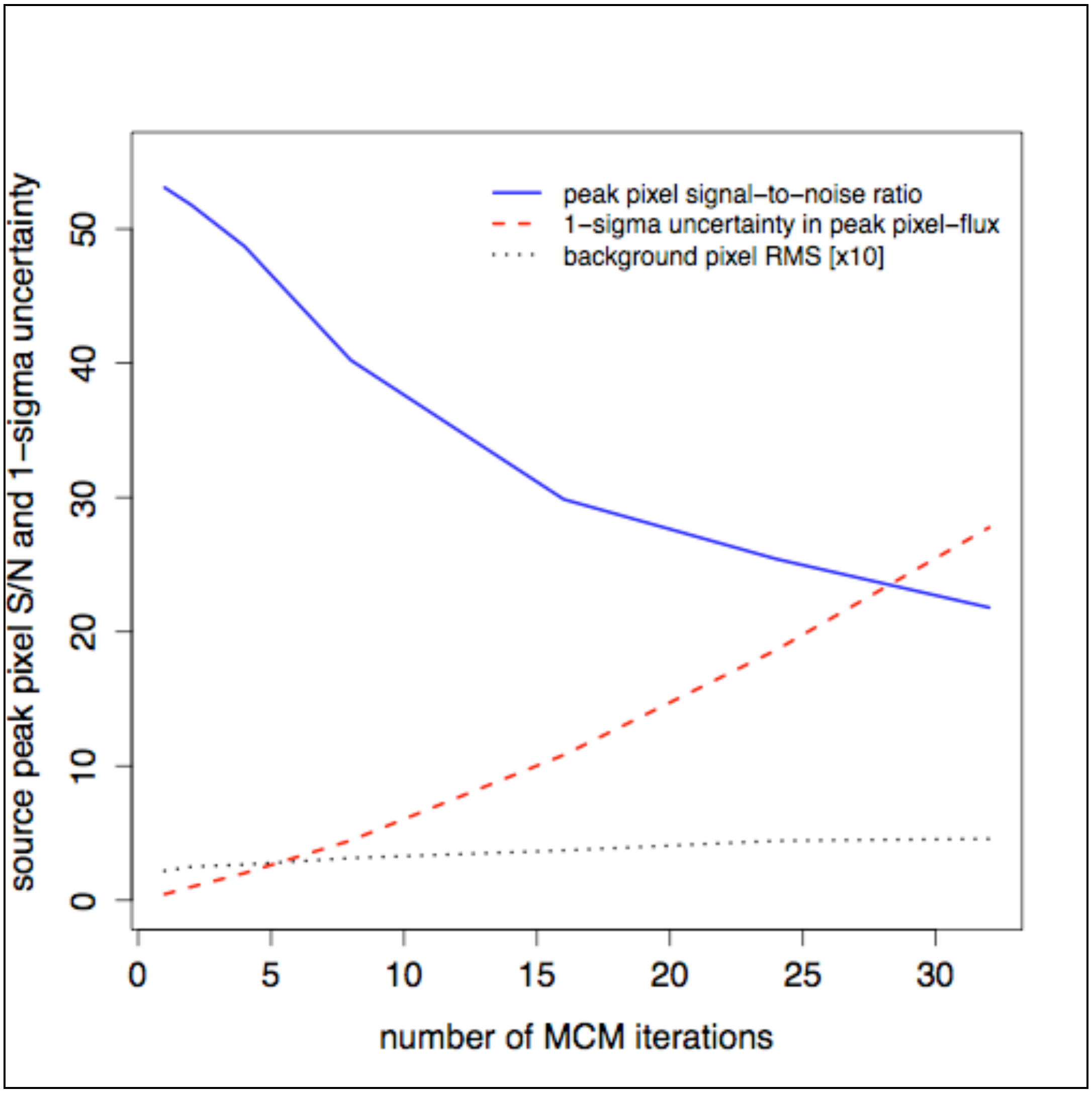

**Figure 22: Evolution of the peak-pixel S/N ratio with MCM iteration number for the single simulation-trial measurements shown in Figure 21 (*blue solid curve*). The 1-σ uncertainty used for these S/N estimates is shown by the (*red dashed curve*). This represents the standard-deviation in the measured *peak-pixel* source flux over 500 independent simulation trials. Also shown is the RMS in background noise fluctuations (per pixel) versus iteration number (*black dotted curve*). The latter were multiplied by 10 for clarity.**

# 9 QUALITY ASSURANCE

Quality Assurance metrics are only generated if the –qa switch is specified. Below we list the output metrics for each product: intensity co-add (*int*); depth-of-coverage map (*cov*); co-add uncertainty (*unc*), as well as ensemble metrics for the input intensity frames (*frm*) and masks (*msk*). These are written to an output table in IPAC format (–qameta). Below we show the metric names and definitions as they appear in the table. Descriptions of some of the metrics and their purpose are given further below.

Metrics are computed for the overall footprint as well as over squares defined by an $N \times N$ grid where $N$ is an input parameter (–qagrid <*N*>; default $N = 3$). This is to facilitate an analysis of the variation in metrics over the co-add footprint. It also provides "independent" sample sets from which to gauge the reliability of a metric (e.g., if a particular grid square happens to contain an extended source, the background-noise estimate therein will be biased).

The various sigma (data-scale) measures shown below are not redundant. It is prudent to measure some of the important metrics using different methods since a specific algorithm may not be robust against unforeseen glitches and irregular behavior in the data. Each sample metric is also subject to uncertainty and bias and it's important not to be mislead by any one of them.

```
\ QA metadata for co-add image products
\ Generated by icore, v.3.8.1 on 2013-04-01 at 13:02:28
\ Definitions of metric identifiers:
\ int: co-add intensity values
\ cov: depth-of-coverage map values
\ unc: co-add uncertainty values
\ msk: statistics on frame-pixels omitted, including temporal outliers
\ frm: input frame-stack ensemble statistics
\ Metric units (excluding enumerates) are in intensity co-add units
\ Names appended with _xy refer to value for partition x,y in co-add footprint: e.g. for 3 x 3
  grid: _11 => square at lower left corner; _33 => square at upper right
\ FITS file products represented:
\ 1: wise/outputs_coad1/mosaic-int.fits
\ 2: wise/outputs_coad1/mosaic-cov.fits
\ 3: wise/outputs_coad1/mosaic-unc.fits
\
|name                    |comment
\
\ Global co-add metrics
\
 coad:int:numframes        Number of input frames overlapping with co-add footprint
 coad:int:NumNaN           Number of NaN pixels over whole intensity co-add
 coad:unc:NumNaN           Number of NaN pixels over whole uncertainty co-add
 coad:int:Min              Minimum pixel value over whole intensity co-add
 coad:int:Max              Maximum pixel value over whole intensity co-add
 coad:int:Mean             Mean pixel value over whole intensity co-add
 coad:int:Median           Median pixel value over whole intensity co-add
 coad:int:StdDev           Standard (RMS) Deviation from mean (unbiased estimate) over co-add
 coad:cov:Min              Minimum pixel value over whole coverage co-add
 coad:cov:Max              Maximum pixel value over whole coverage co-add
 coad:cov:Mean             Mean pixel value over whole coverage co-add
 coad:cov:Median           Median pixel value over whole coverage co-add
 coad:cov:FivePtile        5%-tile value over whole coverage co-add
 coad:cov:NineFivePtile    95%-tile value over whole coverage co-add
\
\ Co-add metrics for square partition 1, 1
\
 coad:int:Min_11           Minimum pixel value
 coad:int:Max_11           Maximum pixel value
```

```
coad:int:Mean_11          Mean pixel value
coad:int:Median_11        Median pixel value
coad:int:Mode_11          Mode pixel value (fuzzy)
coad:int:StdDev_11        Standard (RMS) Deviation from mean (unbiased population estimate)
coad:cov:Min_11           Minimum pixel value
coad:cov:Max_11           Maximum pixel value
coad:cov:Mean_11          Mean pixel value
coad:cov:Median_11        Median pixel value
coad:cov:FivePtile_11     5%-tile value
coad:cov:NineFivePtile_11 95%-tile value
coad:unc:Min_11           Minimum pixel value
coad:unc:Max_11           Maximum pixel value
coad:unc:Mean_11          Mean pixel value
coad:unc:Median_11        Median pixel value
coad:int:NumAtMedCov_11   Number of pixels with Median Coverage
coad:int:SigLTMADMED_11   Sigma from Low-Tail Median Absolute Deviation from Median *at Med Cov*
coad:int:SigLTStdMod_11   Sigma from Trimmed Low-Tail standard deviation from mode *at Med Cov*
coad:int:Med16ptile_11    Sigma from Median - 16%-tile *at Median Coverage*
coad:int:84-16ptile_11    Sigma from [84%-tile - 16%-tile]/2 *at Median Coverage*
coad:unc:MdAtMedCov_11    Median pixel uncertainty *at Median Coverage*
coad:unc:LTMADMED_Unc_11  Ratio: SigLTMADMED_11/MdAtMedCov_11(pseudo Chi2) *at Median Coverage*
coad:unc:Md16ptile_Unc_11 Ratio: Med16ptile_11/MdAtMedCov_11 (pseudo Chi2) *at Median Coverage*
coad:unc:LTStdMod_Unc_11  Ratio: SigLTStdMod_11/MdAtMedCov_11(pseudo Chi2) *at Median Coverage*

...Continued for remaining (N x N) – 1 square partitions over footprint.
\
\ Statistics on frame-pixels omitted over all input frames
\
coad:msk:MinFlag          Minimum number of pixels flagged per frame
coad:msk:MaxFlag          Maximum number of pixels flagged per frame
coad:msk:MeanFlag         Mean number of pixels flagged per frame
coad:msk:MedFlag          Median number of pixels flagged per frame
coad:msk:MinOutlier       Minimum number of temporal outliers per frame
coad:msk:MaxOutlier       Maximum number of temporal outliers per frame
coad:msk:MeanOutlier      Mean number of temporal outliers per frame
coad:msk:MedOutlier       Median number of temporal outliers per frame
\
\ Statistics in frame median backgrounds (bef/aft bckgnd matching) over all input frames
\
coad:frm:MinFrmMedB       Minimum frame median (before bckgnd matching)
coad:frm:MaxFrmMedB       Maximum frame median (before bckgnd matching)
coad:frm:MedFrmMedB       Median of all frame medians (before bckgnd matching)
coad:frm:StdFrmMedB       Std-dev in frame medians from mean (before bckgnd matching)
coad:frm:MinFrmMedA       Minimum frame median (after bckgnd matching)
coad:frm:MaxFrmMedA       Maximum frame median (after bckgnd matching)
coad:frm:MedFrmMedA       Median of all frame medians (after bckgnd matching)
coad:frm:StdFrmMedA       Std-dev in frame medians from mean (after bckgnd matching)
```

Below are descriptions of the not-so-obvious metrics and their purpose. The suffix "_*xy*" means the metric is used for partition *x,y* in the *N* x *N* grid of the co-add footprint (see above).

### coad:int:Mode_*xy*

This metric represents an estimate of the most frequently occurring pixel value. It represents a "fuzzy" measure since it is based on an approximate binning method that requires an appreciable sample size, i.e., >~ 500 pixels. We advise this metric be used with caution since it will be highly inaccurate for small samples. In brief, the method first partitions the histogram of all image pixel values into 10 equal area bins (or "10%-tiles"); it then takes the bin with the smallest width as the one most probable to contain the mode since this bin is located near the peak of the histogram; the median of the data in this bin is then used as an estimate of the mode.

**coad:int:NumAtMedCov_*xy***

This represents the number of pixels in the depth-of-coverage map with values between $m_{cov} \pm 0.25$, where $m_{cov}$ = median depth-of-coverage. The eight metrics following this in the meta-data table only use intensity and uncertainty co-add values corresponding to pixels within this depth-of-coverage range.

**coad:int:SigLTMADMED_*xy***

This is an estimate of sigma using the Median Absolute Deviation (MAD) *from the median* using only pixel values in the *lower tail*. It is rescaled for consistency with the standard deviation of a normal distribution in the large sample limit. This measure is more robust against outliers (including sources) in the upper-tail than the sigma estimated from the sample standard deviation. Therefore it can be used to estimate the background RMS, assuming the background level is stationary. We should mention that this measure can exhibit more variation (less efficiency) than the standard-deviation on average. This metric is defined as:

$$\sigma_{MAD} = 1.4826 median\left\{\left|p_i - median\{p_j\}\right|\right\},$$

where $p_i$ = the set of pixel values < sample median, $p_j$ = all pixel values in original sample.

**coad:int:SigLTStdMod_*xy***

This is another robust measure of sigma that uses the mode measure above as an estimate of the first moment. It is defined as the standard-deviation of the lower-tail pixel values from the sample mode after trimming possible low-tail outliers at $< 5\sigma_{MAD}$. In other words

$$\sigma_{\text{mode}} = \sqrt{\frac{1}{N-1}\sum_i^N \left(p_i - \text{mode}\{p_j\}\right)^2},$$

where $p_i$ = the set of pixel values in the range: $mode - 5\sigma_{MAD} \leq p_i \leq mode$; $p_j$ = all pixels in the original sample; and $\sigma_{MAD}$ was defined above.

**coad:int:Med16ptile_*xy* & coad:int:84-16ptile_*xy***

These represent robust estimates of sigma using the sample quantile differences: $\Delta Q_L = q_{0.5} - q_{0.16}$ and $\Delta Q_R = 0.5(q_{0.84} - q_{0.16})$ respectively. For normally distributed data, these should approximately equal the standard deviation ($\sigma$). When there is severe source contamination or bright extended structure (which affects predominately the high tail), we expect $\Delta Q_R > \Delta Q_L$.

**coad:unc:LTMADMED_Unc_*xy* &**
**coad:unc:Md16ptile_Unc_*xy* &**
**coad:unc:LTStdMod_Unc_*xy***

These represent the respective ratios:

- **coad:int:**SigLTMADMED_*xy* / **coad:unc:**MdAtMedCov_*xy*
- **coad:int:**Med16ptile_*xy* / **coad:unc:**MdAtMedCov_*xy*
- **coad:int:**SigLTStdMod_*xy* / **coad:unc:**MdAtMedCov_*xy*

where the metrics in the numerator were defined above. The denominators contain the median uncertainty of pixels corresponding to the median depth-of-coverage (see discussion under **coad:int:NumAtMedCov_*xy*** above). These ratios can be interpreted as pseudo-$\chi^2$ measures that quantify in a broad sense, the statistical compatibility of uncertainty estimates with the distribution of co-add intensity values. In other words, whether the pixel uncertainties are compatible with the overall RMS fluctuation about the background in a co-add. If so, then these ratios should be $\approx$1. Even though robust sigma estimates are used, it's important to note that high source confusion, bright extended structure, and/or a strongly varying background can make these ratios significantly deviate from 1.

**coad:frm:StdFrmMedB & coad:frm:StdFrmMedA**

These are the two most important metrics under the "Statistics in frame median backgrounds…" section of the meta-data table. They correspond to respectively the standard deviation over all the global medians of the input frames, *before* and *after* background-level matching. The before and after measures are used to assess the quality of background matching in a relative sense. The standard deviation (which quantifies the degree of scatter) is expected to be smaller after background matching. If it is not, it indicates that very little correction to the levels in the original frames was needed, or, that the presence of bright extended structure amongst frames is biasing background levels and affecting the robustness of the background-matching algorithm.

## 10 EXAMPLE SCRIPTS AND RECIPE

Ten test scripts are provided in the ICORE distribution package. Instructions on how to execute any of these are outlined in the README file of the distribution. Here's a summary of the example scripts:

- ***icore_coadd_mips***: creates co-add from *Spitzer* 24 micron observations of NGC 5194/5195 (M51a/b). Uses input data from testing/mips24/. Products get created under: testing/mips24/outputs_coad/.
  *** N.B. This script is optimized for *Spitzer*-MIPS24 data.

- ***icore_hires_mips***: creates a resolution-enhanced (HiRes) coadd from 24 micron observations of NGC 5194/5195 (M51a/b). Uses input data from testing/mips24/. Products get created under: testing/mips24/outputs_hires/.
  *** N.B. This script is optimized for *Spitzer*-MIPS24 data. Two PRFs are available under testing/mips24/psf/, a large one and a smaller one. The file testing/mips24/PRFList.txt (script variable "inpsflist") lists the smaller PRF. For better HiRes quality (but slower runtimes), this can be replaced with the larger PRF.

- ***icore_coadd_irac***: creates co-add from *Spitzer*-IRAC 4.5 micron (band 2) observations of the globular cluster "47 Tucanae". Uses input data from testing/irac/. Products get created under: testing/irac/outputs_coad2/.
  *** N.B. This script is optimized for all *Spitzer*-IRAC bands.

- ***icore_hires_irac***: creates a resolution-enhanced (HiRes) co-add from *Spitzer*-IRAC 4.5 micron (band 2) observations of the globular cluster "47 Tucanae". Uses input data from testing/irac/. Products get created under: testing/irac/outputs_hires2/.
  *** N.B. This script is optimized for *all Spitzer*-IRAC bands.

- ***icore_hires_sim***: creates HiRes coadds from 10 overlapping simulated frames containing a single point source and Gaussian noise. Used to explore uncertainty estimation and HiRes performance. Uses input data from testing/hiressim/. Products get created under: testing/hiressim/outputs/.

- ***icore_coadd_wise***: creates co-adds from WISE observations of the Spiral Galaxy IC342 (any band). Uses input data from testing/wise/. Products get created under: testing/wise/outputs_coad${band}/ where ${band} is the band variable defined in the script.
  *** N.B. This script is optimized for WISE data (all bands).

- ***icore_coaddmove_wise***: creates a co-add from WISE band 3 observations in the rest frame of Comet 67P/Churyumov-Gerasimenko. Uses input data from testing/wise/movingobject/. Products get created under: testing/wise/

movingobject/outputs_coad${band}/ where ${band} is the band variable defined in the script.
*** N.B. This script is optimized for WISE data (all bands), but only band 3 frames are provided for this test.

- ***icore_hires_wise***: creates HiRes co-adds from WISE observations of the Spiral Galaxy IC342 (any band). Uses input data from testing/wise/. Products get created under: testing/wise/outputs_hires${band}/ where ${band} is the band variable defined in the script.
  *** N.B. This script is optimized for WISE data (all bands). Two PRF sizes are available under testing/wise/psf/, large and small. The files testing/wise/PRFList_hires${band}.txt (script variable "inpsflist") list the smaller PRFs for speed. For better HiRes quality (but much slower runtimes), these can be replaced with the larger PRFs.

- ***icore_hiresmove_wise***: creates a HiRes co-add from WISE band 3 observations in the rest frame of Comet 67P/Churyumov-Gerasimenko. Uses input data from testing/wise/movingobject/. Products get created under: testing/wise/ movingobject/outputs_hires${band}/ where ${band} is the band variable defined in the script.
  *** N.B. This script is optimized for WISE data (all bands), but only band 3 frames are provided for this test. Two PRF sizes are available under testing/wise/psf/, large and small. The files testing/wise/PRFList_hires${band}.txt (script variable "inpsflist") list the smaller PRFs for speed. For better HiRes quality (but much slower runtimes), these can be replaced with the larger PRFs.

- ***icore_hires_spire***: creates HiRes'd images from Herschel-SPIRE observations (calibrated 2D maps) of the Spiral Galaxy M33 at either 250 or 500μm. Uses input data from testing/spire/. Products get created under: testing/spire/outputs_hires${band}/ where ${band} is the band variable defined in the script (band 1 => 250μm; band 2 => 500μm).
  *** N.B. This script is optimized for both the 250 and 500μm bands. It is assumed that flux-density timelines have already been calibrated and co-added to create "naive" 2D image maps (according to the SPIRE map-making lexicon). These maps can then be supplied to *icore* for resolution enhancement. Note that this test script (as configured) uses no input prior uncertainties or pixel masks, but they can be provided if available.

### 10.1 Quick Recipe

The best way to proceed with co-adding and/or HiRes'ing your data is to start with one of the above example scripts, study the I/O by consulting Table 1 in §3.1, then modify the script to your needs. Here's a quick recipe for the impatient:

1. Identify your footprint geometry (linear size along X and Y dimension, center RA, Dec, and rotation. The latter assumes a West-of-North convention, i.e., the CROTA2 definition of the standard WCS). It is assumed that the bulk of your input frames overlap with this footprint. The footprint geometry and location is defined by the ra, dec, rot, sx, sy variables in the above example scripts, within the lines delimited by "============".

2. Ensure your input intensity frame list (and optionally, corresponding mask and frame uncertainty lists) have been made and reside in the correct place. The list filenames are defined by the "inimglist", "inmsklist", and "inunclist" variables in any of the above scripts. If you don't have masks and/or uncertainty frames, you can comment out (with a "#") the corresponding lines in the *icore* call further down in the script, e.g:
   # -msklist $inmsklist \
   and/or:
   # -unclist $inunclist \

3. If you are simply co-adding (using any of the icore_coadd_* scripts above), you can create an area-overlap weighted coadd by setting sccoad = 1, otherwise set this to 0 to use the PRF defined by the "inpsflist" variable as the interpolation kernel. Example PRFs are provided for all test scripts and they will be automatically used if you set sccoad = 0. For coadds created using area-weighted interpolation (sccoad = 1), the drizzle factor "-d_coad" in the *icore* call may be of interest [this has default=1 => no drizzle].

4. For optimal processing speed, "set nthreads" to the number of CPU cores available on your machine. This parameter is currently set to 8 in all the example scripts and may not be applicable to your machine. **Note:** some Intel CPUs have hyper-threading and can support two active threads per core, e.g., a machine with 8 physical cores could run 16 simultaneous threads (= 16 "logical" cores). In this case, nthreads = 16.

5. To be safe, check that you have sufficient memory and a fast enough CPU before pulling the trigger. See Section 11 to estimate the approximate memory and expected execution time.

6. Lastly, have a glance at the "Advisories and Caveats" in Section 12 to avoid possible frustration.

## 10.2 Restarting From a Partially Processed Run

An important detail is that the pre-processing steps (background and outlier-detection) generate intermediate products (new images and updated masks) in the user-specified directory "outdir". Depending on the *icore* execution mode, ancillary background images and images with backgrounds removed from the input frames may be generated. If you would like to recreate a co-add (or HiRes'd version) without re-running the pre-processing steps (since they can be time consuming), here's the procedure using the *Spitzer* 24μm co-add example script:

1. Take your original output directory (variable "outdir"), e.g., "mips24/outputs_coad" and prepend the string "mips24/outputs_coad/New_" to your existing file list variables: "inimglist", "inmsklist", and "inunclist", e.g., the "inimglist" variable becomes: "mips24/outputs_coad/New_ImageList.txt". Do the same for the "inunclist" and "inmsklist" names.

2. Reset your "outdir" to a new output directory name if you don't want the previous products over-written.

3. Comment-out with a "#" (or switch off) the pre-processing steps, i.e., -tmatch, -bmatch, -odet, and -cpmsk.

4. Execute your script. New products will be written to the new directory specified in 2 if a new output destination was specified.

# 11 DISK, MEMORY USAGE, AND RUNTIME

## 11.1 Disk Space

You should allow for enough disk space to hold an additional copy of your input frames for coadding, or, an additional two copies of your input data for HiRes'ing. If the "-mcmprod" diagnostic switch is specified for HiRes'ing, you'll need additional space to hold ~155*[n_coad*sx*sy/(pa_coad*pc_coad)$^2$] MB of intermediate data, where n_coad, sx, sy, pa_coad, and pc_coad are your *icore* command-line parameter inputs.

## 11.2 Memory (dynamic RAM)

This will be dominated by the temporal outlier detection step if executed (controlled by the "-odet" switch). In general, the memory usage for this step (in GigaBytes) can be calculated from your input parameters using the following:

If the "-partition" switch is NOT set (for memory optimization):

$$\mathrm{Mem} \sim \left[\left(\frac{0.052}{2^{\mathrm{ip_odet}}}\right)\left(\frac{N_{frames}}{\mathrm{nx_odet\ \ ny_odet}}\right)+1\right]\left[\frac{(\mathrm{sx}+0.00028\mathrm{pb_odet})(\mathrm{sy}+0.00028\mathrm{pb_odet})}{\mathrm{pa_odet}^2}\right]\ \mathrm{GB},$$

where $N_{frames}$ is the number of input intensity frames overlapping with your co-add footprint. Note that all input frames must overlap with your footprint. No provision is made to check this (yet). ip_odet is the "integer processing" flag where ip_odet=1 means processing is performed using I*2 integer arithmetic as opposed to R*4 floating point (ip_odet=0). Currently, ip_odet=1 is only set in the WISE test scripts since the input data ranges are well known [see §5.2.3 to tune for other cases]. All other parameters in the above equation are as set in your *icore* script.

If the "-partition" switch was set AND $N_{frames}$ > nmaxodet [*icore* input parameter with default = 300], we have the following conservative upper limit:

$$\mathrm{Mem} <\sim \left[\mathrm{nmaxodet}\left(\frac{0.052}{2^{\mathrm{ip_odet}}}\right)+1\right]\left[\frac{(\mathrm{sx}+0.00028\mathrm{pb_odet})(\mathrm{sy}+0.00028\mathrm{pb_odet})}{\mathrm{pa_odet}^2}\right]\ \mathrm{GB},$$

On completion of the outlier detection step, memory is returned to the system. The total memory consumed by the co-addition step is as follows:

For co-adds using *area-overlap weighted interpolation* (sccoad=1):

$$\mathrm{Mem} \sim 0.42\left[\frac{\mathrm{sx\ \ sy}}{\mathrm{pa_coad}^2}\right]\ \mathrm{GB},$$

For co-adds using *PRF-weighted interpolation* (sccoad=0):

$$\mathrm{Mem} \sim 0.35\left[\frac{(\mathrm{sx} + 0.00028\ \mathrm{prfX\ pa_coad\ pc_coad})(\mathrm{sy} + 0.00028\ \mathrm{prfY\ pa_coad\ pc_coad})}{(\mathrm{pa_coad\ pc_coad})^2}\right]\ \mathrm{GB},$$

where prfX, prfY are the PRF NAXIS1,NAXIS2 dimensions respectively, obtained from the input PRF FITS header.

*For HiRes'ing*, the total memory consumed can be obtained by multiplying the PRF-weighted co-add case above by 2.

### 11.3 Runtime

The runtime primarily depends on the number of CPU cores available (-nt_coad parameter), the processor clock speed, and total number of input pixels, including the number of PRF pixels if creating a PRF-interpolated co-add or HiRes'd product. For HiRes'ing, the runtime also depends on the total number of HiRes iterations ($N_{iter}$ = n_coad + n_coadn, where the latter are parameters in the *icore* script). Assuming no other CPU-intensive jobs are concurrently running, the total runtime (including all pre-processing steps) is expected to approximately scale as:

$$\mathrm{RunTime}\ \sim\ 544\left(\frac{3.33\mathrm{GHz}}{\mathrm{CPU}}\right)\left(\frac{N_{frames}}{100}\right)\left(\frac{\mathrm{imgNAXIS}^2}{1016^2}\right)\left[1 + 0.38 N_{iter}\left(\frac{\mathrm{prfNAXIS}^2}{27^2}\right)^p\left(\frac{16}{nthreads}\right)\right]\ \mathrm{seconds},$$

where:

CPU = processor speed in GHz (here scaled relative to an Intel Xeon 3.33 GHz processor);

$N_{frames}$ = number of input intensity frames;

imgNAXIS = input linear dimension in pixels of an input frame;

prfNAXIS = input linear dimension in pixels of input PRF;

$N_{iter}$ = n_coad + n_coadn for HiRes'ing;
= 1 for plain co-addition using either PRF-interpolation or area-overlap weighting (with optional drizzling);

$p$ = 0 only for co-addition using area-overlap weighting (with optional drizzling);
= 1 otherwise;

*nthreads* = number of concurrent threads to run (-nt_coad *icore* parameter). Ideally set to the number of CPU cores available. **Note:** some Intel CPUs have hyper-threading and can support two active threads per core, e.g., a machine with 8 physical cores could run 16 simultaneous threads (= 16 “logical” cores). It’s worth experimenting by monitoring the CPU load.

## 12 ADVISORIES AND CAVEATS

Here are a few advisories and traps to be aware of. More will be added as they become known.

- First, to state the obvious: the quality of co-added or HiRes’d image products depends on the instrumental and astrometric calibration accuracy of your input frames (this can’t be stressed enough). Instrumental calibration accuracy refers to how well the frames have been corrected for instrumental signatures (e.g., flat-fielding, linearity, etc.). This includes cosmic rays and other outliers with respect to the inertial sky, e.g., moving objects if these are not specifically targeted. Astrometric accuracy determines the precision at which your data can be mapped onto the co-add footprint. This depends on both the absolute pointing accuracy of your input frames, and the degree to which field-of-view distortion has been corrected. Inaccurately calibrated astrometry and/or distortion will affect relative registration and co-add image quality. The consequences for HiRes'ing are more severe. See below.

- If your input frames are in pure DN (count) units and not normalized by exposure-time (e.g., not already in DN/sec or some flux-related unit such as MJy/sr), ICORE will assume they all correspond to the same exposure-time unless the different exposures are reflected in different zero-point (MAGZP) values so the throughput-matching step (§4.1) can handle the rescaling. If exposure times differ and the input pixels are in un-normalized count units, the user can either: (i) rescale the input frames themselves before executing ICORE, or (ii) set *relative* MAGZP values in each FITS header so the throughput matching step can perform the rescaling, i.e., MAGZP = $2.5\log_{10}[t/t_b]$, where $t = t_b$ is some baseline exposure-time assigned MAGZP = 0.

- To determine whether there will be any gain in HiRes’ing, it is advised that you first determine if your data is well sampled, i.e., if the native PSF of the optical system with which your frames were acquired is sampled at the Nyquist rate or better: typically with detector pixel size <~ $\mathrm{FWHM_{PSF}}/2$ or <~ $\lambda/(2D)$ for *diffraction limited imaging through an unobscured circular aperture*. This includes the effective sampling provided by all dithered overlapping frames. Even if you cannot meet this criterion for a single frame exposure, you may with a stack dithered frames of the same piece of sky. An example is data acquired with the *Spitzer*-IRAC detectors where the native frames are under-sampled. With IRAC, we recommend at least ten sub-pixel-dithered overlapping frames to adequately sample the PSF and obtain a HiRes image that exceeds the native resolution by a factor of >~ 2 per axis (see below for scaling relation). This assumes your target of interest has a reasonably high SNR after combining all the available frames. The more overlapping frames you have, the higher the resolution that can be achieved. As a rule of thumb, I've found that the attainable resolution from HiRes'ing is typically $R \sim 2*(native_pixel_scale)/\sqrt{N_f}$ where $N_f$ is the number of overlapping frames with dithers that sample the detector pixels at the *sub-pixel level as uniformly as possible*. For example, to obtain a resolution of <0.2 arcsec from HiRes'ing IRAC band 1 frames (with pixel scale ~1.21 arcsec) would require <u>at least</u> $N_f = (2*1.21/0.2)^2 = 146$ frames!

- For optimal HiRes'ing it is safest to use a PRF sampled at a finer pixel scale and containing information at higher spatial frequencies than the maximum frequency effectively sampled by all your measurements, i.e., after accounting for all dithered overlapping frames. However, beware of runtime when using a finely sampled PRF (see §11.3).

- Be warned that badly calibrated input data such as a bad non-linearity correction (leading to erroneous peak-to-wing intensities for point source profiles) may not be adequately deconvolved using your template PRF unless its shape is adjusted to do so. Such effects will modify the observed profiles of sources and therefore limit the ability to recover information at high spatial frequencies. Furthermore, bad frame-to-frame registration will modify the effective observed PRF inherent in the combined data. Excessive ringing will occur in areas where there's a bad PRF match. HiRes'ing can therefore be used to test for input PRF accuracy (i.e., how well it matches the data), registration quality, and/or the accuracy of some instrumental calibrations (e.g., non-linearity).

- Two remarks about HiRes'ing:
    - It can be slow if the user is not aware of the potential bottle-necks. E.g., it is not designed to work on fields much larger than about 1 square degree. The goal is to bring out small-scale structure selected from co-adds or observations at other wavelengths. Furthermore, it is advised that you don't use a PRF with too many pixels or one that's too finely sampled, i.e., more than what you need to recover the maximum spatial frequency effectively sampled by your data. Take a look at the RunTime scaling relation in §11.3 to find out why. It is suggested you isolate the most significant structure and/or highest relative peak-to-wing signal in your PRF and crop it. The gain in resolution may be compromised, but this is a small penalty since most of the power in PRFs from most modern instruments resides within a radius of ~2.5 FWHM of the core anyway. You're bound to still get significant gains in resolution. If you're willing to use a large/extended PRF, then that's fine too. But be prepared to wait.

    - It can be a laborious and finicky process to tune the background removal process to mitigate ringing, in particular when one has extended structure (see §8.5 for methods). With the correct choice of background filter size and thresholds to retain the desired signal, it's all possible. You'll be rewarded in the end.

- If input uncertainty frames are used for co-addition (not HiRes'ing), it is assumed these have been appropriately rescaled to match the noise-sigma inherent in the data. The uncertainty frames are usually estimated beforehand using a *Poisson* + *ReadNoise* error model for the detector. It is up to the user to check that these model predictions are consistent with noise fluctuations in the data, e.g., from examining the chi-square distributions from PSF-fit photometry to point sources, from statistics computed within spatially uniform background regions or on the residuals from background surface-fit subtracted images, or from low-pass filtered images.

- For HiRes'ing only, uncertainty products are dynamically rescaled to ensure consistency with local RMS fluctuations in regions selected to have minimal variation due to sources and/or extended structure. It is advised that these products be checked and that uncertainties make sense before using them to estimate photometric uncertainties (see §8.4.1-8.4.2 for further discussion).

- Before embarking on a full-blown HiRes run that could take hours or days, it is recommended you first make a simple co-add of your field to assess the accuracy of the outlier rejection and background matching steps, and explore any artifacts in your data. This will allow you to tune any parameters in the pre-processing steps (§§4,5). Small glitches, artifacts, saturated cores, latents, and bad pixels (that can be tagged in the input masks) will be amplified during the HiRes process. A simple area overlap-weighted co-add (see example scripts in §10) will accentuate cosmic rays and saturated cores into "hard-edged" rim features that can be easily identified and masked.

- Furthermore, a quick simple co-add of your field can be used to validate the photometry in your final HiRes'd image. There's more that can go wrong with HiRes'ing - e.g., biases from improperly tuned parameters to support ringing suppression, convergence of flux levels, and/or inaccurate restoration of the noise distribution. Biases are not expected if everything was tuned correctly, but a sanity check is encouraged. If a significant global bias is found in your HiRes'd fluxes above some local background relative to those in the co-add (integrated on some equivalent scale and "aperture corrected" if necessary), this bias can be corrected assuming the co-add as an "unbiased pseudo-truth".

- Sources with saturated or bright (non-linear) cores may need to have their central regions masked to avoid excessive ringing while HiRes'ing. Recall that the HiRes process will attempt to reconstruct any masked flux in the cores of sources according to the input PRF profile. It is recommended that the diameter of any masked (circularly symmetric) contiguous region not be larger than the linear extent of the input PRF if the goal is to reconstruct the flux therein. The amount reconstructed will depend on the number of iterations. Masking a region considerably smaller than the input PRF is generally preferred.

- Beware of correlated pixel-noise when estimating uncertainties in photometry from co-adds or HiRes'd image products (both for aperture and PSF-fit photometry). See §13 for methods using aperture photometry.

## 13 APERTURE PHOTOMETRY AND UNCERTAINTY ESTIMATION

Below we briefly review the formalism for performing aperture photometry with uncertainty estimation off co-added or HiRes'd image products in the presence of correlated pixel-noise. Here we quote the main equations. Details and derivations can be found in:
*http://web.ipac.caltech.edu/staff/fmasci/home/mystats/ApPhotUncert_corr.pdf*
Furthermore, a tutorial and example of how to apply the methodology below to WISE Atlas Images using *SExtractor* can be found in:
*http://wise2.ipac.caltech.edu/staff/fmasci/SEx_WPhot.html*

First, the flux of a source (in image pixel units) within an aperture containing $N_A$ pixels can be computed from:

$$F_{src} = f_{apcor}\left(F_{tot} - N_A\overline{B}\right), \qquad (24)$$

where $F_{tot}$ is the sum of all pixel fluxes in the source aperture and $\overline{B}$ is a robust measure of the local background level per pixel (robust against outliers and other contaminating sources, e.g., a mode, trimmed mean, or median [with trimming]), usually estimated from an annulus around the source aperture. $f_{apcor}$ is the aperture correction factor, whose equivalent in magnitudes is $m_{apcor} = -2.5\log_{10}(f_{apcor})$

The corresponding 1-σ uncertainty in $F_{src}$ (Eq. 24) is given by:

$$\sigma_{src} = \left[f^2_{apcor}F_{corr}\left(\sum_i^{N_A}\sigma_i^2 + k\frac{N_A^2}{N_B}\sigma^2_{\overline{B}/pix}\right) + \sigma^2_{conf}\right]^{1/2} \qquad (25)$$

where

$N_A$ = number of pixels in source aperture

$N_B$ = number of pixels in background annulus

$\sigma_i$ = flux uncertainty for pixel $i$ from uncertainty image, rescaled if necessary

$F_{corr}$ = correlated noise correction factor for flux variance (see below)

$\overline{B}$ = robust background estimate per pixel in annulus (e.g., trimmed mean or median)

set $k = 1$ if $\overline{B}$ = mean background/pixel

set $k = \pi/2$ if $\overline{B}$ = median background/pixel

set $k = 0$ if assume $\overline{B} = 0$ or if no background is subtracted

$\sigma^2_{\bar{B}/pix}$ = Variance in sky background annulus in [image units]$^2$/pixel.

Can compute from square of RMS deviation from mean or median, or trimmed versions thereof. Can also approximate using robust estimators of scale:

$\approx \left[0.5\left(q_{0.84} - q_{0.16}\right)\right]^2 \approx \left[\left(q_{0.5} - q_{0.16}\right)\right]^2$ where the $q$ are quantiles, or the MAD measure:

$\approx \left[1.4826\ median\left|p_i - median\{p_i\}\right|\right]^2$, the Median Absolute Deviation from the median

$\sigma^2_{conf}$ = Confusion noise variance on scale represented by measurement aperture. This is not discussed here. For estimation details, see:

http://wise2.ipac.caltech.edu/docs/release/allsky/expsup/sec2_3f.html#confnoise

We have assumed the uncertainty in the aperture correction ($f_{apcor}$) is negligible since usually it can be determined to good accuracy. The source flux and its uncertainty as reported by Equations 24 and 25 respectively will be in the same units as the image pixels used, e.g., DN (as found in WISE and 2MASS image products). These can be converted to magnitude units as follows:

$$MAG = MAGZP - 2.5\log_{10}\left(F_{src}\right)$$

$$\sigma_{MAG} = \left[\sigma^2_{MAGZP} + 1.179\frac{\sigma^2_{src}}{F^2_{src}}\right]^{1/2},$$

where $F_{src}$ and $\sigma_{src}$ are computed using Equations 24 and 25. The "1.179" factor comes from $(2.5/\log_e[10])^2$. *MAGZP* is the magnitude zero-point with corresponding uncertainty $\sigma_{MAGZP}$ (described in §4.1). You can then convert to absolute flux density units (e.g., Jy) using the absolute flux $f_0$ for a $MAG = 0$ source: $f = f_0\ 10^{-0.4MAG}$. Alternatively, you could simply apply a DN-to-Jy conversion factor to Equations 24 and 25 directly if known. If however the image pixels were in absolute surface-brightness units (e.g., MJy/sr as found in *Spitzer* image products), you need only multiply the outputs from Equations 24 and 25 by the pixel area, e.g., in *steradians* if the pixel units were in MJy/sr.

The correlated-noise correction factor $F_{corr}$ in Eq. 25 will have to be estimated from your image directly or via a simulation (see below). It is not computed by *icore*. Correlated noise usually occurs in re-sampled and interpolated images with the degree of correlation depending on the extent of the interpolation kernel and ratio of input-to-output pixel size. For HiRes products, it will also depend on the iteration number (being minimal at convergence), and whether ringing suppression was invoked, in which case correlated noise is further reduced. The smoothing (interpolation) kernel moves noise-power from high to low spatial frequencies and therefore needs to be recaptured to properly quantify the uncertainty in the flux summed over a region. Ignorance of correlated-noise will lead to an underestimate of the final photometric uncertainty, sometimes by an appreciable amount. This is the purpose of the correlated-noise correction factor $F_{corr}$ in Eq. 25.

For the case of simple co-addition using overlap-area weighting (§7 - i.e., a top-hat PRF kernel), $F_{corr}$ is approximately the ratio of input-to-output pixel areas, or the ratio of squares of the input-to-output pixel scale, $(s_{in}/s_{out})^2$. For overlap-area weighted co-adds with drizzling, $F_{corr}$ can be reduced to ≈1 with careful tuning of the drizzle factor "–d_coad" (§7).

For co-adds created using the detector PRF as the interpolation kernel (§6), it can be shown that $F_{corr} \approx N_p(s_{in}/s_{out})^2$, where $N_p$ = the effective number of "noise-pixels[5]" characterizing the PRF, in terms of the number of native detector pixels. This is a good approximation for apertures containing $N_A$ >~ 50$N_p$ pixels. For significantly smaller apertures, $F_{corr}$ is smaller and can be approximated to ~10% accuracy by scaling from Figure 23. **Note:** this figure is specific to WISE-band PRFs only and an output co-add pixel size, but the dependence with aperture radius can provide some guidance.

Estimation of $F_{corr}$ for HiRes image products is more difficult since more parameters are involved. Figure 24 shows $F_{corr}$ vs. aperture radius for **WISE** HiRes products for each band using the WISE PRFs and an output pixel scale of 0.6875 arcsec/pixel. These were derived using a Monte-Carlo simulation with Gaussian input noise as described in *http://web.ipac.caltech.edu/staff/fmasci/home/mystats/ApPhotUncert_corr.pdf*. The values of $F_{corr}$ in Figure 23 are conservative and should be interpreted as upper limits since ringing suppression was turned off in the simulation runs. Ringing residuals can amplify pixel-to-pixel correlations.

If you're pedantic on getting the photometric uncertainty right in general for any image product (including HiRes), we advise performing a simulation in the spatial domain of the intensity image. This involves first throwing many apertures of fixed size (or each covering an area of $N_A$ pixels) at random. We then retain those apertures that don't significantly overlap with another and which fall within regions with a spatially-uniform and source-free background. We then compute the variance in their summed pixel-fluxes $V_A$. $F_{corr}$ is then the ratio $V_A/(N_A\sigma_i^2)$, where $\sigma_i$ is either the local pixel-to-pixel variance (square of the RMS) over the same spatially-uniform region, or some median value from the uncertainty image product (propagated from priors if available). The difficulty here is having enough random samples (apertures) that are not significantly contaminated by sources and cover a spatially-uniform background. This method will automatically account for any confusion noise from unresolved sources. If no spatially uniform background region(s) exist about which to compute *unbiased* statistics, this can be created by subtracting a low-pass filtered or surface-fit image from the original intensity image. The statistics are then computed on the residual image.

[5] For definition of 'noise pixels', see: http://wise2.ipac.caltech.edu/docs/release/allsky/expsup/sec4_6ci.html

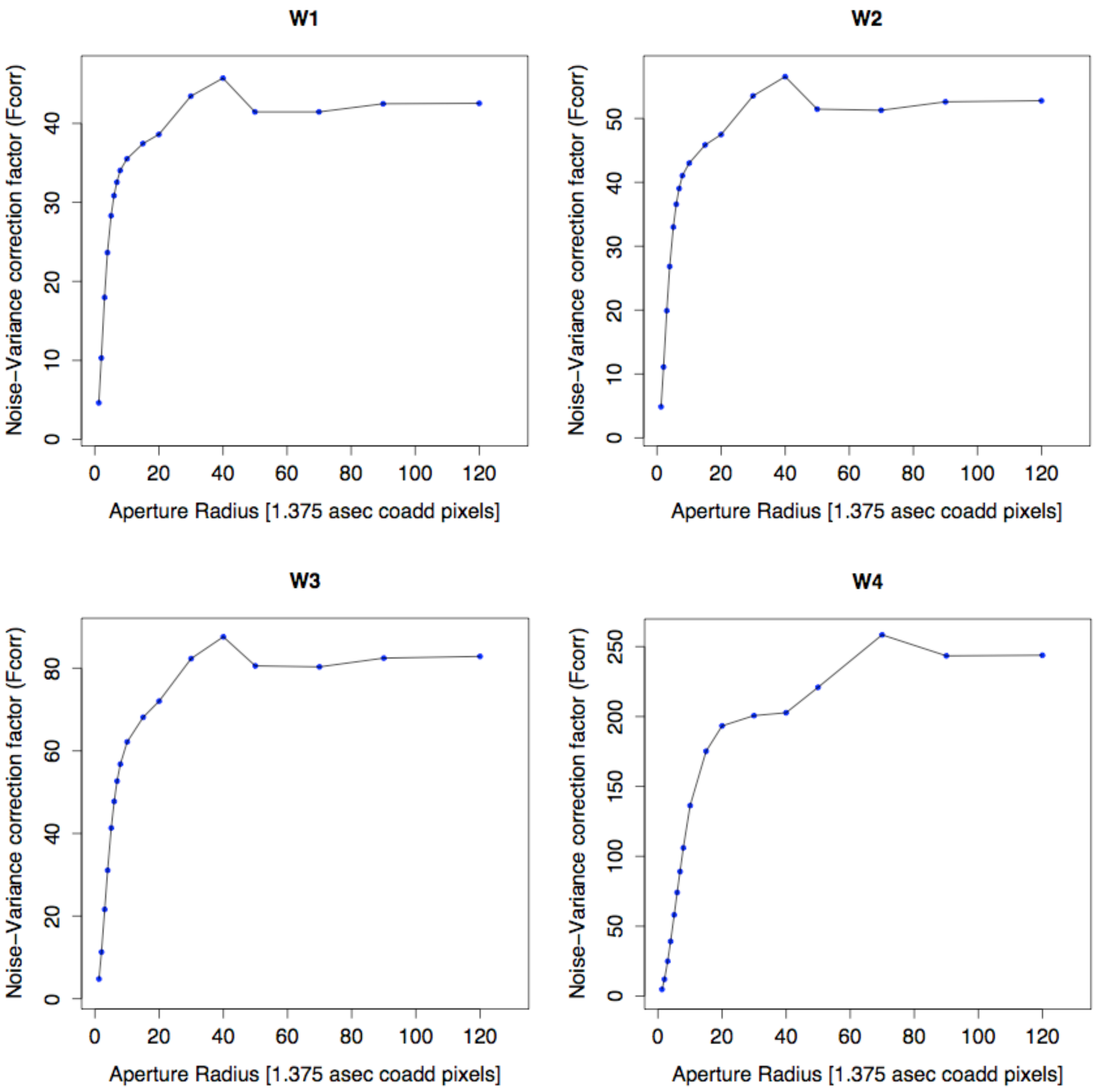

**Figure 23: Correction factor $F_{corr}$ (for Eq. 25) vs aperture radius for summed pixel-variance for co-adds made using the PRF-interpolation method (§6) for all WISE bands and 1.375 arcsec pixels.**

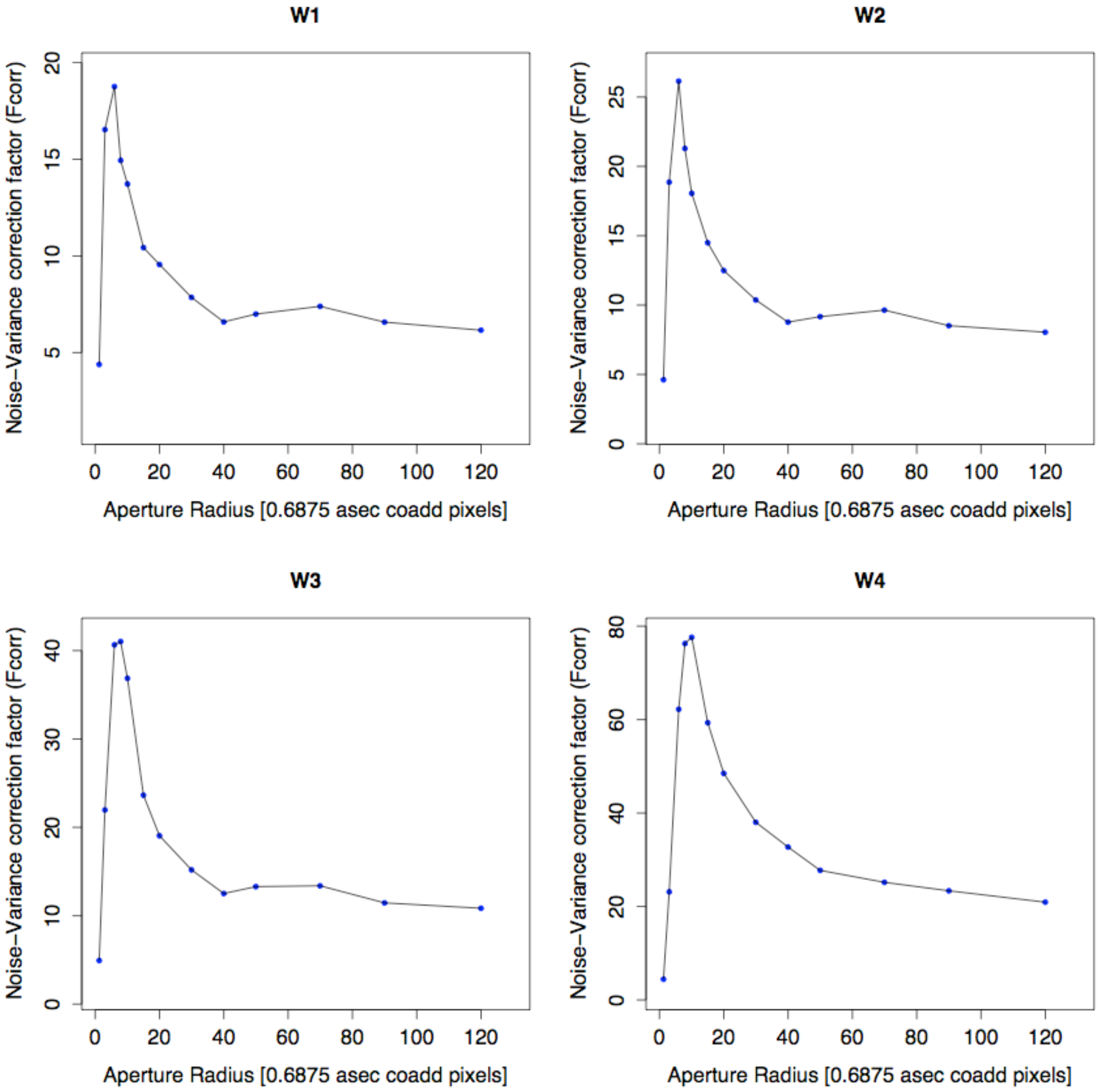

**Figure 24: Correction factor $F_{corr}$ (for Eq. 25) vs aperture radius for summed pixel-variance <u>for HiRes products</u> up to iteration 20 for all <u>WISE</u> bands and 0.6875 arcsec pixels.**

## 14 REFERENCES AND FURTHER READING

- Aumann, H.H., Fowler, J.W., and Melnyk, M., 1990, *A Maximum Correlation Method for Image Construction of IRAS Survey Data*, AJ, **99**, 1674

- Cornwell, T.J., and Evans, K.F., 1985, A&A, 143, 77

- Fowler, J.W., and Aumann, H.H., 1994, *HiRes and Beyond*, in Science with High Spatial Resolution Far-Infrared Data, ed. S. Terebey and J. Mazzarella, (Pasadena, JPL), 1

- Lucy, L.B., 1994, in The Restoration of HST Images and Spectra - II, ed. R.J. Hanisch & R.L. White, 79

- Lucy, L.B., 1974, AJ, 79, 745

- Masci, F.J., and Fowler, J.W., 2009, in ASP Conf. Ser. 411, ADASS XVIII, ed. D.A. Bohlender, D. Durand, & P. Dowler, (San Francisco: ASP), 67: *http://wise2.ipac.caltech.edu/staff/fmasci/awaic_adass08.pdf*

- Shupe, D. L. et al. 2005, in ASP Conf. Ser. 347, ADASS XIV, ed. P. Shopbell, M. Britton, & R. Ebert (San Francisco: ASP), 491: *http://wise2.ipac.caltech.edu/staff/fmasci/figures_codeVdist/SIP_distortion.pdf*

- Wells, B.S., 1997, *Far infrared to radio continuum correlation for nearby star forming galaxies,* Chapter 4: IRAS and the HIRES Simulator.

## 15 REFERENCING ICORE IN PUBLICATIONS

If you use ICORE in your research, we appreciate if you could cite the following paper:

Masci, F.J., and Fowler, J.W., 2009, in ASP Conf. Ser. 411, ADASS XVIII, ed. D.A. Bohlender, D. Durand, & P. Dowler, (San Francisco: ASP), 67

## 16 ACRONYMS

| | |
|---|---|
| 2-D | Two-Dimensional |
| 3-D | Three-Dimensional |
| ADASS | Astronomical Data Analysis Software and Systems |
| ANSI | American National Standards Institute |
| AWAIC | A WISE Astronomical Image Co-adder |
| AWOD | A WISE Outlier Detector |
| Bckgnd | Background |
| Bmatch | Background matching |
| CDR | Critical Design Review |
| CFV | Correction Factor Variance |
| COADD | Co-Adder subsystem |
| COV | depth-of-COVerage |
| CPU | Central Processing Unit |
| CROTA2 | Coordinate ROTAtion about axis 2 (W.of.N convention) |
| DEC | Declination (or Dec) |
| DN | Data Number |
| DRAM | Dynamic Random Access Memory |
| E-W | East-West |
| FDD | Functional Design Document |
| FRD | Functional Requirements Document |
| FITS | Flexible Image Transport System |
| FOV | Field-Of-View |
| FPA | Focal Plane Array |
| FWHM | Full Width at Half Maximum |
| GB | Giga-Byte |
| HIRES | HIgh RESolution (sometimes written as HiRes) |
| HST | Hubble Space Telescope |
| I/O | Input / Output |
| IC | Index Catalog |
| ICal | Instrumental Calibration |
| ICORE | Image Co-addition with Optional Resolution Enhancement |
| ICRS | International Celestial Reference System |
| INT | INTensity |
| IPAC | Infrared Processing and Analysis Center |
| IR | Infra-Red |
| IRAC | Infra-Red Array Camera |
| IRAS | Infra-Red Astronomical Satellite |
| IRSA | NASA/IPAC Infra-Red Science Archive |
| ISO | International Organization for Standardization |
| JPL | Jet Propulsion Laboratory |
| JPEG | Joint Photographic Experts Group format |
| Jy | Jansky |
| MAD | Median Absolute Deviation |
| MAGZP | MAGnitude Zero Point |
| MAGZPUNC | MAGnitude Zero Point UNCertainty |
| MB | Mega-Byte |
| MCM | Maximum Correlation Method |
| MED | MEDian |
| M33,M51 | Messier objects 33 and 51 |
| MJy | MegaJansky |
| MOPEX | Mosaicking and Point source EXtraction |
| N-S | North-South |
| NaN | Not-a-Number |
| N.B. | Nota Bene (Note Well!) |
| NEP | North Ecliptic Pole |
| NGC | New General Catalog |
| ODET | Outlier Detection |
| PA | Position Angle (E.of.N convention) |
| PDL | Perl Data Language |
| PRF | Point Response Function |
| PSF | Point Spread Function |
| PTILE | PercentTILE |
| QA | Quality Assurance |
| QAP | Quality Assurance Plan |
| RA | Right Ascension |
| RAM | Random Access Memory |
| RL | Richardson-Lucy algorithm |
| RMS | Root-Mean-Square fluctuation |
| RSS | Root-Sum-Squared |
| SAA | South Atlantic Anomaly |
| SDS | Subsystem Design Specification |
| SEP | South Ecliptic Pole |
| SIP | Simple Imaging Polynomial |
| SIS | Subsystem Interface Specification |
| S/N | Signal-to-Noise |
| SNR | Signal-to-Noise Ratio (same as S/N) |
| SPIRE | Spectral and Photometric Imaging REceiver (on Herschel) |
| sr | Steradian |
| SUTR | Sample-Up-The-Ramp |
| SVB | Slowly Varying Background |
| SVG | Scalable Vector Graphics format |
| TBD | To Be Determined |
| TBR | To Be Resolved |
| Tmatch | Throughput matching |
| 2MASS | Two Micron All Sky Survey |
| UNC | UNCertainty |
| WCS | World Coordinate System |
| W? | WISE band number: ? = 1, 2, 3, or 4 (~3.4, 4.6, 12.1, 22.2 μm for $\nu f_\nu$=constant) |
| WISE | Wide-field Infrared Survey Explorer |
| WSDC | WISE Science Data Center |
| WSDS | WISE Science Data System |

## Acknowledgments

The author is indebted to John Fowler for his enthusiasm, amusing stories from the IRAS days, and for convincing me that the MCM-HiRes algorithm is worth implementing *now* because CPUs can only get faster. This work was carried out at the California Institute of Technology, with funding from the National Aeronautics and Space Administration, under contract to the Jet Propulsion Laboratory.